\documentclass[reprint,pre,twocolumn,footinbib,aps,superscriptaddress,nobalancelastpage]{revtex4-2}

\usepackage[usenames, dvipsnames ]{xcolor}
\usepackage{amsmath, amsfonts, amssymb, amsthm, amstext, bm, bbm, multirow}
\usepackage[colorlinks=true, linkcolor=blue, citecolor=red, 
urlcolor=blue, anchorcolor = blue, filecolor = blue, hypertexnames=true]{hyperref}
\usepackage{graphicx, caption, subcaption}
\usepackage{ragged2e}
\DeclareCaptionJustification{justified}{\justifying}
\newcommand{\bec}[1]{\mbox{\boldmath $ #1$}}
\newcommand{\Fig}[1]{Fig.~\ref{#1}}
\newcommand{\Figs}[2]{Figs.~\ref{#1} and \ref{#2}}
\newcommand{\Eq}[1]{Eq.~(\ref{#1})}
\newcommand{\Eqs}[2]{Eqs.~(\ref{#1}) and~(\ref{#2})}

\newcommand{\mnras}{Mon. Not. R. Astron. Soc.}
\newcommand{\apjl}{Astrophys. J. Lett.}
\newcommand{\aap}{Astron. Astrophys.}
\newcommand{\pasj}{PASJ}
\newcommand{\physrep}{Phys. Rep.}
\newcommand{\araa}{ARA\&A}

\begin{document}

\title{Unified treatment of mean-field dynamo and angular-momentum transport in magnetorotational instability-driven turbulence}

\author{Tushar Mondal}
\email{tushar.mondal@icts.res.in}
\affiliation{International Centre for Theoretical Sciences, Tata Institute of Fundamental Research, Bangalore 560089, India}
\author{Pallavi Bhat}
\email{pallavi.bhat@icts.res.in}
\affiliation{International Centre for Theoretical Sciences, Tata Institute of Fundamental Research, Bangalore 560089, India}


\begin{abstract}
Magnetorotational instability (MRI)-driven turbulence and dynamo phenomena are analyzed using direct statistical simulations. Our approach begins by developing a unified mean-field model that combines the traditionally decoupled problems of the large-scale dynamo and angular-momentum transport in accretion disks. The model consists of a hierarchical set of equations, capturing up to the second-order cumulants, while a statistical closure approximation is employed to model the three-point correlators. We highlight the web of interactions that connect different components of stress tensors---Maxwell, Reynolds, and Faraday---through shear, rotation, correlators associated with mean fields, and nonlinear terms. We determine the dominant interactions crucial for the development and sustenance of MRI turbulence. Our general mean field model for the MRI-driven system allows for a self-consistent construction of the electromotive force, inclusive of inhomogeneities and anisotropies. Within the realm of large-scale magnetic field dynamo, we identify two key mechanisms---the rotation-shear-current effect and the rotation-shear-vorticity effect---that are responsible for generating the radial and vertical magnetic fields, respectively. We provide the explicit (nonperturbative) form of the transport coefficients associated with each of these dynamo effects. Notably, both of these mechanisms rely on the intrinsic presence of large-scale vorticity dynamo within MRI turbulence.
\end{abstract}

\maketitle

\section{Introduction}

The origin of angular momentum transport is a central problem in accretion disk theory. It is now widely accepted that magnetorotational instability (MRI) \citep{1998RvMP...70....1B} is responsible for generating turbulent motions and facilitating the outward transport of angular momentum in accretion disks. For MRI to manifest, the disk must possess sufficient ionization levels to allow for effective coupling with magnetic field lines. In its original form, it appears as a linear instability in differentially rotating flows threaded by vertical magnetic fields. However, a purely toroidal field is also capable of initiating an instability \citep{1992ApJ...400..610B}. The MRI continues to operate in the nonlinear regime, and eventually leads to a fully nonlinear, turbulent state \citep{1995ApJ...440..742H}.

In general, MRI-driven magnetohydrodynamic (MHD) turbulence requires sufficiently coherent magnetic fields 
for sustenance \citep{2017MNRAS.472.2569B}. For a given initial magnetic field configuration, the MRI can be initiated locally, but it has the opportunity to dissipate the large-scale fields via the generated turbulence, which then affects the ability of MRI to further sustain the turbulence. To perpetuate the turbulent motions, one needs to regenerate and sustain large-scale magnetic fields against dissipation through a dynamo mechanism \citep{1995ApJ...446..741B, 1996ApJ...464..690H, 2008A&A...488..451L, 2010MNRAS.405...41G}. Since the discovery of MRI, numerous direct numerical simulations (DNS), both local \citep{1995ApJ...440..742H, 2010ApJ...713...52D, 2016MNRAS.456.2273S, 2016MNRAS.462..818B, 2022MNRAS.517.2639Z} and global \citep{2011ApJ...736..107O, 2011MNRAS.416..361B, 2013ApJ...772..102H, 2013MNRAS.435.2281P, 2018ApJ...861...24H, 2020MNRAS.494.4854D}, have confirmed the sustenance of MRI turbulence along with the coexistence of large-scale magnetic fields. These studies have demonstrated that turbulent angular momentum transport is primarily driven by the correlated magnetic fluctuations (Maxwell stress) rather than their kinetic counterpart (Reynolds stress).

Various physically motivated models have been developed to describe the mechanism of angular momentum transport in accretion disks. Notably, Kato and Yoshizawa \cite{1995PASJ...47..629K} and Ogilvie \cite{2003MNRAS.340..969O} derived a set of closed dynamical equations describing the evolution of the mean Reynolds and Maxwell stress tensors in a rotating shear flow, assuming the absence of any mean magnetic fields. For the MRI to be operative, Pessah, Chan and Psaltis \cite{2006PhRvL..97v1103P} developed a local model for the dynamical evolution of the Reynolds and Maxwell tensors in a differentially rotating flow, threaded by a mean vertical magnetic field. All of these models successfully capture the initial exponential growth and subsequent saturation of the Reynolds and Maxwell stresses. However, an important limitation of these models is the assumption that the Faraday tensor, denoted as $\bar F_{ij} \equiv \langle u_i b_j \rangle$, vanishes, thereby resulting in the absence of mean magnetic field generation. Here, $u$ and $b$ represent the fluctuating components of velocity and magnetic field, respectively. While these studies represent valuable foundations for understanding the MRI mechanism, particularly the angular momentum transport phenomena, they fall short of capturing the practical aspects of MRI-driven turbulence, primarily due to the neglect of large-scale dynamos.

In principle, large-scale magnetic fields are expected to emerge through the stretching and twisting of field lines by small-scale turbulence. A commonly adopted framework to study large-scale dynamos is mean-field electrodynamics \citep{1978mfge.book.....M, 1980opp..bookR....K, 2005PhR...417....1B}. Within this framework, the evolution of mean magnetic fields is described in terms of transport coefficients derived from statistically averaged properties of small-scale velocity and magnetic fields. A prominent mechanism responsible for the amplification of large-scale magnetic fields is known as the $\alpha$-effect \citep{1978mfge.book.....M, 1979cmft.book.....P, 1980opp..bookR....K}, where the small-scale turbulence generates an electromotive force (EMF, represented by $\mathcal{\bar E}$ ) that is directly proportional to large-scale magnetic fields, $\mathcal{\bar E}_i = \alpha_{ij} \bar B_j$. For the $\alpha$-effect to operate effectively, the turbulence must break statistical symmetry in some way, either through the presence of a net helicity or through stratification and rotation. 
An important part of the dynamo mechanism is the $\Omega$-effect, which arises from the presence of large-scale velocity shear commonly found in astrophysical systems under the influence of gravitational forces. In the $\Omega$-effect, shear stretches the mean magnetic fields, facilitating their amplification and evolution. Specifically, the $\Omega$-effect converts a radial field component into a toroidal one. However, one of the most challenging aspects of mean-field theory is to close the dynamo cycle through the sustained generation of poloidal fields (both radial and vertical components) by mechanisms that require a detailed understanding. In this context, the traditional $\alpha$-effect has demonstrated great success in flows that lack reflectional symmetry, and its existence has been well established through numerical simulations of helically forced turbulence. 
While stratified shearing-box simulations of MRI-driven turbulence also show some support for an $\alpha-\Omega$ dynamo, it is possible that a different mechanism is more fundamental to the evolution of large-scale magnetic fields in accretion disks \citep{2015ApJ...810...59G, 2016MNRAS.456.2273S, 2022A&A...659A..91W, 2022MNRAS.517.2639Z}.

In fact, studies conducted on unstratified, zero-net flux simulations of MRI turbulence have revealed the presence of large-scale dynamo action in the absence of an $\alpha$-effect \citep{2016MNRAS.456.2273S, 2022A&A...659A..91W, 2022MNRAS.517.2639Z}. Moreover, in different contexts, it has been demonstrated that the combined effects of shear and turbulent rotating convection can give rise to large-scale dynamo action, where the driving mechanism is different from the classical $\alpha$-effect \citep{2009PhRvL.102d4501H}. In the context of the shear dynamo, Yousef et al. \cite{2008PhRvL.100r4501Y} showed that forced small-scale nonhelical turbulence in non-rotating linear shear flows can lead to the exponential growth of large-scale magnetic fields. These findings highlight the importance of considering alternative dynamo mechanisms beyond the traditional $\alpha$-effect in systems characterized by shear and turbulence, providing insights into the diverse range of processes contributing to the generation and amplification of large-scale magnetic fields.

The underlying process of dynamo generation is multifaceted, as several potential mechanisms have been proposed to explain the generation of large-scale fields without a net $\alpha$-effect. One such possibility is the `stochastic $\alpha$-effect' in turbulent flows, where the mean $\alpha$ coefficients are zero. In this approach, sufficiently strong fluctuations of $\alpha$ in interaction with shear can lead to the growth of mean magnetic fields \citep{1997ApJ...475..263V, 2000A&A...364..339S, 2007MNRAS.382L..39P, 2011PhRvL.107y5004H, 2012MNRAS.420.2170M, 2014MNRAS.445.3770S}. Another explanation lies in the `shear-current effect' \citep{2003PhRvE..68c6301R, 2004PhRvE..70d6310R} and the `magnetic shear-current effect' \citep{2015PhRvL.115q5003S}, emerging from the off-diagonal turbulent resistivity in the presence of large-scale velocity shear. In the case of the magnetic shear-current effect, magnetic fluctuations arising from small-scale dynamo action can generate large-scale magnetic fields. A third possibility is the `cross-helicity effect' \citep{1993ApJ...407..540Y, 1998A&A...332L..41B, 2000ApJ...529..138B, 2013GApFD.107..114Y}. This mechanism involves augmenting the induction equation for the mean magnetic field with an inhomogeneous term proportional to the product of cross-helicity and mean vorticity. The interplay of cross-helicity and vorticity provides an additional avenue for the generation and evolution of large-scale magnetic fields.
It has to be noted that stochastic $\alpha$-effect and the original shear-current effect are kinematic in nature and the MRI-driven dynamo is expected to be intrinsically nonlinear. But also in all of the above mentioned mechanisms, the role of rotation has not been considered actively. In differentially rotating turbulent flows, an EMF proportional to ${\bec \Omega}\times (\nabla \times {\bec B})$ can drive dynamo action, referred to as the ${\bec \Omega} \times {\bec J}$ or R\text{\"{a}}dler-effect \cite{1980opp..bookR....K, 1986AN....307...89R}.

In theoretical investigations of the MRI-driven system, it was found that turbulence is not required for large-scale dynamo action \cite{2016MNRAS.462..818B} and the non-normality in the system allows for self-coupling of non-axisymmetric modes \cite{2008AN....329..750R, 2007A&A...463..817R} leading to a nontrivial electromotive force on a quasi-linear analysis \cite{2016MNRAS.459.1422E}.
More recently, a calculation of the triple correlation term in the small scale magnetic helicity equation 
has indicated the possibility of helicity fluxes leading to localization of helicity in space (rather than spectrally) and large-scale vorticity features prominently in the arising new helicity flux \cite{2023ApJ...943...66G}.

Thus far, the mean-field dynamo and angular momentum transport problems have been approached independently in a decoupled manner. The mean-field dynamo theory has traditionally disregarded the transport dynamics, while angular momentum transport theory has overlooked the evolution of large-scale magnetic fields \cite{2010AN....331..101B,2015JPlPh..81e3905B}. However, the large-scale behavior of velocity and magnetic fields depends on the interaction at smaller scales. Moreover, it is crucial to account for the back reaction of the large-scale dynamics on the small-scale environments. This inherent complexity renders both the problems highly nonlinear and poses a substantial challenge in formulating a comprehensive coupled theory for MRI-driven MHD turbulence.

Another route to investigating the dynamo problem in the MRI-driven system (in both stratified and unstratified domains) is via the measurement of turbulent transport co-efficients in the mean field theory \citep{2022ApJ...932....8K,2022MNRAS.511.4454B,2020MNRAS.494.4854D,2016MNRAS.456.2273S,2015ApJ...810...59G,2008ApJ...676..740B}. However, many of these studies use mean-field models based on homogeneous and isotropic small-scale turbulence, which is not justified for an MRI driven system. Others that use a more general mean field model, employ methods for inversion which are either unsuitable for nonlinear systems or set some of the coefficients to zero which is somewhat questionable or have to deal with complexity of correlated noise, nonlocality, degeneracies, overconstraining, etc. By using direct statistical simulations with a general model, we have been able to overcome most of these issues, and are able to directly determine the terms and transport coefficients (and their exact expressions) central to MRI dynamo action.

In this paper, we construct a unified mean-field model that combines dynamo and transport phenomena self-consistently, and perform direct statistical simulations (DSS) in a zero net-flux unstratified shearing box. Our methodology begins by developing a mean-field model that consists of a hierarchical set of equations, capturing up to the second-order cumulants. To close these hierarchical equations, we express the third-order cumulants in terms of second-order cumulants using the CE2.5 statistical closure model, which lies between the second-order (CE2) and third-order cumulant expansion (CE3) methods \citep{2021RSPSA.47710427L}. We also apply the two-scale approach \cite{1975AN....296...49R} to model second-order correlators involving the spatial gradient of a fluctuating field. The general nature of our model allows us to capture the effects of inhomogenieties and anisotropies in the system. Further it has been useful to determine directly the dominant terms/effects responsible for dynamo and transport in tandem. This path allows us to explore the possibilities of developing sub-grid models which can be used in global simulations and possibly also in general relativistic MHD simulations of accretion disks aimed towards understanding data from Event Horizon Telescope \cite{2021ApJ...910L..13E}.
In order to numerically solve this model consisting of coupled equations for the mean field and stress tensors, we develop a special module within the \textsc{Pencil-Code} \cite{2021JOSS....6.2807P} framework.

Our unified mean-field model addresses two key challenges: $(a)$ disentangling the diverse physical processes involved in sustaining MRI turbulence and dynamo activity in accretion disks, and $(b)$ identifying the dominant mechanisms responsible for generating large-scale magnetic fields. We present a comprehensive framework that elucidates the intricate network of interactions connecting different mean fields and stress components through shear, rotation, correlators associated with mean fields, and nonlinear terms. By studying the induction equation, we investigate the role of different EMFs in the evolution of large-scale magnetic fields. Our DSS results demonstrate good agreement with those obtained from DNS \citep{2008A&A...488..451L, 2016MNRAS.462..818B}. Specifically, the radial EMF has a resistive effect, reducing the energy of the azimuthal field. The azimuthal EMF, $\mathcal{\bar E}_y$, generates a radial field, which, in turn, drives the azimuthal field through the $\Omega$-effect. Notably, $\mathcal{\bar E}_y$ is also responsible for generating the vertical magnetic field. 
Next, with our model, we construct the EMF for an MRI driven system. We find that the constructed EMF is a linear combination of not only the usual terms proportional to mean magnetic fields and the gradient of mean magnetic fields, but also the gradient of mean velocity fields, and a nonlinear term. Thus, our EMF is not just an ansatz but an expression that naturally arises out of our model. The proportionality coefficients depend on shear, rotation, and statistical correlators associated with fluctuating fields. In our search for large-scale magnetic field dynamo, we identify two crucial mechanisms---the ``\textit{rotation-shear-current effect}'' and the ``\textit{rotation-shear-vorticity effect}''---that are responsible for generating the radial and vertical magnetic fields, respectively. Remarkably, both of these mechanisms rely on the presence of large-scale velocity dynamo in the self-sustaining MRI-driven turbulence.

The paper is organized as follows. In Section~\ref{sec:model} and its subsections, we present our unified mean-field model for MRI-driven turbulence and dynamo. Section~\ref{sec:closure_model} discusses the requirements of a high-order closure model and provides a statistical closure model to facilitate our analysis. The numerical simulation set-up is described in Section~\ref{sec:simulation set-up}. In Section~\ref{sec:results} and its subsections, we present the results obtained from the simulations. Section~\ref{sec:turbulent_transport} focuses on phenomena associated with the outward transport of angular momentum and the interplay of various Maxwell and Reynolds stresses. The role of large-scale magnetic fields in turbulent transport is also highlighted. Section~\ref{sec:Large-scale dynamo} addresses the large-scale dynamo associated with magnetic fields. Different planar-averaged large-scale fields are distinguished, and the role of various terms in generating mean magnetic fields is analyzed using planar-averaged induction equations. The construction of the EMF for an MRI-driven system is discussed in Section~\ref{sec:EMF}, followed by a detailed analysis of the dynamo mechanisms responsible for generating radial and vertical large-scale magnetic fields in Sections~\ref{sec:radial_B_fied} and \ref{sec:vertical_B_fied}, respectively. In Section~\ref{sec:discussion}, we discuss our newly discovered dynamo mechanisms, namely the rotation-shear-current effect and the rotation-shear-vorticity effect, and provide a comprehensive discussion comparing them with existing dynamo mechanisms. Finally, the paper concludes with a summary in Section~\ref{sec:conclusion}.

\section{Model}
\label{sec:model}

We adopt the local shearing box model to investigate MRI turbulence and dynamo in a three-dimensional, zero net-flux configuration, employing novel DSS methods.  In order to simplify the analysis, we assume an isothermal, unstratified, and weakly compressible fluid. The DSS method has proven to be a valuable computational technique and has shown promising results. However, applying the DSS method to study self-sustaining MRI-driven turbulence presents a significant challenge compared to forced turbulence, primarily due to the complexity of turbulent flows. The presence of an unlimited number of statistical properties that cannot be directly calculated from first principles complicates the analysis. Furthermore, a closure model is necessary to handle the high-order nonlinear terms of the statistically-averaged equations. Our investigation begins with a statistical averaging approach applied to the standard MHD equations, which describe the mean flow and flow statistics in an accretion disk. First, we write down the standard MHD equations in a shearing background in the rotating frame, as given by
\begin{equation}
	\frac{\mathcal{D} {\bec A} }{ \mathcal{D} t } = -SA_y \hat{x} + {\bec U} \times {\bec B} - \eta \mu_0 \bec J \;,
	\label{eq:induction}
\end{equation}
\begin{multline}
	\frac{\mathcal{D} {\bec U} }{ \mathcal{D} t } = - ({\bec U } \cdot \nabla )
	{\bec U} -SU_x \hat{y} - \frac{1}{\rho} \nabla P
	+ \frac{1}{\rho} {\bec J } \times {\bec B} \\
	- 2 {\bec \Omega} \times {\bec U} 
	+ \frac{1}{\rho} \nabla \cdot 2\nu \rho \mathbf{S} \;,
	\label{eq:navier_stokes}
\end{multline}
\begin{equation}
	\frac{\mathcal{D} {\ln \rho} }{ \mathcal{D} t } = - ({\bec U} \cdot \nabla) \ln \rho - \nabla \cdot {\bec U} \;.
	\label{eq:continuity}
\end{equation}
Here, $\mathcal{D}/\mathcal{D}t \equiv \partial / \partial t - q\Omega x \partial / \partial y$ includes the advective transport by a uniform shear flow, ${\bec U^0} = -q\Omega x \hat{y}$. ${\bec \Omega }= \Omega \hat{z}$ is the background rotational velocity. The constant $q \equiv - d\ln \Omega / d\ln R$; for a Keplerian disk $q=3/2$. The magnetic field $\bec B$ is related to the magnetic vector potential $\bec A$ by $\bec B = \nabla \times \bec A$, and $\bec J = \nabla \times \bec B/ \mu_0$ is the current density, where $\mu_0$ is the vacuum permeability. 
The constraint $\nabla \cdot \bec B = 0$ is enforced by solving the evolution equation for $\bec A$ \cite{1995ApJ...446..741B, 2020Galax...8...68Y}.
The other quantities have their usual meanings: $\bec U$ is the velocity, $P$ the pressure, $\rho$ the density, $\eta$ the magnetic diffusivity, $\nu$ the microscopic viscosity, and $\mathbf{S}_{ij}=\frac{1}{2}(\bec U_{i,j}+\bec U_{j,i} - \frac{2}{3} \delta_{ij} \nabla \cdot \bec U )$ the rate of strain tensor. We use an isothermal equation of state $P = \rho c_s^2$, characterized by a constant sound speed, $c_s$.

In the conventional mean-field theory, one solves the Reynolds averaged equations. We thus consider a Reynolds decomposition of the dynamical flow variables, expressing them as the sum of a mean component (denoted by over-bars) and a fluctuating component (represented by small letters): ${\bec U} = \bar{U} + u , \ {\bec A} = \bar{A} + a $, and so on. It satisfies the Reynolds averaging rules, i.e., $\bar{a} = 0, \ \bar{\bar{A}} = \bar{A}$. Here, we consider the ensemble averaging to derive the cumulant equations. For weakly compressible fluids, where the density remains approximately constant, the mean-field equations in ensemble averaging can be written as:
\begin{equation}
	\mathcal{D}_t \bar A_i = \bar S_i^A + \epsilon_{ijk} \bar U_j \bar B_k  + \mathcal{\bar E}_i - \eta \bar J_i \; ,
	\label{eq:meanA}
\end{equation}
\begin{multline}
	\mathcal{D}_t \bar U_i = - \bar U_j\partial_j \bar U_i + \bar S_i^U - 2\epsilon_{ijk}\Omega_j\bar U_k -\frac{1}{\rho}\partial_i \bar P \\
	+ \frac{1}{\rho} \epsilon_{ijk} \bar J_j \bar B_k +  \frac{1}{\rho}\partial_j(\bar M_{ij}-\bar R_{ij}) - \frac{1}{2\rho}\partial_i \bar M + \nu\partial_{jj}\bar U_i \;,
	\label{eq:meanU}
\end{multline}
\begin{equation}
	\frac{\mathcal{D} {\ln \rho} }{ \mathcal{D} t } = - ({\bec {\bar U}} \cdot \nabla) \ln \rho - \nabla \cdot {\bec {\bar U}} \;.
	\label{eq:mean_continuity}
\end{equation}
Here, $ \bar S^A = (-S \bar A_y,0,0)$, and $ \bar S^U = (0,-S \bar U_x, 0)$. $\mathcal{\bar E}_i = \left<u\times b \right>_i = \epsilon_{ijk} \bar F_{jk}$ is the mean electromotive force. $M_{ij}=b_i b_j / \mu_0$, $R_{ij}=\rho u_i u_j$, and $F_{ij}=u_i b_j$ are the Maxwell, Reynolds, and Faraday tensors, respectively. The effect of turbulence on the mean-field evolution is captured through the mean stress tensors $\bar M_{ij}$, $\bar R_{ij}$, and $\bar F_{ij}$. We require knowledge of the evolution of such stress tensors to close the mean-field equations (\ref{eq:meanA}) and (\ref{eq:meanU}). 
	
By subtracting the ensemble-averaged equation from the total equation, we derive the evolution equations for the fluctuating velocity and magnetic fields:
\begin{eqnarray}
	\mathcal{D}_t u_i &=& -\bar U_j\partial_j u_i -u_j\partial_j \bar U_i - 2\epsilon_{ijk}\Omega_j u_k \nonumber\\
	&&+ \frac{1}{\mu_0 \rho}(\bar B_j\partial_j b_i + b_j\partial_j\bar B_i) - \frac{1}{\rho}\partial_i \Pi' \nonumber\\
	&&+ \frac{1}{\rho}\partial_j(M_{ij}-R_{ij}-\bar M_{ij}+\bar R_{ij}) 
	 + \nu\partial_{jj} u_i \;,
	\label{eq:fluctuatingU} \\
	\mathcal{D}_t b_i  &=& \bar B_j\partial_j u_i + b_j\partial_j\bar U_i -\bar U_j\partial_j b_i - u_j\partial_j\bar B_i \nonumber\\
	&& + \partial_j (F_{ij}-F_{ji}-\bar F_{ij}+\bar F_{ji}) +\eta\partial_{jj} b_i \;.
	\label{eq:fluctuatingB}
\end{eqnarray}
We combine different fluctuating equations and apply Reynolds average rule to construct the governing equations for the $\bar M_{ij}$, $\bar R_{ij}$, and $\bar F_{ij}$. The resulting equations for the mean Maxwell, Reynolds, and Faraday tensors are, respectively,
\begin{multline}
	\mathcal{D}_t \bar M_{ij} + \bar U_k\partial_k \bar M_{ij} -
	\bar M_{ik}\partial_k\bar U_j - \bar M_{jk}\partial_k\bar U_i + \bar S_{ij}^M \\
	+ \frac{1}{\mu_0} \left(\bar F_{ki} \partial_k \bar B_j + \bar F_{kj} \partial_k \bar B_i \right) \\
	= \frac{1}{\mu_0}\Big[ \bar B_k \langle b_i\partial_k u_j+b_j\partial_k u_i \rangle  
	+ \mathcal{\bar T}^M_{ij}
	+ \eta\langle b_i\partial_{kk}b_j+b_j\partial_{kk}b_i\rangle \Big] \; ,
	\label{eq:Mij_exact}
\end{multline}
\begin{multline}
	\mathcal{D}_t \bar R_{ij} + \bar U_k\partial_k \bar R_{ij} + \bar R_{ik}\partial_k\bar U_j + \bar R_{jk}\partial_k\bar U_i + \bar S_{ij}^R  \\
	+ 2\epsilon_{jkl}\Omega_k\bar R_{il} + 2\epsilon_{ikl}\Omega_k\bar R_{jl}
	- \frac{1}{\mu_0} \left(\bar F_{ik} \partial_k \bar B_j + \bar F_{jk} \partial_k \bar B_i \right) \\
	= - \langle u_i\partial_j \Pi' + u_j\partial_i \Pi' \rangle 
	+\frac{1}{\mu_0}\Big[ \bar B_k \langle u_i\partial_k b_j+u_j\partial_k b_i \rangle \Big] 
	+ \mathcal{\bar T}^R_{ij} \\
	+ \rho\nu\langle u_i\partial_{kk}u_j+u_j\partial_{kk}u_i\rangle \; ,
	\label{eq:Rij_exact}
\end{multline}
\begin{multline}
	\mathcal{D}_t \bar F_{ij} + \bar U_k\partial_k \bar F_{ij} - \bar F_{ik}\partial_k\bar U_j + \bar F_{kj}\partial_k\bar U_i + 2\epsilon_{ikl}\Omega_k\bar F_{lj} \\
	+ \bar S_{ij}^F - \frac{1}{\rho}( \bar M_{jk}\partial_k\bar B_i-\bar R_{ik}\partial_k\bar B_j) \\ = -\frac{1}{\rho}\langle b_j\partial_i\Pi'\rangle + \bar B_k \langle u_i\partial_k u_j \rangle  
	+ \frac{\bar B_k}{\mu_0 \rho}  \langle b_j\partial_k b_i \rangle + \mathcal{\bar T}^F_{ij} \\
	+ \eta \langle u_i\partial_{kk}b_j \rangle + \nu \langle b_j\partial_{kk}u_i\rangle. 
	\label{eq:Fij_exact} 
\end{multline}
The left-hand side of these equations describes the linear dynamics of the respective stress tensors. The terms $\bar S_{ij}^M$, $\bar S_{ij}^R$, and $\bar S_{ij}^F$ represent how the Maxwell, Reynolds, and Faraday tensors are `stretched' by the gradients of the background shear flow, ${\bec U^0}$, respectively. They are expressed as $\bar S_{ij}^M = - \bar M_{ik}\partial_k\bar U^0_j - \bar M_{jk}\partial_k\bar U^0_i $, $\bar S_{ij}^R = \bar R_{ik}\partial_k\bar U^0_j + \bar R_{jk}\partial_k\bar U^0_i $, and $\bar S_{ij}^F =  - \bar F_{ik}\partial_k\bar U^0_j + \bar F_{kj}\partial_k\bar U^0_i $. Note that different stress tensors interact with the background velocity gradient in distinct ways. The terms $\mathcal{\bar T}^M_{ij}$, $\mathcal{\bar T}^R_{ij}$, and $\mathcal{\bar T}^F_{ij}$ represent the nonlinear three-point terms that appear in the evolution equations for the Maxwell, Reynolds, and Faraday tensors, respectively. Mathematically, they are given by 
\begin{subequations}
	\begin{equation}
	\mathcal{\bar T}^M_{ij} = \langle b_i b_k \partial_k u_j + b_j b_k \partial_k u_i - u_k \partial_k M_{ij} \rangle \; ,
	\label{eq: 3p_Mij}
	\end{equation}
	\begin{equation}
	\mathcal{\bar T}^R_{ij} = \langle u_i b_k \partial_k b_j + u_j b_k \partial_k b_i - u_k \partial_k R_{ij} \rangle \; ,
	\label{eq: 3p_Rij}	
	\end{equation}
	\begin{equation}
	\mathcal{\bar T}^F_{ij} = \langle u_i b_k \partial_k u_j + b_j b_k \partial_k b_i - u_k \partial_k F_{ij} \rangle \; .
	\label{eq: 3p_Fij}	
	\end{equation}
\end{subequations}
	
The right-hand side of the stress equations (\ref{eq:Mij_exact}--\ref{eq:Fij_exact}) poses significant challenges due to the presence of four distinct types of terms. These terms include $(a)$ the triple correlation of fluctuating quantities, $(b)$ second-order correlations involving the spatial gradient of a fluctuating field, $(c)$ pressure-strain correlators in the evaluation equations for $\bar R_{ij}$ and $\bar F_{ij}$, and $(d)$ terms associated with the microscopic diffusion process. It is essential to develop closure models for each of these challenging terms to make progress in our analysis.
	
\subsection{The Closure Model} \label{sec:closure_model}
	
In the framework of a cumulant hierarchy, the expansion of the MHD equations (\ref{eq:induction}--\ref{eq:navier_stokes}) results in an infinite set of coupled partial differential equations. Due to the quadratic nonlinearities present in the standard MHD equations, the first-order cumulant equations for the coherent components (Eqs.~\ref{eq:meanA}--\ref{eq:meanU}) involve terms that are second-order, such as the Maxwell, Reynolds, and Faraday tensors. Similarly, the second-order cumulant equations (Eqs.~\ref{eq:Mij_exact}--\ref{eq:Fij_exact}) contain terms up to third order, and so on. Therefore, in order to make progress in the analysis, it is necessary to select an appropriate statistical closure that truncates the cumulant expansion at the lowest feasible order.

Among the well-studied formalisms in DSS, the truncation of the cumulant hierarchy at second order (CE2) stands out as a simple yet effective approach \citep{2008JAtS...65.1955M}. In CE2, all statistics of order greater than two are zero. This truncation scheme selectively preserves the mean-eddy interactions in the eddy (or fluctuation) equations and the eddy-eddy interactions in the mean equations, while disregarding the eddy-eddy interactions in the eddy equations. Consequently, CE2 is considered to be weakly nonlinear or quasilinear. From a theoretical perspective, CE2 can be interpreted as the exact solution of a linear model driven by stochastic forces. This method has been successfully applied to study MRI turbulence and dynamo in the zero net-flux unstratified shearing box \citep{2015PhRvL.114h5002S}. In this approach, the mean fields are assumed to depend solely on the vertical coordinate, thereby simplifying the system representation. The nonlinearity that is neglected in CE2 is approximated by incorporating white-in-time driving noise, allowing for the exploration of essential aspects of MRI turbulence and dynamo effects.

The third-order cumulant expansion (CE3) includes the eddy-eddy interactions in the eddy equations. However, extending the analysis to third order and beyond presents technical challenges in deriving and solving the DSS system as it involves numerous interactions. To address this complexity, a simplified model called the CE2.5 approximation has been proposed as a practical alternative. The CE2.5 approximation makes several key assumptions to simplify the analysis. First, it sets all time derivatives for the third cumulants to zero, assuming that the third cumulant evolves more rapidly compared to the first and second cumulants. Second, the CE2.5 approximation neglects all terms in the equations for the third cumulant that involve the first-order cumulants. Finally, the fourth-order cumulants are replaced by an eddy-damping parameter or a diffusion process.
	
In our statistical closure model for the three-point interactions, we employ an approach inspired by the CE2.5 approximation. The nonlinear three-point terms (Eqs.~\ref{eq: 3p_Mij}--\ref{eq: 3p_Fij}) can be expressed as (see Appendix~\ref{sec: app_three_point}):
\begin{subequations}
	\begin{align}
	\mathcal{\bar T}^M_{ij} = \frac{1}{L}\Bigg[& 2c_1 \sqrt{\bar M} \bar R_{ij} - 2c_2 \sqrt{\bar M} \bar M_{ij} - 2c_3 \sqrt{\bar R} \bar M_{ij}  \nonumber \\
	& - c_4 \sqrt{\bar F} \left( \bar F_{ij}+\bar F_{ji} \right)\Bigg] ,
	\label{eq: 3p_Mij_approx}	 \\
	\mathcal{\bar T}^R_{ij} = \frac{1}{L}\Bigg[& 2c_2 \sqrt{\bar M} \bar M_{ij} - 2c_1 \sqrt{\bar M} \bar R_{ij} - 2c_5 \sqrt{\bar R} \bar R_{ij} \ \nonumber \\
	 & - c_4 \sqrt{\bar F} \left( \bar F_{ij}+ \bar F_{ji} \right) - c_6 \sqrt{\bar R} \left(\bar R_{ij} - \frac{1}{3} \bar R \delta_{ij} \right) \Bigg] ,
	\label{eq: 3p_Rij_approx}	\\
	\mathcal{\bar T}^F_{ij} = \frac{1}{L}\Bigg[& 2c_7 \sqrt{\bar M} \bar F_{ji} - 2c_8 \sqrt{\bar M} \bar F_{ij} \nonumber \\
	& - 2c_9 \sqrt{\bar R}\bar F_{ij}  
	- c_{10} \sqrt{\bar F} \left( \bar M_{ij}+\bar R_{ij} \right) \nonumber \\
	& - \frac{c_{11}}{2} \sqrt{\bar F} \left(\bar F_{ij} +\bar F_{ji} - \frac{2}{3} \bar F \delta_{ij} \right) \Bigg] ,
	\label{eq: 3p_Fij_approx}	
	\end{align}
\end{subequations}
where, $c_1, \ldots, c_{11}$ are positive dimensionless constants of the order of unity, and $L$ represents a vertical characteristic length (such as the disk thickness or the height of the simulation box). Throughout our computations, we have set $c_1 = \cdots = c_{11} =1$. The quantities $\bar M$ and $\bar R$ denote the traces of the Maxwell and Reynolds tensors, respectively, while $\bar F = (\bar F_{xx}^2 + \bar F_{yy}^2 + \bar F_{zz}^2)^{1/2}$. It is important to highlight that the last term in the $\mathcal{\bar T}^R_{ij}$ equation (Eq.~\ref{eq: 3p_Rij_approx}), involving the constant $c_6$, and the last term in the $\mathcal{\bar T}^F_{ij}$ equation (Eq.~\ref{eq: 3p_Fij_approx}), involving the constant $c_{11}$, correspond to the isotropization terms arising from the pressure-strain nonlinearity \cite{1995PASJ...47..629K, 1997PASJ...49..659N, 2003MNRAS.340..969O}. The pressure-strain correlation is a third-order quantity, appearing in \Eqs{eq:Rij_exact}{eq:Fij_exact}, and necessitates a non-deductive closure. Henceforth, we incorporate them into the three-point correlators, $\mathcal{\bar T}_{ij}$. Furthermore, various other terms arising from the three-point correlators have significant implications. The terms $c_3, c_5,$ and $c_9$ correspond to the turbulent dissipation of the Maxwell, Reynolds, and Faraday tensors, respectively. The terms $c_1$ and $c_2$ represent the interaction between the Maxwell and Reynolds stresses. In terms of energy transfer, the net rate of transfer from turbulent kinetic energy to magnetic energy can be expressed as $L^{-1} \sqrt{\bar M} (c_1 \bar R - c_2 \bar M)$. The sign of this expression determines whether the kinetic or magnetic energy dominates the energy transfer process. Similarly, the terms $c_7$ and $c_8$ describe the interaction between the Faraday tensor and its transpose. Finally, the Faraday tensor interacts with the Maxwell and Reynolds stresses via $c_4$ and $c_{10}$ terms.  

In addition to the three-point correlators, the right-hand side of the stress equations (Eqs.~\ref{eq:Mij_exact}--\ref{eq:Fij_exact}) contains terms that are directly proportional to $\bar B$. These proportionality coefficients correspond to second-order correlators involving the spatial gradient of a fluctuating field. To proceed with our analysis, it is necessary to close these terms. To accomplish this, we employ a two-scale approach \cite{1975AN....296...49R} to determine the second-order correlators associated with the spatial gradient. These correlators can be expressed as (see Appendix~\ref{sec: app_two_scale}): 
\begin{subequations}
	\begin{eqnarray}
	\bar B_m \langle u_i \partial_m b_j \rangle &=& - \mathrm{Tr}({\bf \bar B}) l^{-1} \bar F_{ij} + \frac{1}{2} (\bar {\bf B} \cdot {\nabla}) \bar F_{ij},  \\
	\bar B_m \langle u_j \partial_m b_i \rangle &=& - \mathrm{Tr}({\bf \bar B}) l^{-1} \bar F_{ji} + \frac{1}{2} (\bar {\bf B} \cdot {\nabla}) \bar F_{ji},  \\
	\bar B_m \langle b_i \partial_m u_j \rangle &=& \mathrm{Tr}({\bf \bar B}) l^{-1} \bar F_{ji} + \frac{1}{2} (\bar {\bf B} \cdot {\nabla}) \bar F_{ji},  \\
	\bar B_m \langle b_j \partial_m u_i \rangle &=& \mathrm{Tr}({\bf \bar B}) l^{-1} \bar F_{ij} + \frac{1}{2} (\bar {\bf B} \cdot {\nabla}) \bar F_{ij},  \\
	\bar B_m \langle u_i \partial_m u_j \rangle &=& - \mathrm{Tr}({\bf \bar B}) l^{-1} \bar R_{ij} + \frac{1}{2} (\bar {\bf B} \cdot {\nabla}) \bar R_{ij}, 
	\label{eq: ui_dm_uj}  \\
	\bar B_m \langle b_j \partial_m b_i \rangle &=& \mathrm{Tr}({\bf \bar B}) l^{-1} \bar M_{ij} + \frac{1}{2} (\bar {\bf B} \cdot {\nabla}) \bar M_{ij}. 
	\label{eq: bj_dm_bi}
	\end{eqnarray}
\end{subequations}
Note that \Eqs{eq: ui_dm_uj}{eq: bj_dm_bi} hold for $i \neq j$. Here, we introduce the inverse of length scale as $l^{-1} = s ( \Omega / \sqrt{\bar B^2 / \mu_0 \rho})$, where $s$ is a constant. In our computations, we have set $s = 0.25$, as physical solutions are obtained for $s < 0.3$. However, further studies are required to determine a unique value of $s$ based on DNS results. Similarly, we utilize the two-scale approach to determine terms associated with the microscopic diffusion process present in the stress equations (Eqs.~\ref{eq:Mij_exact}--\ref{eq:Fij_exact}).

\subsection{Simulation Set-up} \label{sec:simulation set-up}
	
We have developed a special module within the framework of the \textsc{Pencil Code} \citep{2021JOSS....6.2807P}, which is a high-order (sixth order in space and third order in time) finite-difference code, to numerically solve the model described by Eqs.~(\ref{eq:meanA})--(\ref{eq:mean_continuity}), and (\ref{eq:Mij_exact})--(\ref{eq:Fij_exact}). This model consists of a set of 28 coupled partial differential equations, involving variables such as $\bar A_i$, $\bar U_i$, $\rho$, $\bar M_{ij}$, $\bar R_{ij}$, and $\bar F_{ij}$. The numerical simulations are performed on a Cartesian grid with dimensions $N_x \times N_y \times N_z$, and a size of $L_x$, $L_y$, and $L_z$ along the three Cartesian directions. For our simulations, we have employed an aspect ratio of $(L_x: L_y: L_z) = (L: L: L)$, with a resolution of $256^3$. The boundary conditions are periodic in the azimuthal ($y$) and vertical ($z$) directions, while being shearing-periodic in the radial ($x$) direction. In the code, all quantities are expressed in dimensionless units, where length is scaled by $L$, velocity by the isothermal sound speed $c_s$, density by the initial value $\rho_0$, magnetic field by $(\mu_0 \rho_0 c_s^2)^{1/2}$, and so on. For convenience, we have set the reference values as $L = \rho_0 = c_s = \mu_0 = 1$.
	
The mean velocity field is initialized with Gaussian random noise, with an amplitude of $10^{-4}$. Similarly, the initial conditions for the stress tensors, namely the Maxwell stress, Reynolds stress, and Faraday stress, are also set as Gaussian random noise with an amplitude of $10^{-4}$, except for the diagonal components of the Maxwell and Reynolds stresses. To preserve the positive definiteness of $\bar M_{ii}$ and $\bar R_{ii}$, we initialize these stress components with positive random noise of amplitude $10^{-4}$.
			
The set-up we have adopted is similar to the one used in Ref.~\cite{2016MNRAS.462..818B}. The initial magnetic field configuration is given by $\bec{\bar B} = \bar B_0 \sin (k_x x) \hat{z}$, which can be written in terms of the vector potential as $\bec{\bar A} = \bar A_0 \cos (k_x x) \hat{y}$, so that the magnitude of the magnetic field is related to the vector potential through $|\bar B_0| = k_x \bar A_0$, where $k_x = 2\pi / L_x$. We choose a rotation rate of $\Omega = 1$ and $\bar A_0 = 0.005$, resulting in $k_{\text{max}} / k_1 = \sqrt{15/16} (\Omega / \bar U_{A,0}) / k_1 \approx 5$. Here, $\bar U_{A,0} = \bar B_0 / \sqrt{\mu_0 \rho_0}$ represents the initial Alfven velocity, $k_{\text{max}}$ corresponds to the wavenumber associated with the maximum growth rate predicted by linear MRI analysis, and $k_1 = 2\pi / L$ is the wavenumber associated with the box size $L$. These choices ensure that the most unstable mode of the MRI, $k_{\text{max}}$, is well resolved by the numerical grid. Additionally, the initial conditions satisfy the condition for the onset of MRI, namely $\beta > 1$, where $\beta = 2\mu_0 P / \bar B_0^2$ is the ratio of thermal to magnetic pressure. In our case, $\beta \simeq 1014$ for the maximum values of the initial magnetic field. With these parameters, the resulting steady-state turbulence driven by the MRI exhibits a characteristic root mean square velocity of $\bar U_{\text{rms}} \sim 0.1 c_s$. Consequently, the Mach number remains of the order of 0.1, ensuring that compressibility effects are negligible. The fluid and magnetic Reynolds number are defined as $\text{Re} \equiv \bar U_{\text{rms}} L / \nu$ and $\text{Rm} \equiv \bar U_{\text{rms}} L / \eta$, respectively, where $\nu$ and $\eta$ represent the microscopic viscosity and resistivity. In our study, we utilize values of $\nu = 3.2\times 10^{-4}$ and $\eta = 8.0 \times 10^{-5}$, yielding a magnetic Reynolds number of $\text{Rm} = 1250$ and a magnetic Prandtl number of $\text{Pm} \equiv \text{Rm} / \text{Re} = 4$.

\section{Results}
\label{sec:results}

We discuss the results from our fiducial statistical simulation of 
the local shearing box MRI system. 
We provide an exposition on the problems of turbulent transport and turbulent large-scale dynamo in different subsections below. 
In each subsection, we first  provide the time evolution of the relevant quantities. 
Then we set out to investigate the sources and sinks involved in the 
evolution of the transport terms or the large-scale fields. 
We show how the terms in the statistical equations compare 
with each other allowing us to deduce the dominant effects. 
In this manner, we establish connections between the mean fields and 
the cumulants. 

In the fiducial simulation used to draw inferences from, the linear stage is up to $t/T_{orb}\sim 5$. 
The total length of the simulation is about $t/T_{orb}\sim 150$, which includes cyclic patterns in 
the evolution of the mean fields and cumulants. However, in this work, we do not address the cyclic behaviour 
as we focus on uncovering the main effects responsible for driving the MRI transport and dynamo.


\subsection{Turbulent Angular Momentum Transport} \label{sec:turbulent_transport}

We consider the problem of turbulent transport first to demonstrate that the results from our statistical simulations display the standard behaviors that agree with the theory or direct numerical simulations of MRI turbulence. In the latter part of this subsection, we present our findings related to the generation of the Reynolds and Maxwell stresses, previously unexplained. It is worth noting that existing local models that address the generation process of these stresses have either neglected mean magnetic fields \cite{1995PASJ...47..629K, 2003MNRAS.340..969O} or considered only constant vertical magnetic fields \cite{2006PhRvL..97v1103P}, thereby disregarding several significant interactions.

\begin{figure*}
	\includegraphics[width=0.9\columnwidth]{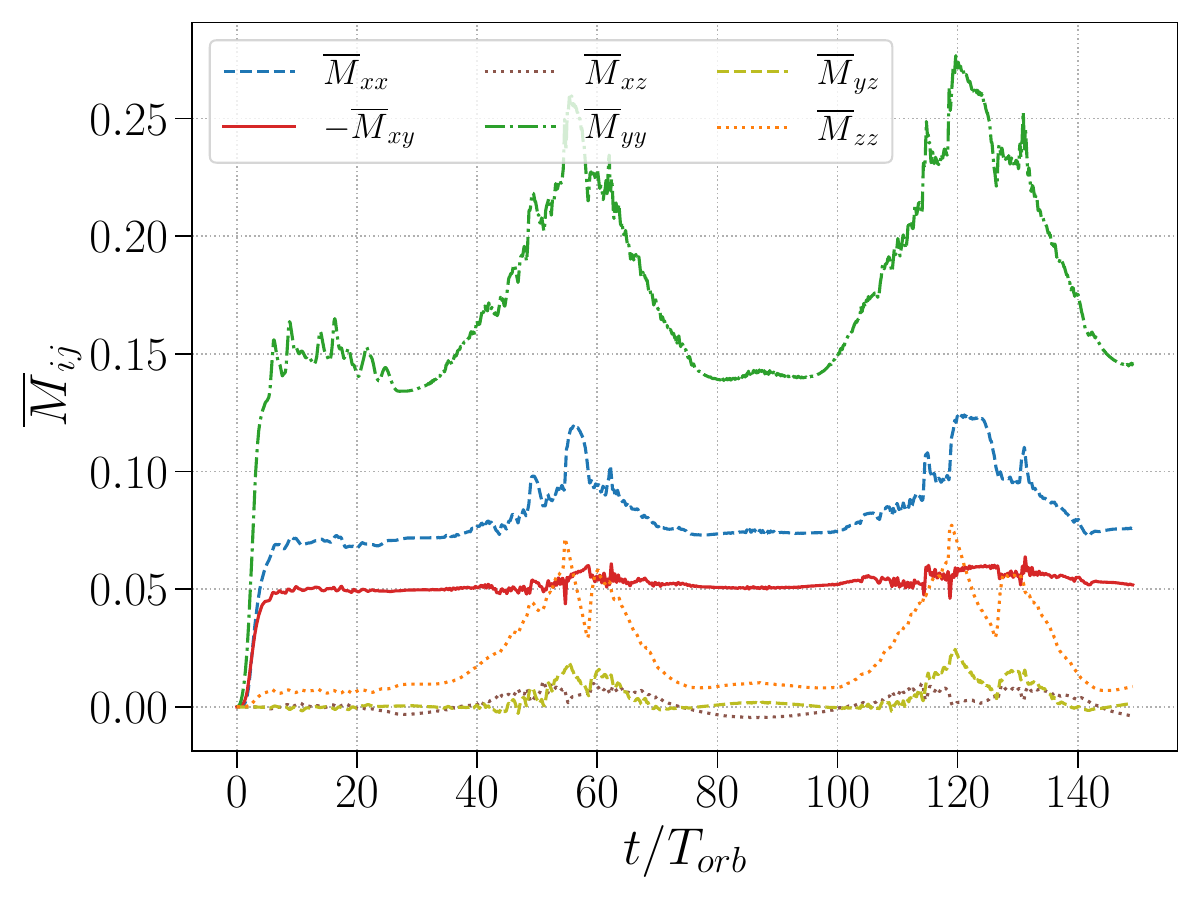} 
	\includegraphics[width=0.9\columnwidth]{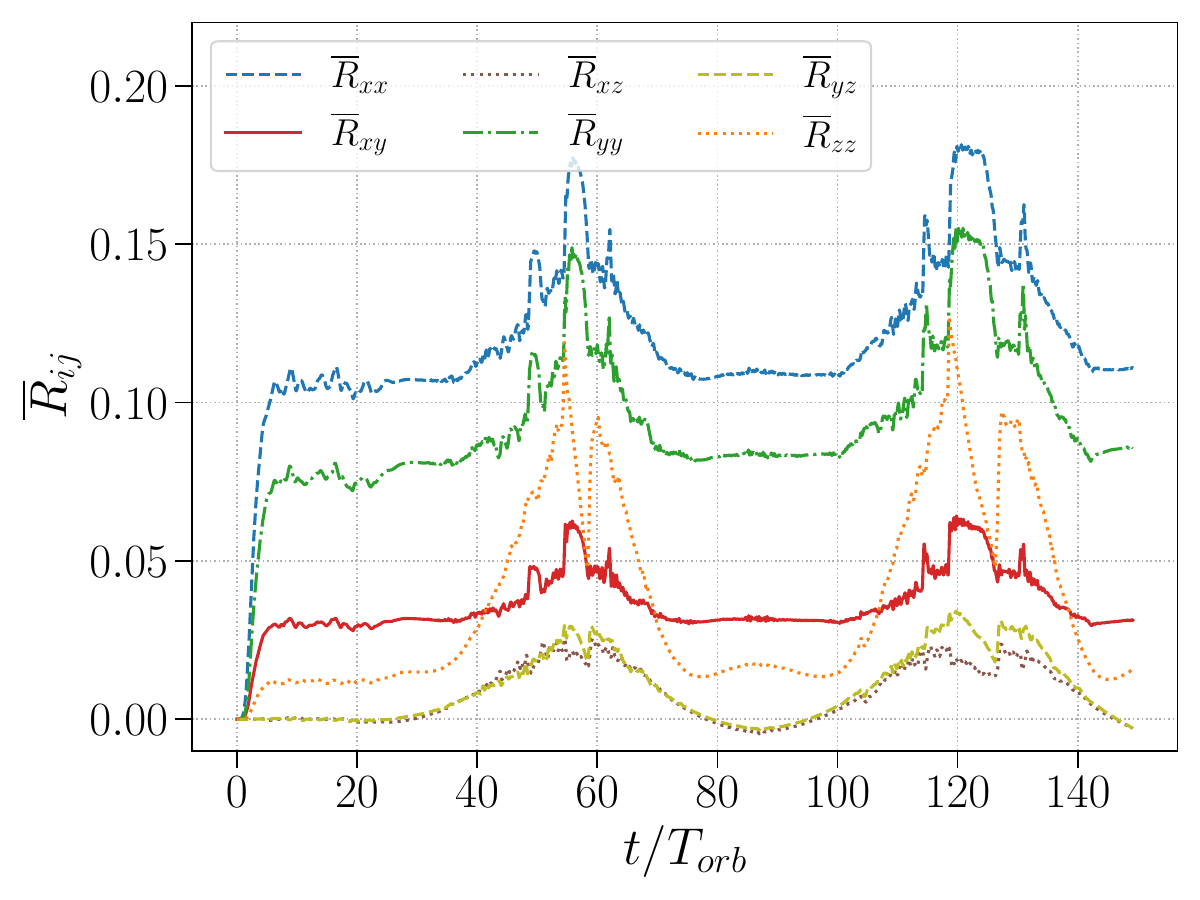}  
	\caption{(Color online)		 
	Time-evolution of the volume-averaged Maxwell (left panel) and Reynolds tensors (right panel). The $xx-$, $xy-$, $xz-$, $yy-$, $yz-$ and $zz-$components are distinguished by dashed blue, solid red, dotted brown, dash-dotted green, dashed olive, and dotted orange lines, respectively. }
	\label{fig:ts_Mij_Rij}
\end{figure*}
Consider the evolution of volume-averaged components of the Maxwell $(\bar M_{ij})$ and Reynolds $(\bar R_{ij})$ tensors in the left and right panels of Fig.~\ref{fig:ts_Mij_Rij}, respectively. The $xy-$components of the stress tensors are mainly responsible for the (radially) outward angular momentum transport. For the matter in accretion disks to accrete, i.e., to lose angular momentum, the sign of the mean total stress, $\bar W_{xy} = \bar R_{xy} - \bar M_{xy}$, must be positive. 
This can be inferred straightforwardly from the radial component of the angular momentum flux,  $-\partial_j (\bar R_{yj} - \bar M_{yj})$, in \Eq{eq:meanU} with $i=y$ and $j=x$. 
From the \Fig{fig:ts_Mij_Rij}, we see that the components of Maxwell and Reynolds stresses responsible for the outward angular momentum transport are always
negative and positive, respectively, i.e., $\bar M_{xy}<0$ and $\bar R_{xy}>0$. 
This naturally leads to a net (radially) outward angular momentum flux mediated by total positive mean stress, $\bar W_{xy} = \bar R_{xy} - \bar M_{xy} >0$. Furthermore, the dominant contribution to the total stress arises from the correlated magnetic fluctuations, rather than from their kinetic counterpart, i.e., $-\bar M_{xy} > \bar R_{xy}$, as expected. Note that the vertically outward angular momentum transport through $\bar W_{yz} = \bar R_{yz} - \bar M_{yz}$ is smaller. 

In Fig.~\ref{fig:ts_Mij_Rij}, we also highlight the turbulent energy densities along three directions. The diagonal components ($xx$, $yy$, and $zz$) of the Maxwell and Reynolds stresses indicate the turbulent magnetic and kinetic energy densities with a multiplication factor of two, respectively. The total turbulent energy is $(\bar M + \bar R)/2$, where $\bar M = \bar M_{ii}$ and $\bar R = \bar R_{ii}$ are the traces of the Maxwell and Reynolds tensors, respectively. As expected, the turbulent magnetic energy dominates over the kinetic counterpart. 
In the magnetic counterpart of the total energy, the azimuthal component is the most significant one followed by the radial and vertical contributions, i.e., $\bar M_{yy} > \bar M_{xx} > \bar M_{zz}$. In the kinetic counterpart of the total energy, the radial component is the most dominant one followed by the azimuthal and vertical contributions, i.e., $\bar R_{xx} > \bar R_{yy} > \bar R_{zz}$. 
All of these features are in agreement with exiting local models \cite{1995PASJ...47..629K, 2003MNRAS.340..969O, 2008MNRAS.383..683P, 2009AN....330...92L}. 
Thus, we are reassured that our statistical simulations are reliable to use for investigations related to turbulent transport. 

Next, we describe the generation mechanism of different components of stress tensors to understand the turbulent transport in more detail. Below we provide the comprehensive web by which the stress components connect to each other through 
(i) shear, (ii) rotation, (iii) mean fields, (iv) other small-scale correlators, and/or (v) nonlinear three-point terms. 
\begin{figure*}
	\includegraphics[width=2\columnwidth]{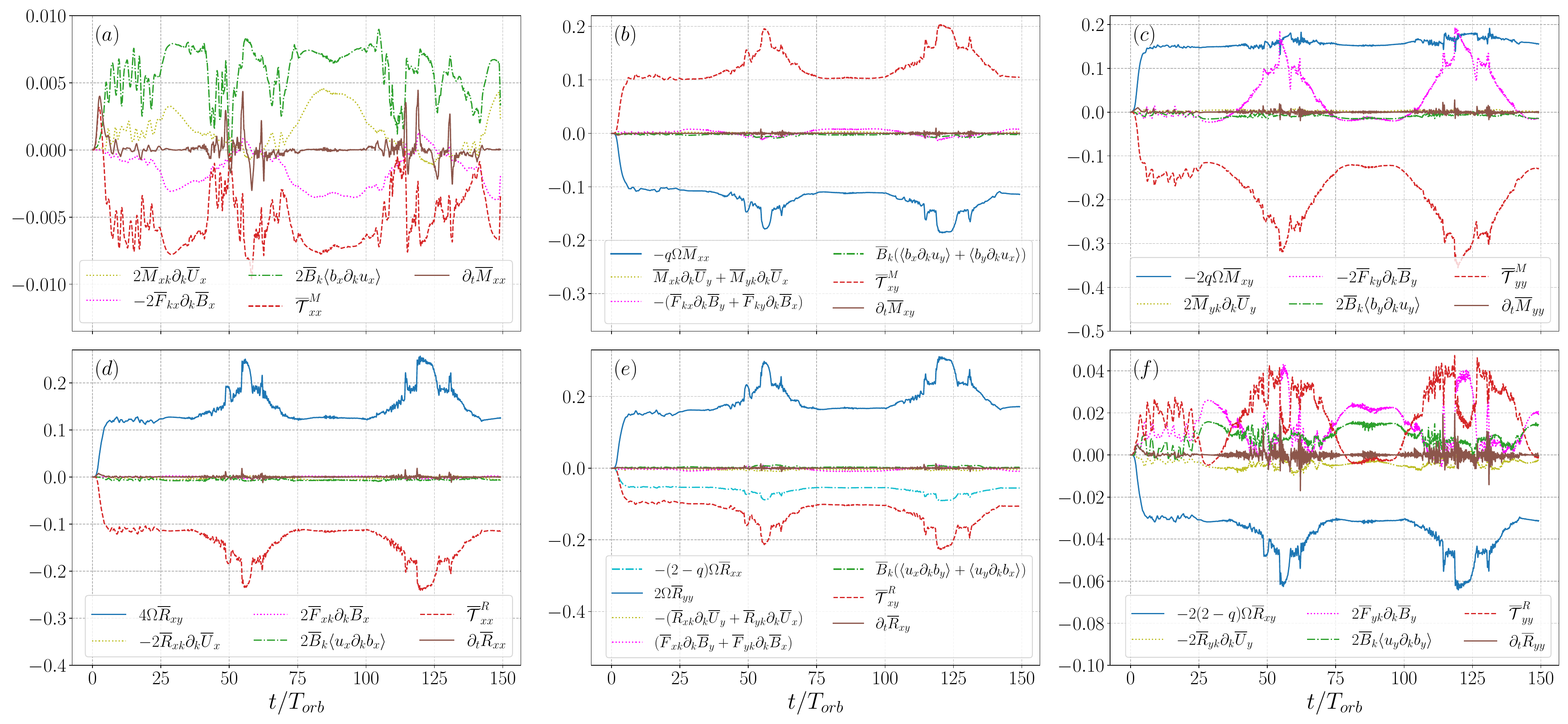}
	\caption{(Color online)
		\textit{Time-evolution of individual terms appeared in the volume-averaged equations for Maxwell stress} (upper panels) \textit{and Reynolds stress} (lower panels) \textit{to explain the turbulent angular momentum transport problem.} Different panels correspond to different stress components: upper panels---$(a)$ $\bar M_{xx}$, $(b)$ $\bar M_{xy}$, $(c)$ $\bar M_{yy}$, and lower panels---$(d)$ $\bar R_{xx}$, $(e)$ $\bar R_{xy}$, $(f)$ $\bar R_{yy}$.}  
	\label{fig:terms_delt_Mij_Rij}
\end{figure*}
In Fig.~\ref{fig:terms_delt_Mij_Rij}, we plot the volume-averaged terms that appeared in the equations for Maxwell (\Eq{eq:Mij_exact}) and Reynolds (\Eq{eq:Rij_exact}) stresses. 
In next few paragraphs, we compare the amplitude and phase of the various terms in these equations to work out the chain of production leading to efficient turbulent transport. 

\begin{figure}
	\centering
	\includegraphics[width=\columnwidth]{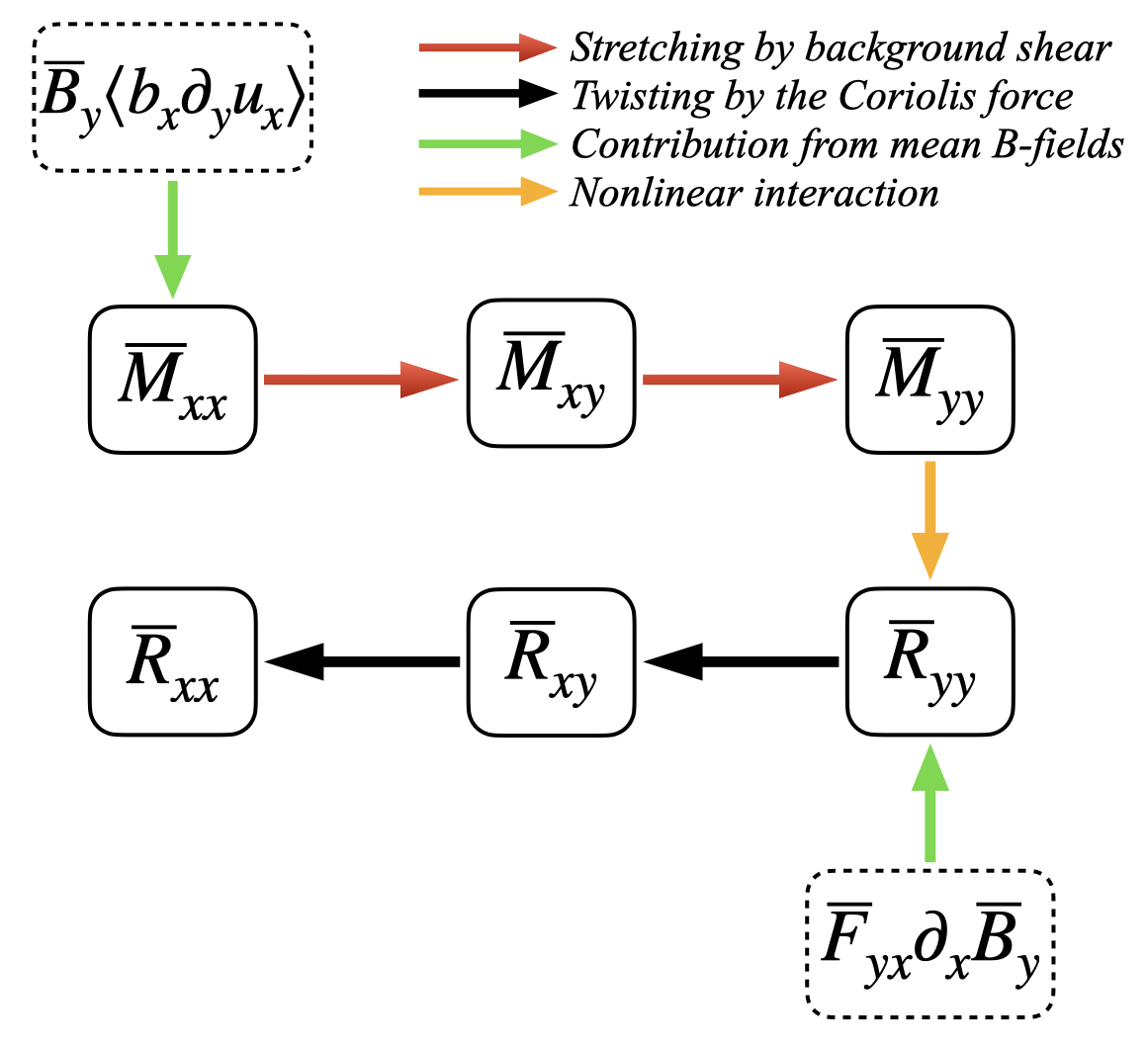} 
	\caption{(Color online)		 
		\textit{A schematic representation of the MRI-driven turbulent angular momentum transport.} Different arrow colors correspond to the paths by which the stress components connect to each other through shear, rotation, mean fields, other small-scale correlators, and nonlinear three-point interactions. Note that we have highlighted only the dominant source terms.}
	\label{fig:mri_transport}
\end{figure}

Those readers who are interested in the final summary immediately can skip to the last paragraph in this subsection and/or to a summary schematic in \Fig{fig:mri_transport}.

To understand the process of outward angular momentum transport, we examine the time evolutions for the $xx-$, $xy-$, and $yy-$components of stress tensors. This is because the $xy-$components of stress tensors are directly connected to the $xx-$ and $yy-$components of stress tensors via shear and/or rotation (more specifically, Coriolis force appears in the Reynolds stress equations). 
Since $\bar M_{xx}$ is positive throughout the evolution, the positive term in $\partial_t \bar M_{xx}$ acts as a source, whereas the negative term behaves like a sink. 
The same is true for all the stress tensors, which are positive throughout their evolution. For negative $\bar M_{xy}$, the roles of different terms are opposite: the positive term in $\partial_t \bar M_{xy}$ acts as a sink, whereas the negative term behaves like a source. 

\begin{figure*}
	\includegraphics[width=1.5\columnwidth]{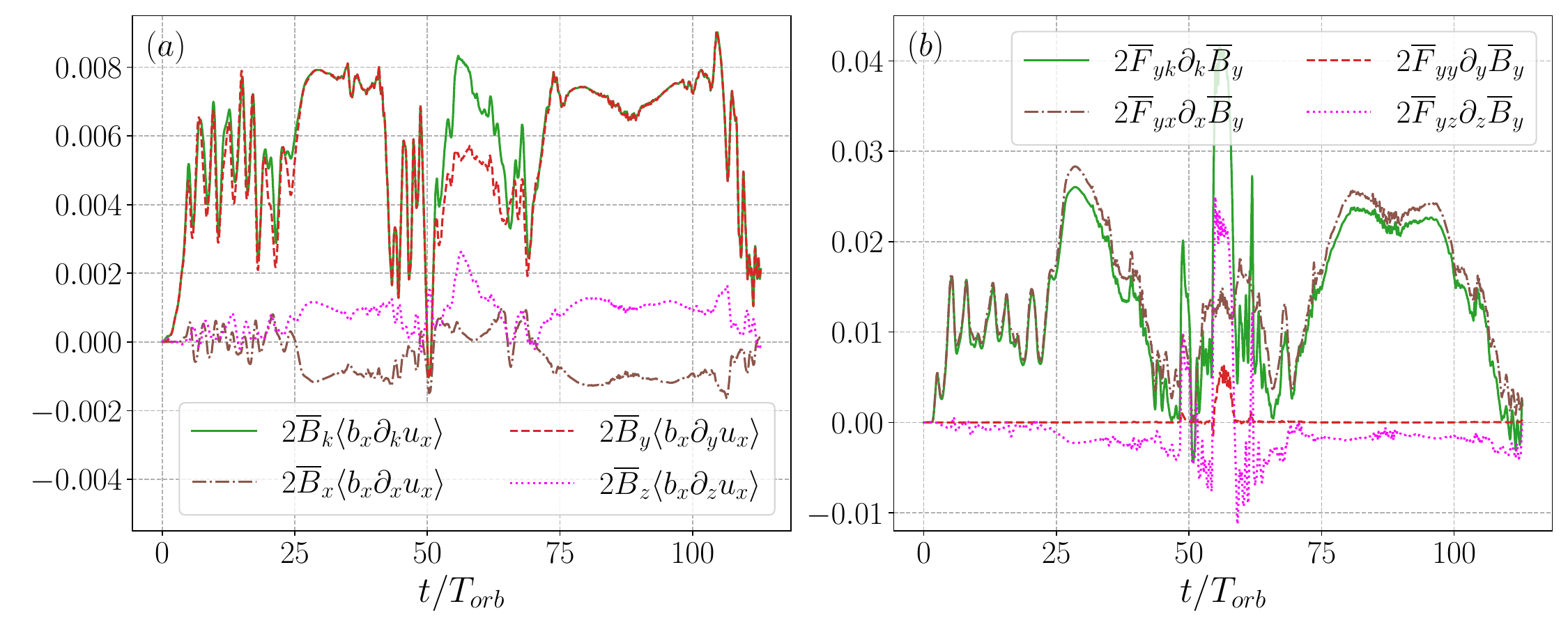} 
	\caption{(Color online)		 
		\textit{The time-evolution of specific terms,  which are proportional to the mean magnetic fields and the gradient of mean magnetic fields}, is observed in the volume-averaged equations for $\bar M_{xx}$ (left panel) and $\bar R_{yy}$ (right panel), respectively. These terms are analyzed to explain the interconnection between the turbulent angular momentum transport and mean field dynamo. This analysis serves as a continuation of the findings presented in Fig.~\ref{fig:terms_delt_Mij_Rij}.}
	\label{fig:terms_delt_Mxx_Ryy}
\end{figure*}

The upper panels of Fig.~\ref{fig:terms_delt_Mij_Rij} are for the Maxwell stress: $(a)$ $\bar M_{xx}$, $(b)$ $\bar M_{xy}$, and $(c)$ $\bar M_{yy}$.
In the evolution of the latter two, $\bar M_{xy}$ and $\bar M_{yy}$, the shear terms (solid blue line) act as the dominant source term, whereas the nonlinear three-point terms (dashed red line) act as the sink. Here, the ``stretching" of the positive stress component $\bar M_{xx}$ via the shear produces $\bar M_{xy}$ at a rate of $-q\Omega$. This renders $\bar M_{xy}$ negative.
Similarly, shear acts on $\bar M_{xy}$ to produce the positive stress component $\bar M_{yy}$ at a rate of $-2q\Omega$. 
There is no shear term in the time evolution of $\bar M_{xx}$. 
Thus, it is evident that the turbulent transport via $-\bar M_{xy}$ can not work with Keplerian shear alone---$\bar M_{xx}$ component is needed for Keplerian shear to act on. The generation mechanism of $\bar M_{xx}$ is critically important here. For $\bar M_{xx}$, the dominant source term is $\bar B_k \langle b_x \partial_k u_x \rangle$ (dash-dotted green line) with $k=y$ (left panel of Fig.~\ref{fig:terms_delt_Mxx_Ryy}). 
The nonlinear three-point term (dashed red line) acts as the dominant sink. 
The other two subdominant terms, one proportional to $\partial_k \bar U_x$ (dotted olive line) and the other proportional to $\partial_k \bar B_x$ (dotted magenta line), behave like a source and a sink, respectively. 

The lower panels of Fig.~\ref{fig:terms_delt_Mij_Rij} correspond to the Reynolds stress: $(d)$ $\bar R_{xx}$, $(e)$ $\bar R_{xy}$, and $(f)$ $\bar R_{yy}$. 
For Reynolds stress, shear acts similarly as in the case of Maxwell stress but with an opposite sign. 
In addition, the Coriolis force plays a significant role in the evolution of Reynolds stresses. The ``stretching" of the positive stress component
$\bar R_{xx}$ produces $\bar R_{xy}$ via shear at a rate of $q\Omega$. However, Coriolis force makes the positive stresses, $\bar R_{xx}$ and $\bar R_{yy}$, act oppositely in the evolution of $\bar R_{xy}$ with the same weighting factor of $2\Omega$. Since $q<2$, the combined effects of shear and Coriolis force make the term with $\bar R_{xx}$ (dash-dotted cyan line) behave as a sink in the evolution of $\bar R_{xy}$. 
The nonlinear three-point term (dashed red line) acts as a sink here as well. 
Hence, the term with $\bar R_{yy}$ is the only source term (solid blue line) in the $\bar R_{xy}$ evolution via the Coriolis force. This finding is illustrated in \Fig{fig:terms_delt_Mij_Rij}(e). 
In \Fig{fig:terms_delt_Mij_Rij}(d), for $\bar R_{xx}$, the source term is $4\Omega \bar R_{xy}$ (solid blue line) arising through the Coriolis force, whereas the nonlinear three-point term (dashed red line) acts as a sink. 

Finally in \Fig{fig:terms_delt_Mij_Rij}(f), we see that the shear acting on $\bar R_{xy}$ produces $\bar R_{yy}$ at a rate of $2q\Omega$. However, the term with $\bar R_{xy}$, overall, acts as a sink in the evolution of $\bar R_{yy}$. Since $q<2$, the combined effects of shear and Coriolis force make the term with $\bar R_{xy}$ behave as a sink (solid blue line). Consequently, the question that arises is how is $\bar R_{yy}$ generated. 
We find that the most dominant source terms in the $\bar R_{yy}$ evolution are the nonlinear three-point term (dashed red line), the term associated with the gradient of the azimuthal magnetic fields, $2\bar F_{yk} \partial_k \bar B_y$ (dotted magenta line) with $k=x$ (right panel of \Fig{fig:terms_delt_Mxx_Ryy}). The other two subdominant terms, one proportional to $ \bar B_k$ (dash-dotted green line) and the other proportional to $\partial_k \bar U_y$ (dotted olive line), behave like a source and a sink, respectively. 

The overall findings associated with the turbulent transport are summarized schematically in Fig.~\ref{fig:mri_transport}. We remind the reader that we are able to delineate this chain of production because of the structure of our model being used in this statistical simulation which helps in making the connections directly between mean fields and the cumulants. The stretching of $\bar M_{xx}$ via shear produces $\bar M_{xy}$, whose stretching by shear further produces $\bar M_{yy}$. The large-scale field (here, $\bar B_y$) acts in conjunction with $\langle b_x \partial_k u_x \rangle$ to generate $\bar M_{xx}$ (this can be interpreted essentially as tangling of the mean magnetic field leading to the generation of small-scale fields). 
For the Reynolds stress, the Coriolis force is responsible for generating $\bar R_{xx}$ from $\bar R_{xy}$, and $\bar R_{xy}$ from $\bar R_{yy}$. The outcome of nonlinear interactions between $\bar M_{yy}$ and $\bar R_{yy}$ via the three-point term is the formation of $\bar R_{yy}$ from $\bar M_{yy}$. The other dominant source term for $\bar R_{yy}$ is the term proportional to the radial gradient of the mean azimuthal magnetic field. Hence, turbulent transport is not possible without large-scale fields, i.e., the mean-field dynamo mechanism is necessary.


\subsection{Large-scale Dynamo}  \label{sec:Large-scale dynamo}

We begin by presenting the overall evolution of the relevant quantities, namely the mean magnetic and velocity fields, both as volume averages (of the energy) and planar-averages. 
Then we compare the evolution of the different terms in the dynamical equations for the planar-averaged 
large-scale magnetic fields, to determine which components of the EMF are important for the MRI large-scale dynamo. Thereafter, we specify how we can recover a general expression for the $y$ and $z$-components of the EMF from our model equations in Section~\ref{sec:EMF}. 
We find that a given component of the EMF is a linear combination of terms proportional to mean magnetic fields, the gradient of mean magnetic fields, the gradient of mean velocity fields, and a nonlinear term.
With expressions for the EMFs in hand, we set out to investigate the contribution of the various terms to determine the dominant dynamo effects. To do so, we first examine volume-averages of the terms in time windows from both linear and nonlinear regimes, to get a global picture. Next, we examine the planar average of various terms to study the behaviour locally in space. In the latter analysis, we uncover a more sophisticated behaviour of the large-scale dynamo. But overall, we find both types of analysis lead to the same conclusions. The EMF analysis for radial large-scale field generation is in Section~\ref{sec:radial_B_fied} and for vertical large-scale field generation is in Section~\ref{sec:vertical_B_fied}. 

\subsubsection{Volume averaged large-scale or mean field energies}

\begin{figure}
	\includegraphics[width=\columnwidth]{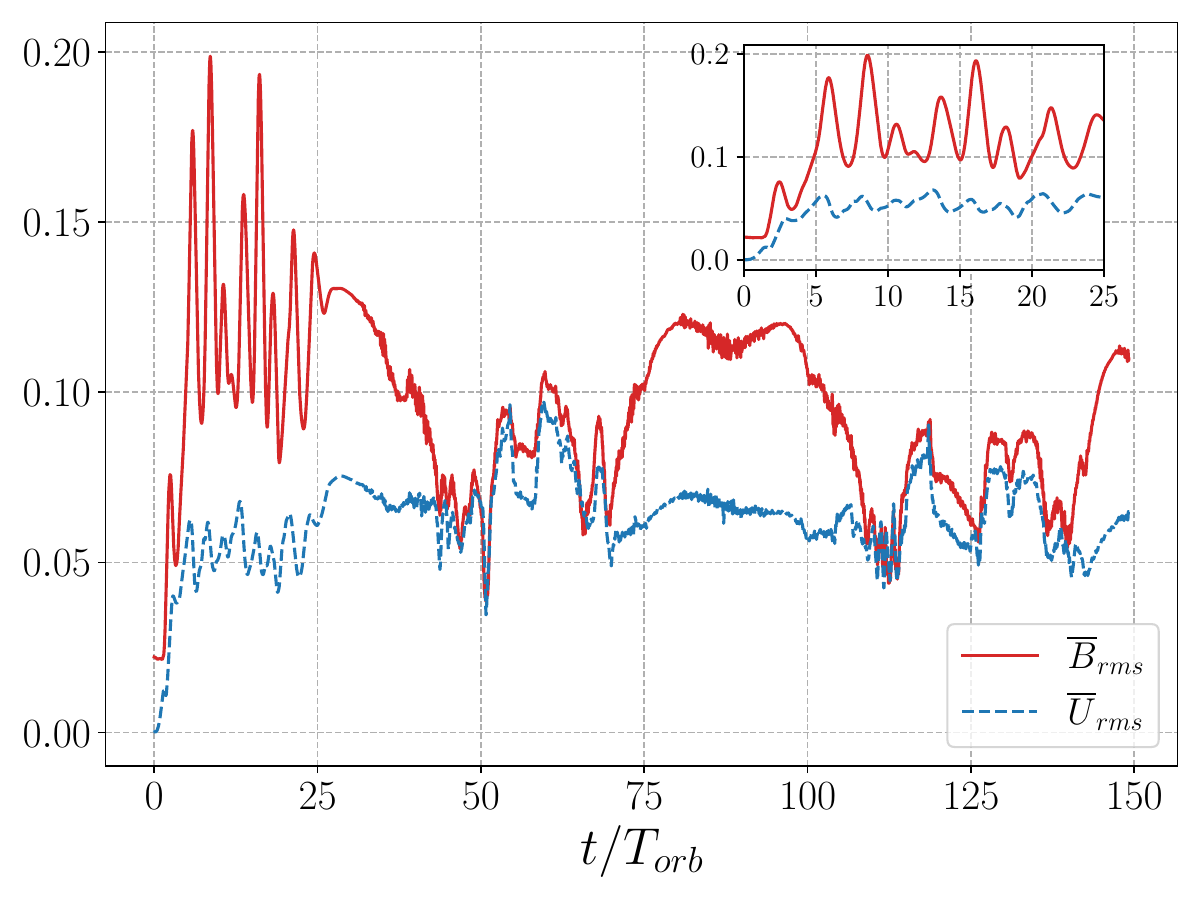}  
	\caption{(Color online)		 
		Time-evolution of volume averaged large-scale fields $\bar B_{\text{rms}}$ (solid red curve) and $\bar U_{\text{rms}}$ (dashed blue curve) are shown. The zoomed-in part indicates the initial growth to the early saturation phase.}
	\label{fig:ts_Brms}
\end{figure}
Consider the evolution of volume-averaged large-scale magnetic and velocity fields. \Fig{fig:ts_Brms} shows the time evolution of the root-mean-square (rms) velocity $(\bar U_{\text{rms}})$ and magnetic $(\bar B_{\text{rms}})$ fields. We find that the MRI-driven turbulence hosts both the large-scale dynamo of velocity and magnetic fields. 
The amplitude of $\bar B_{\text{rms}}$ dominates over that of $\bar U_{\text{rms}}$ throughout the entire duration of our longest simulation run, spanning approximately $150$ orbits. The zoomed-in view of the initial growth to the early saturation phase of the large-scale fields is also depicted in \Fig{fig:ts_Brms}. The initial growth phase of both fields is observed up to a time of $t/T_{\text{orb}} \sim 5.5$. Following this initial growth phase, the fields settle into a steady state, indicating the saturation regime.

\begin{figure}
	\includegraphics[width=\columnwidth]{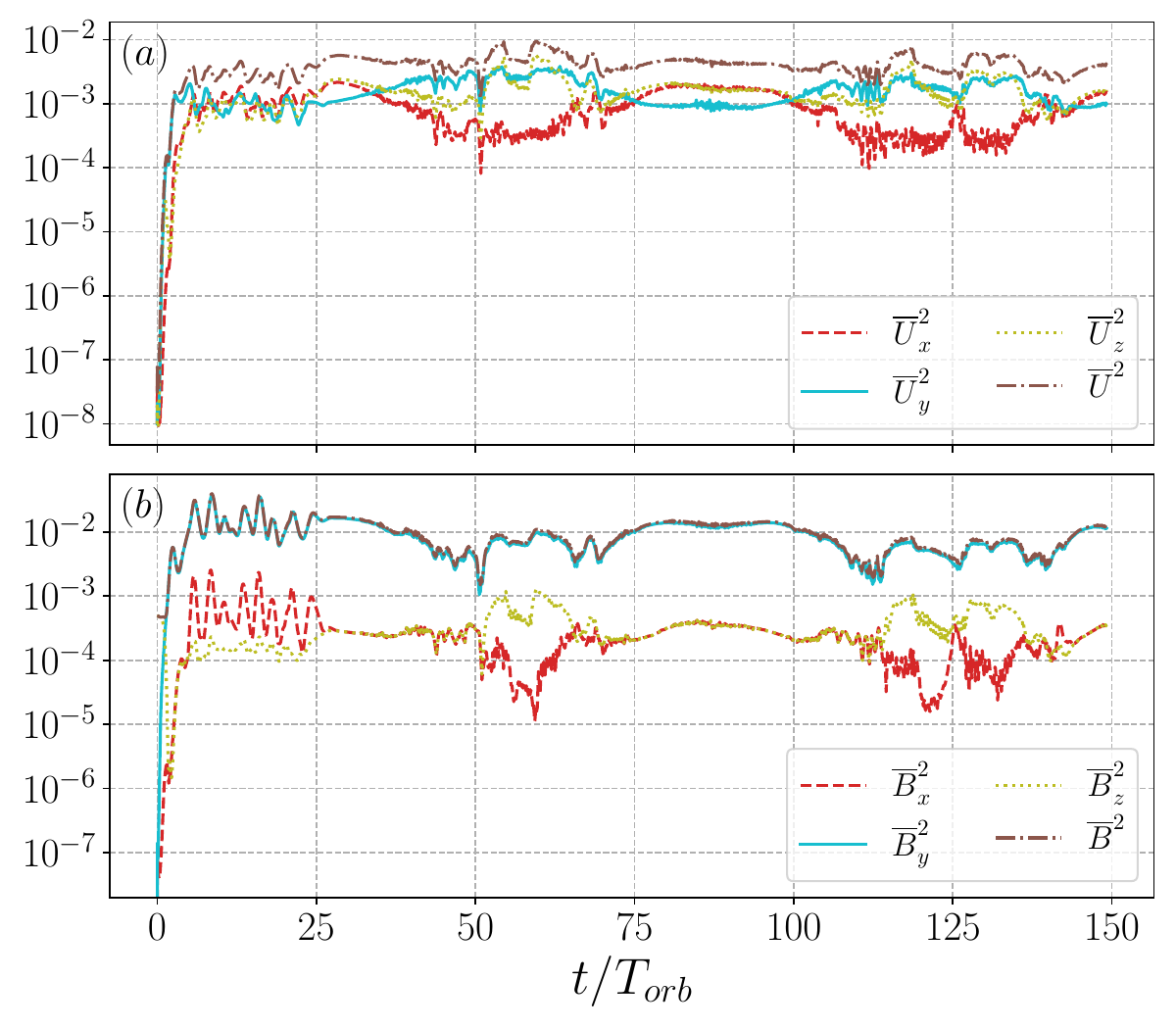}  
	\caption{(Color online)		 
		Time-evolution of volume averaged large-scale kinetic and magnetic energy densities (multiplied by two) are shown in the upper and lower panels, respectively. The energy associated with the $x-$, $y-$, and $z-$components are distinguished by dashed red, solid cyan, and dotted yellow lines, respectively. The brown dash-dotted lines refer to the total energy densities.}
	\label{fig:ts_mean_energy_density}
\end{figure}
Next, we consider the volume averaged energy density associated with large-scale velocity and magnetic fields, $\frac{1}{2} \langle \rho\bar U^2 \rangle$ and $\frac{1}{2}\langle \bar B^2 \rangle$, respectively. Fig.~\ref{fig:ts_mean_energy_density} shows the time evolution of the large-scale kinetic (upper panel) and magnetic (lower panel) energy densities with a multiplication factor of two. Fig.~\ref{fig:ts_mean_energy_density} also demonstrates the contribution from the three components of the fields. Important to note that the large-scale magnetic energy dominates over the kinetic one, indicating that the MRI dynamo in accretion disks is characterized by super-equipartition of magnetic energy relative to kinetic energy. Most of the contribution to the large-scale magnetic energy arises from the toroidal mean magnetic field, $\bar B_y$, while the radial and vertical components of the mean magnetic field are of similar magnitude. In large-scale velocity fields, all three components share almost similar magnitudes. Below we explore the generation mechanism of different large-scale magnetic field components extensively. It may be interesting in future work to examine the generation process of $\bar U$ (i.e., the vorticity dynamo) in more detail.

\subsubsection{Planar averaged large scale or mean fields}

In order to understand the behaviour of the mean fields locally in space, we perform planar averages. We consider three different planar averages, $x\text{--}y$, $y\text{--}z$, and $x\text{--}z$ averaging, to determine the mean magnetic fields $\bec{\bar B}(z)$ or $\langle \bec{B} \rangle_{(x,y)}$, $\bec{\bar B}(x)$ or $\langle \bec{B} \rangle_{(y,z)}$, and $\bec{\bar B}(y)$ or $\langle \bec{B} \rangle_{(x,z)}$ respectively. In the $x\text{--}y$ averaged case, the generated field components are $\bar B_x(z)$ and $\bar B_y(z)$; where $\bar B_z(z)$ vanishes to maintain the divergence-free condition. Similarly, $\bar B_x(x)$ in $y\text{--}z$ averaging and $\bar B_y(y)$ in $x\text{--}z$ averaging are zero.

\begin{figure}
	\includegraphics[width=\columnwidth]{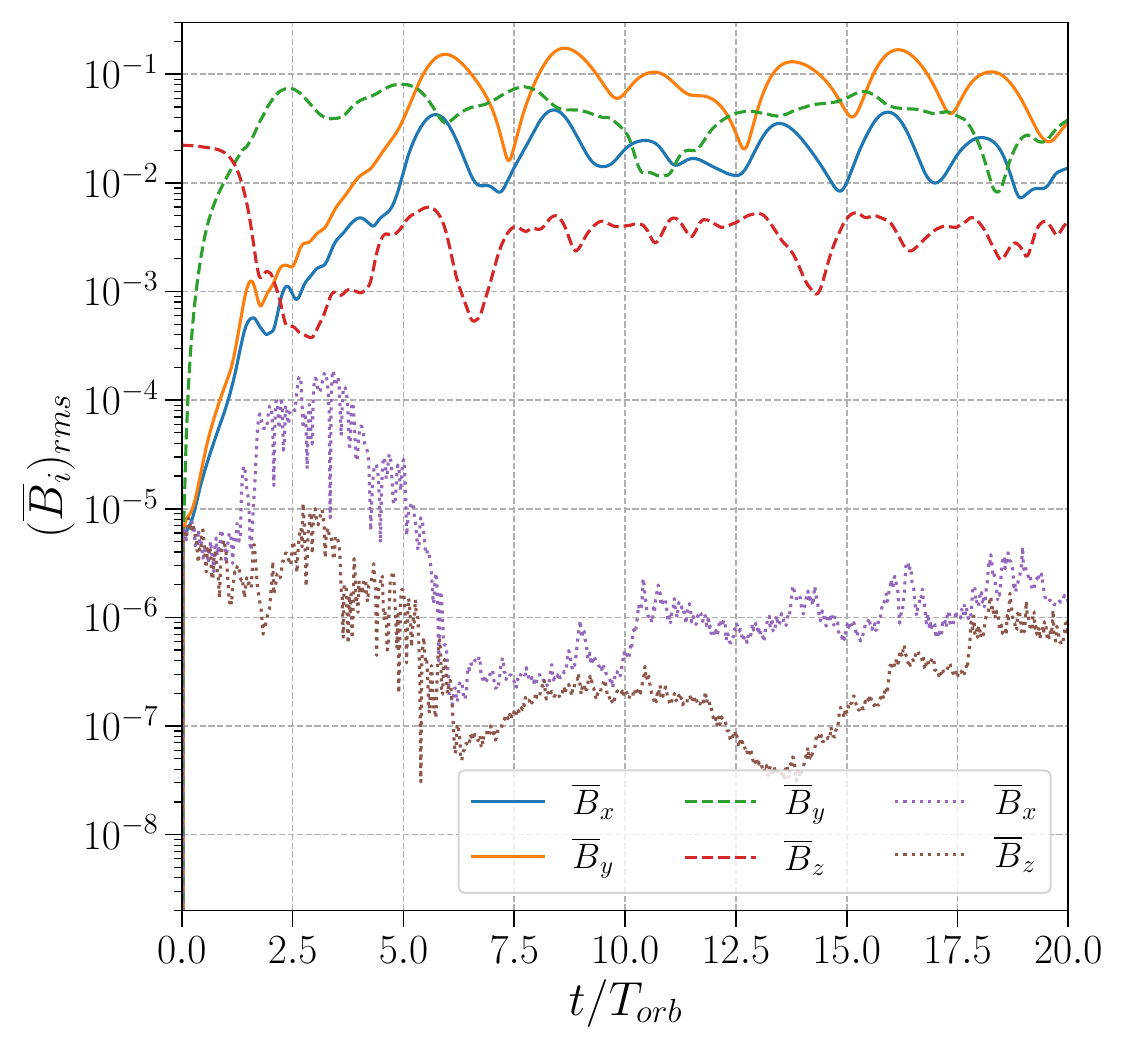}  
	\caption{(Color online)		 
		The time evolution of planar-averaged root-mean-square large-scale magnetic fields is shown. The planar averages are performed along three different planes, $x\text{--}y$, $y\text{--}z$, and $x\text{--}z$, denoted by the solid, dashed, and dotted lines, respectively. Additionally, we perform further averaging of the root-mean-square field over the remaining third direction.}
	\label{fig: Brms_plane_ave}
\end{figure}
In \Fig{fig: Brms_plane_ave}, we present the time evolution of the root-mean-square planar-averaged fields.
To obtain these quantities, we first square the planar-averaged fields. Then we perform further averaging over the remaining third direction and then take the square root. 
The most prominent large-scale field observed is $\bar{B}_y$ (solid orange line), resulting from the $x\text{--}y$-averaging. Notably, in the saturation regime, the rms value of $\bar{B}_y$ is around four times greater than that of $\bar{B}_x$ (solid blue line). The $y\text{--}z$ averaging reveals significant $\bar B_y$ as well, represented by the dashed green line. The $y\text{--}z$ averaged $\bar B_z$ (dashed red line) at the beginning of the growth phase reflects the initial condition of $\bar B_z = B_0 \sin (k_x x)$. During the saturation stage, the $y\text{--}z$ averaged fields show that $|\bar{B}_y|$ is approximately one order stronger compared to $|\bar{B}_z|$. Conversely, when applying the $x\text{--}z$ averaging (represented by the dotted lines), 
resulting mean fields $\bar{B}_x$ and $\bar{B}_z$ do not exhibit a dynamo growth and further decay to small values. 
The weakness of $x$--$z$-averaged fields renders the MRI-generated large-scale fields largely axisymmetric.

\begin{figure}
	\includegraphics[width=\columnwidth]{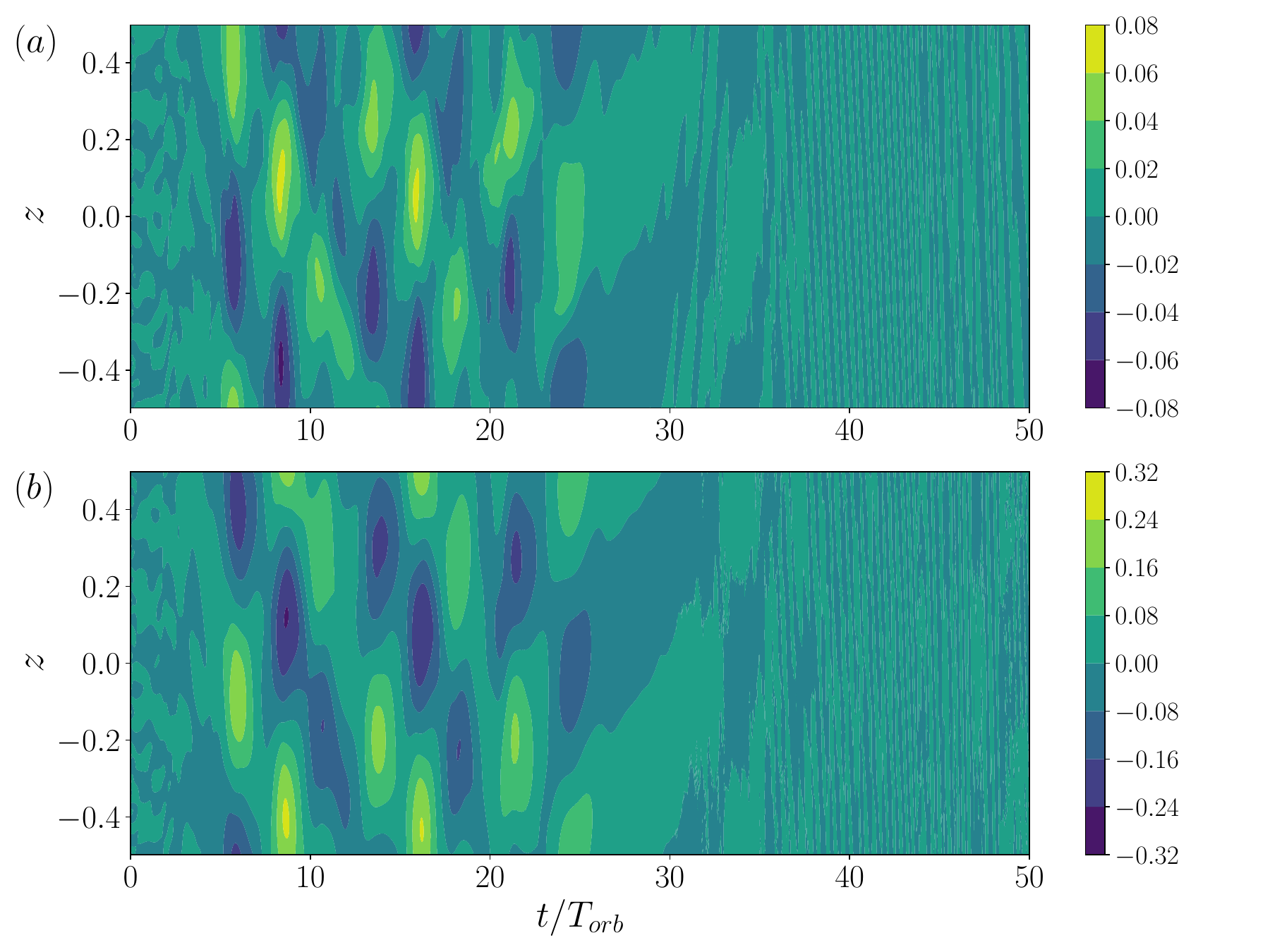} 
	\caption{(Color online) The $x\text{--}y$-averaged magnetic fields, $\bar B_x(z)$ and $\bar B_y(z)$, are shown on the top, and bottom panels, respectively.}
	\label{fig:B_slice}
\end{figure}
We then shift our focus to the most-studied large-scale magnetic fields $\bar B_x(z)$ and $\bar B_y(z)$, obtained from $x\text{--}y$-averaging. \Fig{fig:B_slice} illustrates their temporal evolution along the abscissa and spatial variation along the ordinate, with the color scale representing the field strengths. We identify three distinct stages in this evolution: $(a)$ the initial growth phase extending up to $t/T_{\text{orb}}\sim 5.5$, $(b)$ the intermediate or initial saturation phase from $t/T_{\text{orb}}\sim 5.5$ to $t/T_{\text{orb}}\sim 25$, and $(c)$ the fully nonlinear saturation phase after $t/T_{\text{orb}} \gtrsim 25$. Notably, the appearance of short dynamo cycles during the intermediate phase indicates a   quasi-linear nature of the dynamo.

\begin{figure*}
	\includegraphics[width=1.5\columnwidth]{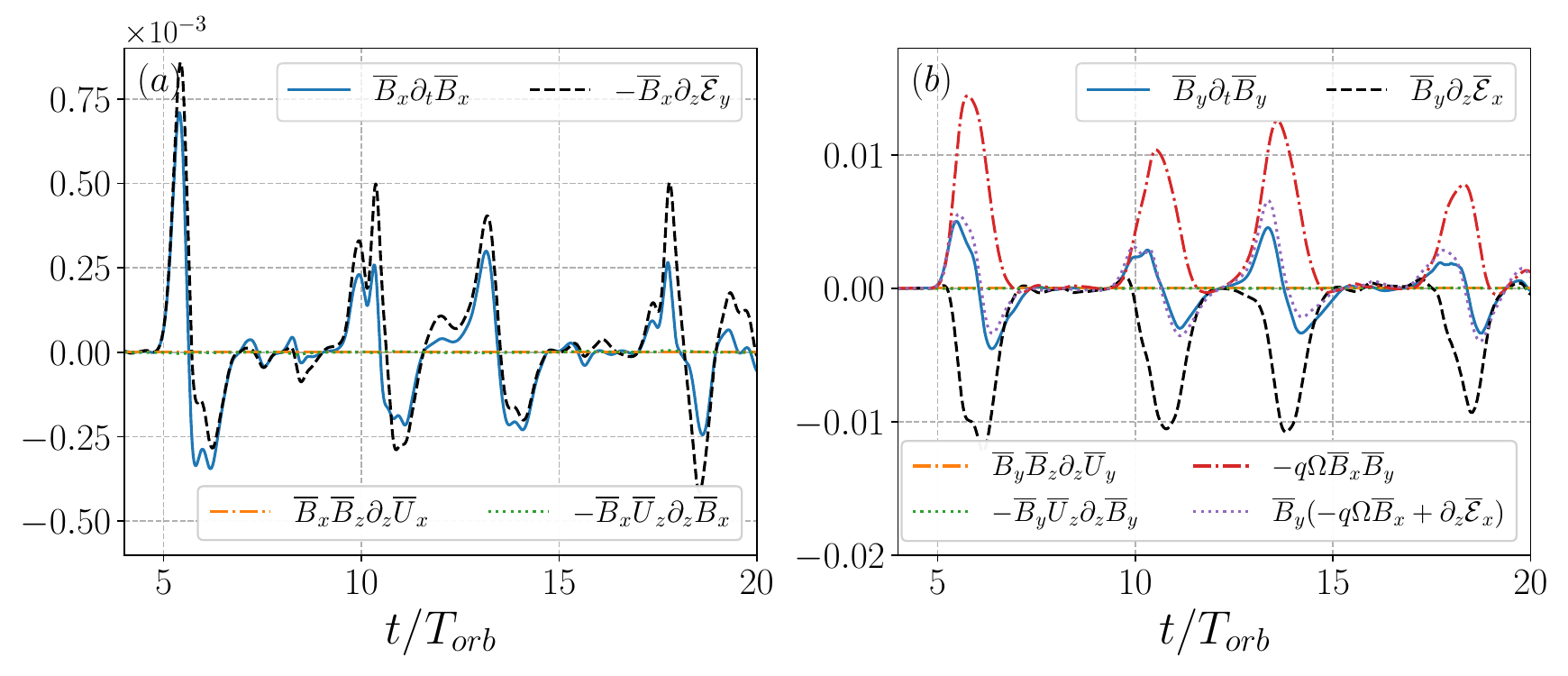} 
	\includegraphics[width=1.5\columnwidth]{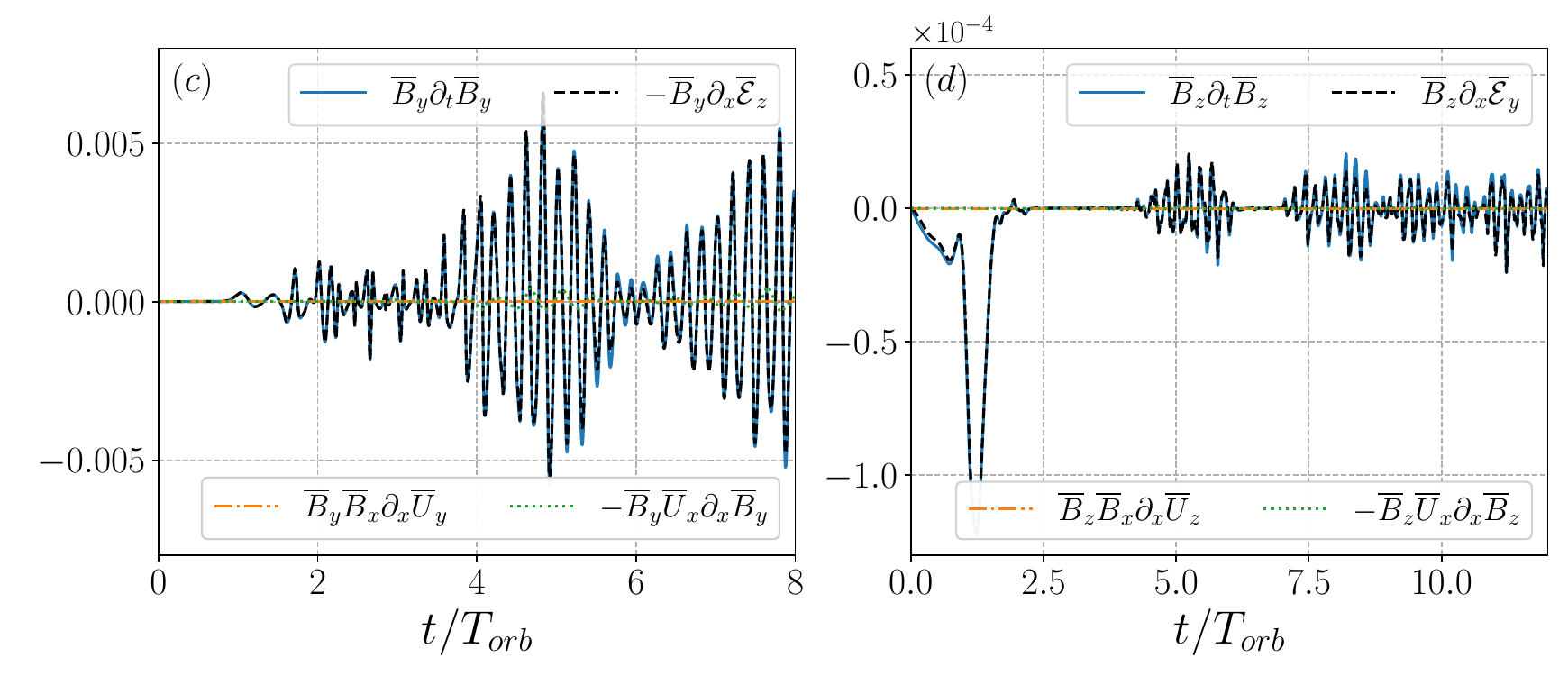} 
	\caption{(Color online)
		The top two panels show the terms from $x\text{--}y$ averaged mean field equation for $\bar B_x(z)$ and $\bar B_y(z)$ on left and right, respectively. The bottom two panels show the terms from $y\text{--}z$ averaged mean field equation for $\bar B_y(x)$ and $\bar B_z(x)$ on left and right, respectively. These are evaluated at $z=-0.15$ (top panels) and at $x=-0.15$ (bottom panels), for $x\text{--}y$ and $y\text{--}z$ averaging, respectively. To understand the contribution of each term to the evolution of mean magnetic fields, we multiply $\bar B_i$ on both sides of the $\partial_t \bar B_i$ equations. 
		The solid blue curve is for the time derivative of the mean field, the dashed black curve is for the corresponding EMF term, the red dash-dotted curve is for the shear term $(-q\Omega \bar B_x)$, and the green dotted and orange dash-dotted lines are, respectively, for the advection and stretching terms involving mean fields.}
	\label{fig:Bi_delt_Bi}
\end{figure*}
\begin{figure*}
	\includegraphics[width=1.5\columnwidth]{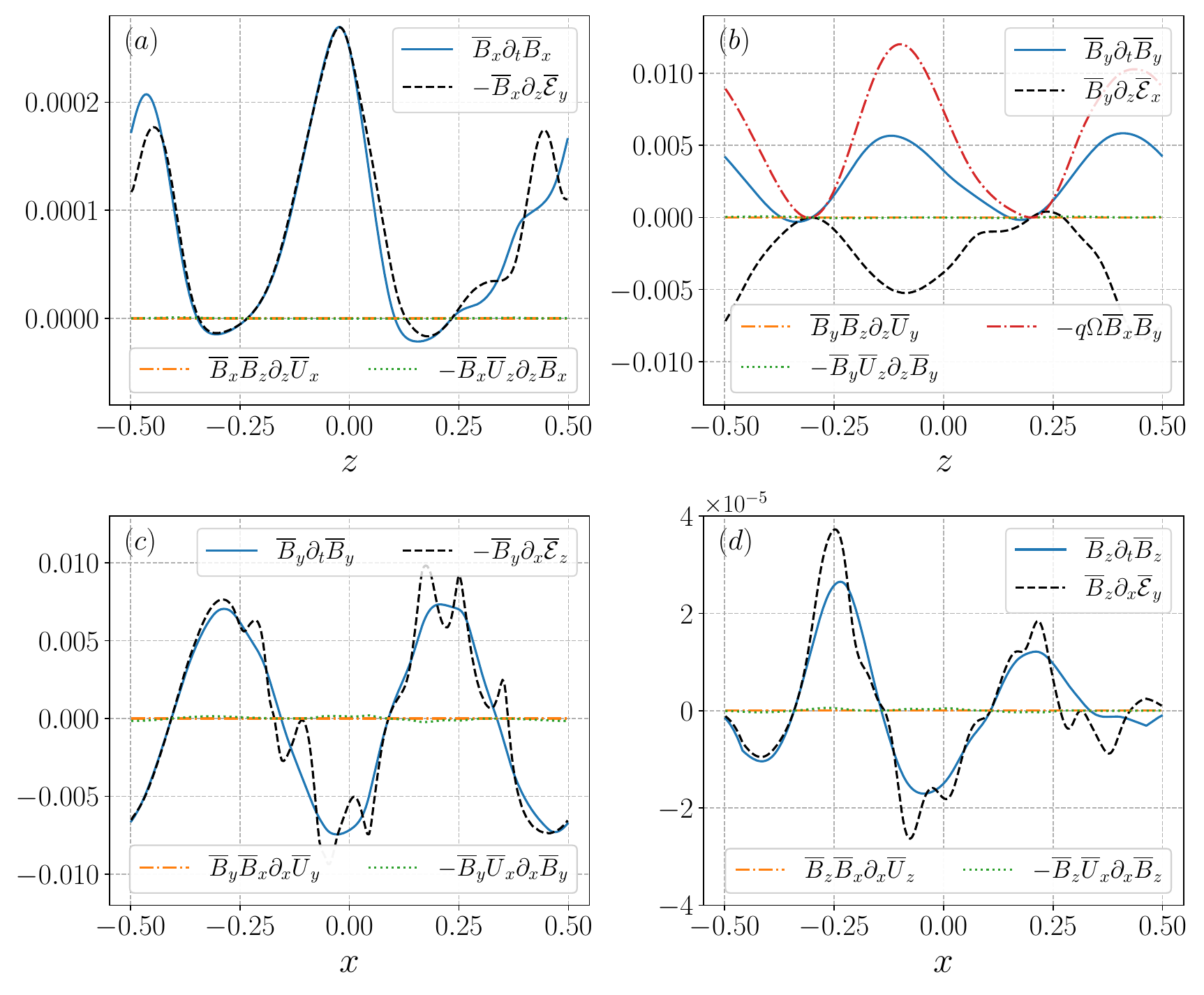} 
	\caption{(Color online)
		The top two panels show the terms from $x\text{--}y$ averaged mean field equation for $\bar B_x(z)$ and $\bar B_y(z)$ on left and right, respectively. The bottom two panels show the terms from $y\text{--}z$ averaged mean field equation for $\bar B_y(x)$ and $\bar B_z(x)$ on left and right, respectively. These are evaluated at $t/T_{\text{orb}} \sim 5$. To understand the contribution of each term to the evolution of mean magnetic fields, we multiply $\bar B_i$ on both sides of the $\partial_t \bar B_i$ equations. 
		The solid blue curve is for the time derivative of the mean field, the dashed black curve is for the corresponding EMF term, the red dash-dotted curve is for the shear term $(-q\Omega \bar B_x)$, and the green dotted and orange dash-dotted lines are, respectively, for the advection and stretching terms involving mean fields.}
	\label{fig:Bi_delt_Bi_t5}
\end{figure*}

To understand the MRI dynamo mechanism, we require both $x\text{--}y$ and $y\text{--}z$ averaging (given the significant mean fields arising from both of these planar averages). Here, we present the time evolution of the mean field equations in both kinds of averaging. The mean field equations in $x\text{--}y$ averaging are given by
\begin{subequations}
	\begin{align}
	\partial_t \bar B_x &= - \partial_z \mathcal{\bar E}_y - \bar U_z\partial_z\bar B_x + \bar B_z\partial_z\bar U_x ,
	\label{eq:meanBx_xy} \\
	\partial_t \bar B_y &= - q\Omega \bar B_x + \partial_z \mathcal{\bar E}_x - \bar U_z\partial_z\bar B_y + \bar B_z\partial_z\bar U_y . 
	\label{eq:meanBy_xy}
	\end{align}
	\label{eq:meanB_xy}
\end{subequations} 
The mean field equations in $y\text{--}z$ averaging are given by
\begin{subequations}
	\begin{align}
	\partial_t \bar B_y &= - \partial_x \mathcal{\bar E}_z - \bar U_x\partial_x\bar B_y + \bar B_x\partial_x\bar U_y ,
	\label{eq:meanBy_yz} \\
	\partial_t \bar B_z &= \partial_x \mathcal{\bar E}_y - \bar U_x\partial_x\bar B_z + \bar B_x\partial_x\bar U_z .
	\label{eq:meanBz_yz}
	\end{align}
	\label{eq:meanB_yz}
\end{subequations} 
Here, the main contributions arise from the shear term $(-q\Omega \bar B_x)$, the advection term $(\bec{\bar U}\cdot \nabla \bec{\bar B})$, the stretching term $(\bec{\bar B}\cdot \nabla \bec{\bar U})$, and the different components of the EMF: $\mathcal{\bar E}_x = (\bar F_{yz} - \bar F_{zy})$, $\mathcal{\bar E}_y = (\bar F_{zx} - \bar F_{xz})$, and $\mathcal{\bar E}_z = (\bar F_{xy} - \bar F_{yx})$. The shear term only appears on the $x\text{--}y$ averaged azimuthal field evolution equation. It will not operate in the $y\text{--}z$ averaged azimuthal field equation because of the divergence-free magnetic field condition, i.e., $\langle B_x \rangle_{(y,z)} \simeq 0$. 
To study the contribution of each term to the evolution of the mean magnetic fields, we multiply $\bar B_i$ on both sides of the $\partial_t \bar B_i $ equations. The resultant individual terms of equations in $\overline{B}_{i} \partial_t \overline{B}_{i}$ are shown in Figs \ref{fig:Bi_delt_Bi} and \ref{fig:Bi_delt_Bi_t5}. 
Here, we use three different terminologies to describe the role of each term: the `source'-term has positive contributions throughout, the `sink'-term contributes negatively, and the `dual'-term can be either positive or negative with time.

In \Fig{fig:Bi_delt_Bi}, we examine the behaviour of the individual terms in \Eqs{eq:meanB_xy}{eq:meanB_yz} in time, 
for both $x\text{--}y$ (top panels) and $y\text{--}z$ averaging (bottom panels). 
These terms are evaluated at $z=-0.15$ (top panels) and at $x=-0.15$ (bottom panels), for $x\text{--}y$ and $y\text{--}z$ averaging, respectively. 
For all the cases, the corresponding advection and stretching terms are negligible. 
The $x\text{--}y$ averaged mean field $\bar B_x(z)$ (top left panel of Fig. \ref{fig:Bi_delt_Bi}) fully arises from 
the vertical variation of the azimuthal EMF, $\mathcal{\bar E}_y$. 
On the other hand, the field $\bar B_y(z)$ (top right panel of Fig. \ref{fig:Bi_delt_Bi}) results from a combination of the shear term 
and the vertical variation of the radial EMF, $\mathcal{\bar E}_x$. The shear term acts as a source (traditionally, known as the $\Omega$-effect), whereas the radial EMF has a sink effect. 
In the $y\text{--}z$ averaged analysis, both $\bar B_y(x)$ and $\bar B_z(x)$ arise due to their respective EMF terms in the induction equation: $\bar B_y(x)$ arises from the radial variation of $\mathcal{\bar E}_z$ (bottom left panel of Fig. \ref{fig:Bi_delt_Bi}), whereas $\bar B_z(x)$ arises from the radial variation of $\mathcal{\bar E}_y$ (bottom right panel of Fig. \ref{fig:Bi_delt_Bi}). The sharp decay of $\bar B_z(x)$ at the beginning indicates the destruction of the initial field configuration $\bar B_z = B_0 \sin (k_x x)$. 

\begin{figure*}
	\includegraphics[width=1.5\columnwidth]{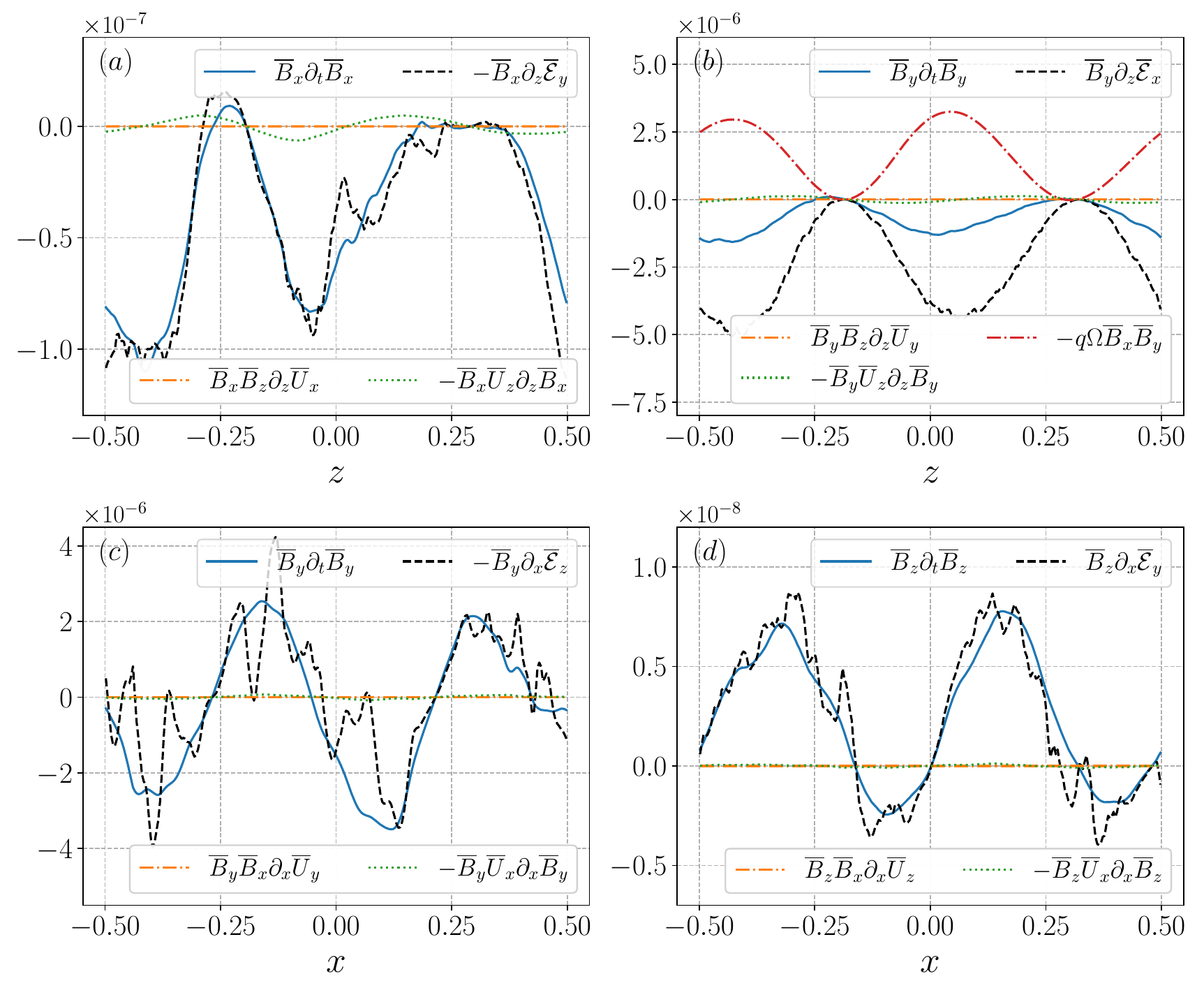} 
	\caption{(Color online)
		The top two panels show the terms from $x\text{--}y$ averaged mean field equation for $\bar B_x(z)$ and $\bar B_y(z)$ on left and right, respectively. The bottom two panels show the terms from $y\text{--}z$ averaged mean field equation for $\bar B_y(x)$ and $\bar B_z(x)$ on left and right, respectively. These are evaluated at $t/T_{\text{orb}} \sim 97$. To understand the contribution of each term to the evolution of mean magnetic fields, we multiply $\bar B_i$ on both sides of the $\partial_t \bar B_i$ equations. 
		The line style and color for each term is the same, as shown in Fig.~\ref{fig:Bi_delt_Bi_t5}.
	}
	\label{fig:Bi_delt_Bi_t600}
\end{figure*}
Next we examine the behaviour of the terms in \Eqs{eq:meanB_xy}{eq:meanB_yz} locally in space instead. 
In Figs~\ref{fig:Bi_delt_Bi_t5} and \ref{fig:Bi_delt_Bi_t600}, we show the individual terms in \Eqs{eq:meanB_xy}{eq:meanB_yz}, for both $x\text{--}y$ (top panels) and $y\text{--}z$ averaging (bottom panels) as function of $z$ and $x$, respectively. 
Fig.~\ref{fig:Bi_delt_Bi_t5} corresponds to the MRI growth phase, evaluated at  
$t/T_{\text{orb}} \sim 5$, and Fig.~\ref{fig:Bi_delt_Bi_t600} corresponds to the fully nonlinear saturation regime, evaluated at $t/T_{\text{orb}} \sim 97$. 
We use the same line style and color for individual terms as that in Fig.~\ref{fig:Bi_delt_Bi}. The overall conclusions remain the same as discussed in the previous paragraph. The azimuthal EMF, $\mathcal{\bar E}_y$, generates the field $\bar B_x(z)$ (top left panel of Fig. \ref{fig:Bi_delt_Bi_t5}), which in turn drives $\bar B_y(z)$ (top right panel of Fig. \ref{fig:Bi_delt_Bi_t5}) through the $\Omega$-effect. The radial EMF, $\mathcal{\bar E}_x$, has a sink effect, reducing the energy of $\bar B_y(z)$. In the nonlinear regime, the radial EMF is seen to dominate over the shear term at the given instance in time, and hence the field $\bar B_y(z)$ decays at that instant (top right panel of Fig.~\ref{fig:Bi_delt_Bi_t600}).
In $y\text{--}z$ averaging, both $\bar B_y(x)$ (bottom left panel of Figs~\ref{fig:Bi_delt_Bi_t5} and \ref{fig:Bi_delt_Bi_t600}) and $\bar B_z(x)$ (bottom right panel of Figs~\ref{fig:Bi_delt_Bi_t5} and \ref{fig:Bi_delt_Bi_t600}) arise due to their respective EMF terms in the induction equation. The contribution from the advection and stretching terms involving only mean fields are negligible for all the cases. Thus, we find that the behaviour of all the terms (in Eqs.~\ref{eq:meanB_xy} and \ref{eq:meanB_yz}) locally in time is consistent with that locally in space. 

In summary, the EMFs $\mathcal{\bar E}_y$ and $\mathcal{\bar E}_z$ play significant roles in dynamo, whereas $\mathcal{\bar E}_x$ acts like a sink on $\bar B_y(z)$. To find the solution to the MRI dynamo problem, we have to formulate the key components of the EMF: $\mathcal{\bar E}_y$ and $\mathcal{\bar E}_z$.

\subsection{Construction of the Electromotive Force} \label{sec:EMF}

In traditional mean-field dynamo theory, the turbulent electromotive force (EMF) is commonly expressed as a linear combination of the mean magnetic field and its derivatives:
\begin{equation}
\mathcal{\bar E}_i = \alpha_{ij} \bar B_j + \beta_{ijk} \bar B_{j,k} \; ,
\end{equation}
where the tensor components $\alpha_{ij}$ and $\beta_{ijk}$ are known as turbulent transport coefficients. However, this assumption of expansion solely with respect to the mean magnetic field may not be sufficient \cite{2010AN....331...14R}, as the form of the EMF directly emerges from the assumption of $\bar U = 0$, disregarding the influence of mean velocity fields. In the context of MRI-driven turbulence, both the large-scale vorticity dynamo and the large-scale magnetic field dynamo are integral components of the overall turbulent behavior \cite{2016MNRAS.462..818B}. Therefore, in constructing the EMF, it is essential to account for the effects of mean velocity fields alongside the mean magnetic field. 
Another challenge in mean-field dynamo theory is determining the numerous unknown transport coefficients involved in the mean EMF. Extracting data from simulations, specifically $\bec{\bar B}$ and $\bec{\mathcal{\bar E}}$, allows for the estimation of these coefficients. However, measurement results often suffer from high levels of noise. To improve the signal and reduce the noise, certain coefficients are typically assumed to be negligible \cite{2015PhRvL.115q5003S, 2016MNRAS.456.2273S, 2020MNRAS.494.4854D}. However, the appropriateness of such fitting assumptions has been a subject of debate \cite{2022ApJ...932....8K}. 

To overcome both limitations, we propose a novel approach that constructs the key components of the EMF in a self-consistent manner without making any assumptions. We utilize the interaction terms arising from the Coriolis force and background shear in the evolution equations for the Faraday tensors to construct the EMF. More detailed information can be found in Appendix~\ref{sec: app_EMF}. Specifically, the azimuthal EMF, $\mathcal{\bar E}_y$, can be expressed as
\begin{widetext}
	\begin{align}
	\mathcal{\bar E}_y = \frac{-1}{q(2-q)\Omega} \Bigg[
	& \left\{ - q \mathcal{D}_t \bar F_{yz} + (2-q) \mathcal{D}_t \bar F_{zy} \right\}
	+ \left \{ q\bar F_{yk} +(2-q)\bar F_{ky} \right \} \partial_k \bar U_z     
	- \left\{ q\bar F_{kz} +(2-q)\bar F_{zk} \right \} \partial_k \bar U_y         \nonumber\\
	& + \frac{1}{\rho}\left \{q\bar M_{zk} +(2-q)\bar R_{zk} \right\} \partial_k \bar B_y
	- \frac{1}{\rho}\left\{q\bar R_{yk} +(2-q)\bar M_{yk} \right\}\partial_k \bar B_z  \nonumber\\
	& + \bar B_k \left\{ q \left(\langle u_y \partial_k u_z \rangle + \frac{\langle b_z \partial_k b_y \rangle}{\mu_0 \rho} \right) - (2-q) \left(\langle u_z \partial_k u_y \rangle + \frac{\langle b_y \partial_k b_z \rangle}{\mu_0 \rho} \right) \right\}   
	+ \mathcal{\bar T}_{y} \Bigg] .
	\label{eq:emfy}
	\end{align}
\end{widetext}
It is worth noting that the EMF consists of terms that are proportional to: $(a)$ mean magnetic fields, $(b)$ gradient of mean magnetic fields, $(c)$ gradient of mean velocity fields, and $(d)$ nonlinear three-point terms $(\mathcal{T})$. The proportionality coefficients depend on factors such as rotation, shear rate, and the correlators associated with different fluctuating fields. Similarly, the vertical component of the EMF, $\mathcal{\bar E}_z$, can be derived, and its mathematical expression is available in Appendix~\ref{sec: app_EMF}.

\begin{figure}
	\includegraphics[width=\columnwidth]{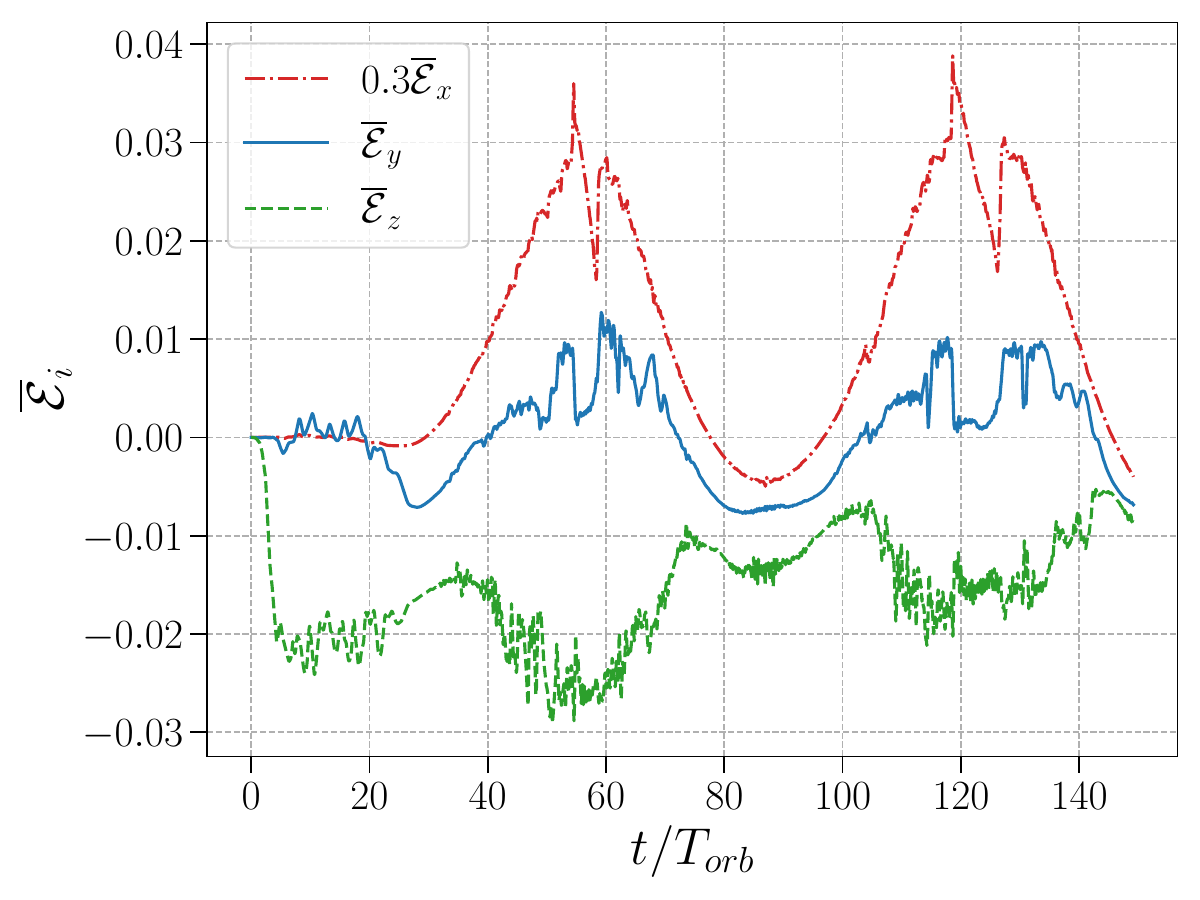}  
	\caption{(Color online)		 
		\textit{The time evolution of the volume-averaged EMF is presented}. The dash-dotted red, solid blue, and dashed green lines correspond to the $x-$, $y-$, and $z-$components of EMF, respectively. To help visualisation, the $x-$component $\mathcal{\bar E}_x$ has been scaled down with a factor of 0.3. }
	\label{fig:ts_emf}
\end{figure}
For a comprehensive understanding, we present the individual components of the EMF obtained from the volume-averaged analysis, as depicted in \Fig{fig:ts_emf}. It is evident that the EMF components $\mathcal{\bar E}_x$ and $\mathcal{\bar E}_y$ exhibit a cyclic pattern over time, with alternating positive and negative values. In contrast, the EMF component $\mathcal{\bar E}_z$ consistently remains negative throughout the entire duration of the analysis.

\subsection{Generation of Radial Magnetic Fields} \label{sec:radial_B_fied}

\begin{figure*}
		\centering
		\includegraphics[width=0.88\columnwidth]{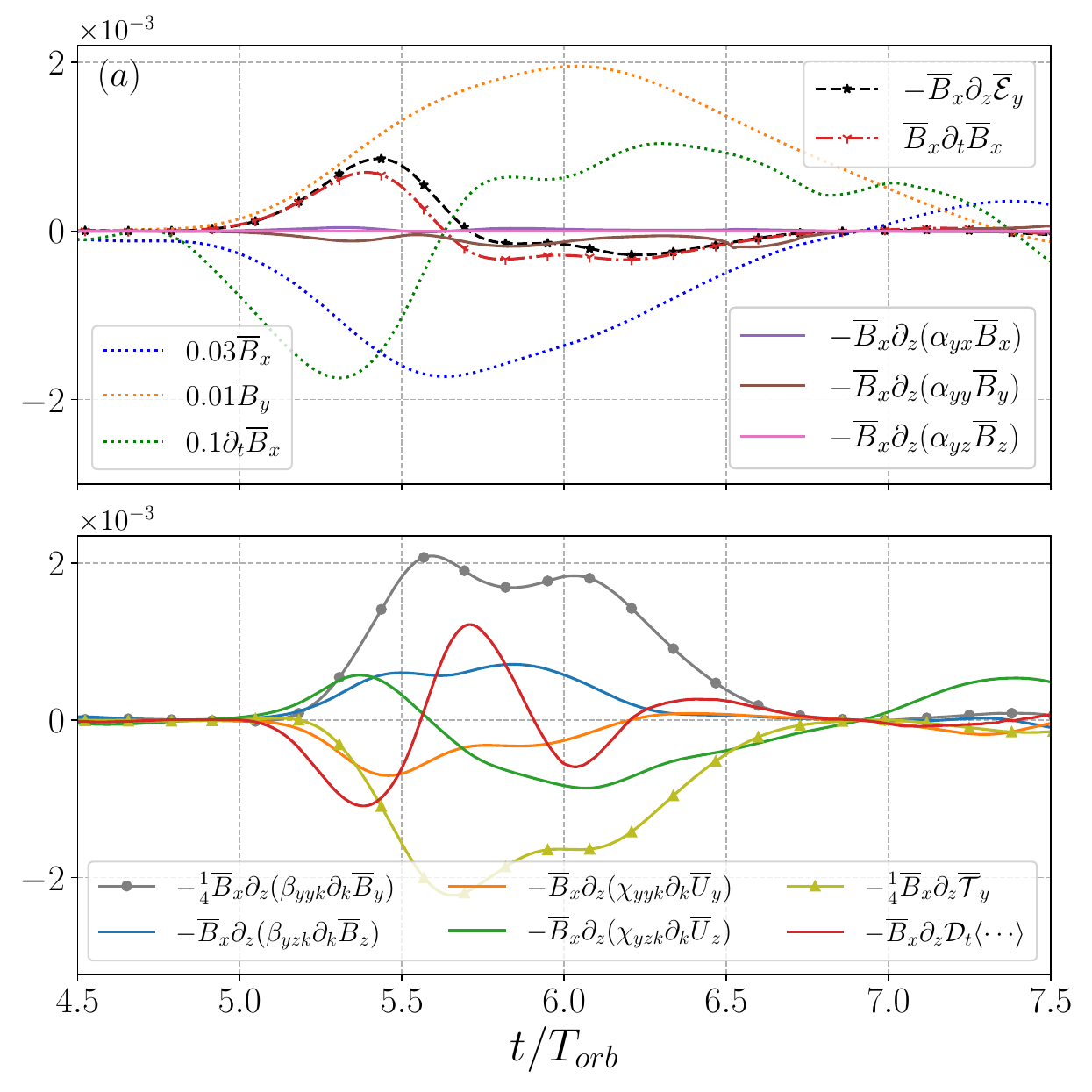} 
		\hspace{-0.3cm}
		\includegraphics[width=1.20\columnwidth]{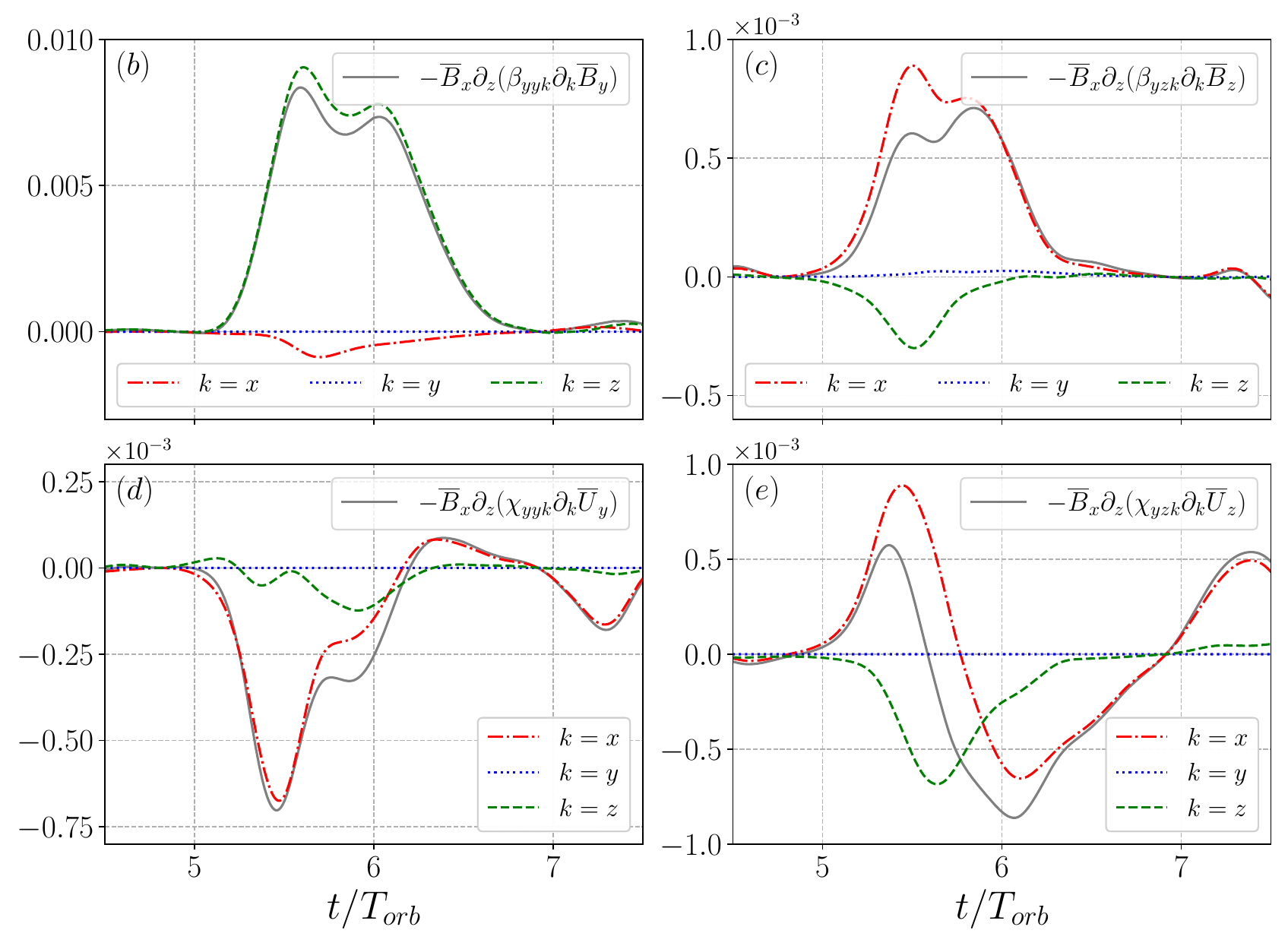} 
		\caption{(Color online)
			$(a)$ \textit{The left panels depict the terms arising from the vertical variation of the azimuthal EMF} (Eq.~\ref{eq:emfy}) \textit{responsible to generate the $x\text{--}y$-averaged field $\bar B_x(z)$ during the growth and initial saturation phase of MRI.}  To improve visual clarity, the numerous terms associated with the EMF are distributed across two left panels. In order to assess the individual contributions of these terms to the evolution of $\bar B_x(z)$, we multiply $-\bar B_x(z)$ on both sides of the $\partial_z \mathcal{\bar E}_y$ equation (as described by Eq.\ref{eq:meanBx_xy}). These are evaluated at $z=-0.15$. The curves represent distinct terms: the dash-dotted red curve corresponds to $\bar B_x \partial_t \bar B_x$, the dashed black curve corresponds to the corresponding EMF term, the solid grey curve corresponds to the term proportional to $\partial_k \bar B_y$, the solid blue curve corresponds to the term proportional to $\partial_k \bar B_z$, the solid orange curve corresponds to the term proportional to $\partial_k \bar U_y$, the solid green curve corresponds to the term proportional to $\partial_k \bar U_z$, the solid olive curve corresponds to the nonlinear three-point term, the solid red curve corresponds to the time derivative of the Faraday-tensor terms, and the solid purple, brown, and pink curves correspond to the terms proportional to the $x$-, $y$-, and $z$-components of the $\bar B$ fields, respectively. Notably, we have utilized markers only for the most significant curves. 
			The middle and right panels illustrate the terms proportional to different components of the field gradients: $(b)$ the term proportional to $\partial_k \bar B_y$, $(c)$ the term proportional to $\partial_k \bar B_z$, $(d)$ the term proportional to $\partial_k \bar U_y$, and $(e)$ the term proportional to $\partial_k \bar U_z$.}
		\label{fig:BxdEydz_xy_t}
	\end{figure*}

We have seen that the $x\text{--}y$ averaged mean field $\bar B_x(z)$ is solely determined by the vertical variation of the azimuthal EMF, $\mathcal{\bar E}_y$. The main challenge in mean-field dynamo theory is to identify the term responsible for generating $\bar B_x$ via $\mathcal{\bar E}_y$. In \Fig{fig:BxdEydz_xy_t}, we present individual terms of $\mathcal{\bar E}_y$ (Eq.~\ref{eq:emfy}) in the MRI growth to the early saturation phase. We compute these terms at $z=-0.15$. To assess the contribution of each term in the evolution of $\bar B_x$ (Eq.~\ref{eq:meanBx_xy}), we multiply $-\bar B_x$ on both sides of the equation for $\partial_z \mathcal{\bar E}_y$. Two crucial curves that aid in determining whether the magnetic field is growing or decaying with time are the EMF term, depicted as the dashed black line with star markers, and $\bar B_x \partial_t \bar B_x$, illustrated as the dash-dotted red line with tri-down markers. 
We observe that the field $\bar B_x(z)$, represented by the dotted blue line, undergoes amplification during the growth phase (with negative growth and in opposite phase to $\bar B_y(z)$, shown as the dotted orange line) in the range of $t/T_{\text{orb}} \approx 5$ to $5.6$. Subsequently, it decays, leading to saturation.	The primary driver for the growth of $\bar B_x$ is the term proportional to $\partial_k \bar B_y$ (depicted by the solid grey line with circle markers). In particular, the $k=z$ component of this term plays a crucial role, as illustrated in \Fig{fig:BxdEydz_xy_t}b. During the growth phase, there are two additional source terms. One originates from the term proportional to $\partial_k \bar B_z$ (illustrated by the solid blue line) with $k=x$ (see \Fig{fig:BxdEydz_xy_t}c), while the other arises from the term proportional to $\partial_k \bar U_z$ (represented by the solid green line) with $k=x$ (\Fig{fig:BxdEydz_xy_t}d). In the initial decay phase (around $t/T_{\text{orb}} \approx 5.6$ to $7$), the most significant role is played by the nonlinear three-point term (depicted by the solid olive line with triangle markers). The term proportional to $\partial_k\bar U_z$ with $k=x$ (\Fig{fig:BxdEydz_xy_t}d), which previously acted as a source during the growth phase, now acts as a sink. These contributions collectively lead to the saturation of $\bar B_x$. Notably, the terms proportional to $\bar B_i$ are negligible in both the growth and initial decay phases.


\begin{figure*}
		\centering
		\includegraphics[width=0.88\columnwidth]{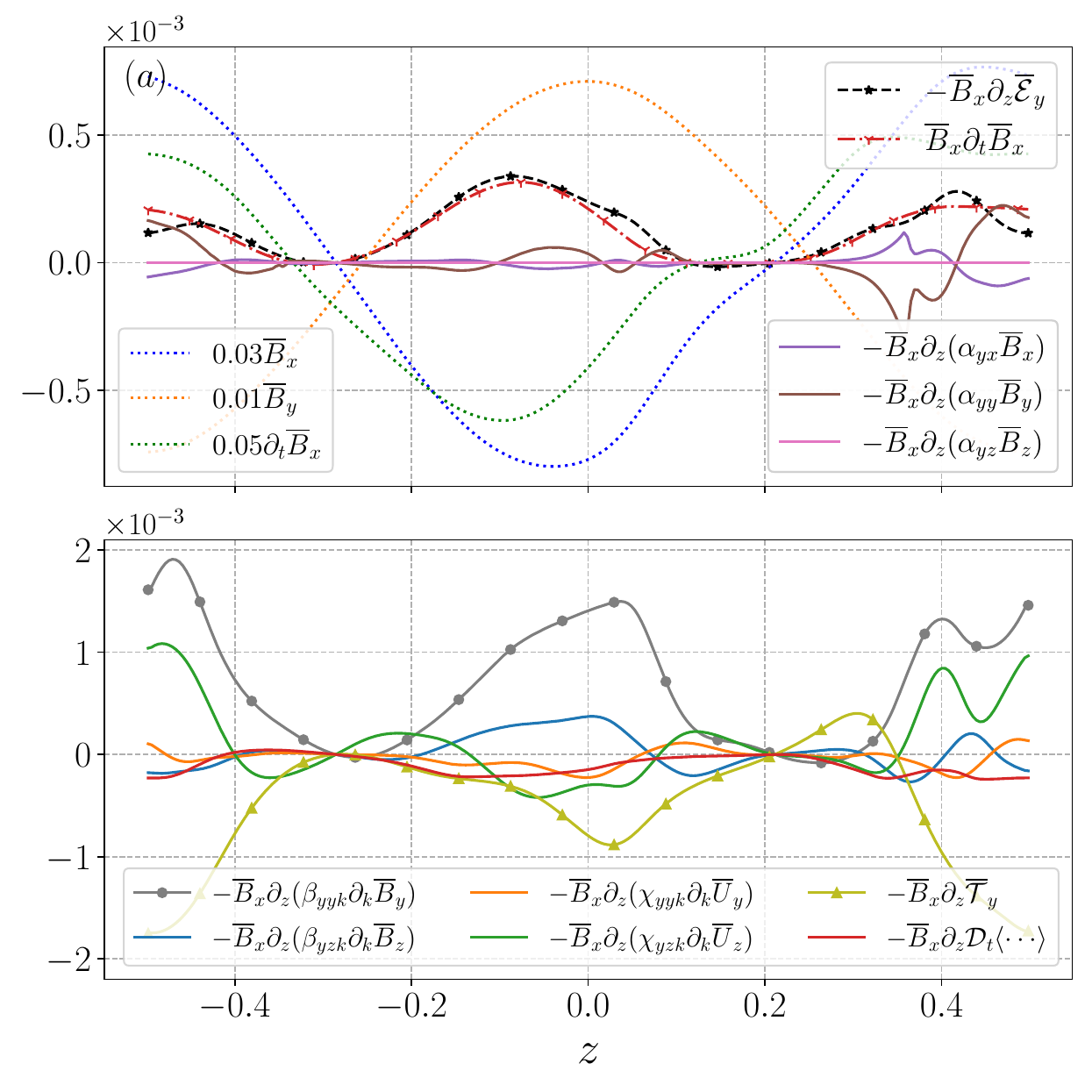} 
		\hspace{-0.3cm}
		\includegraphics[width=1.20\columnwidth]{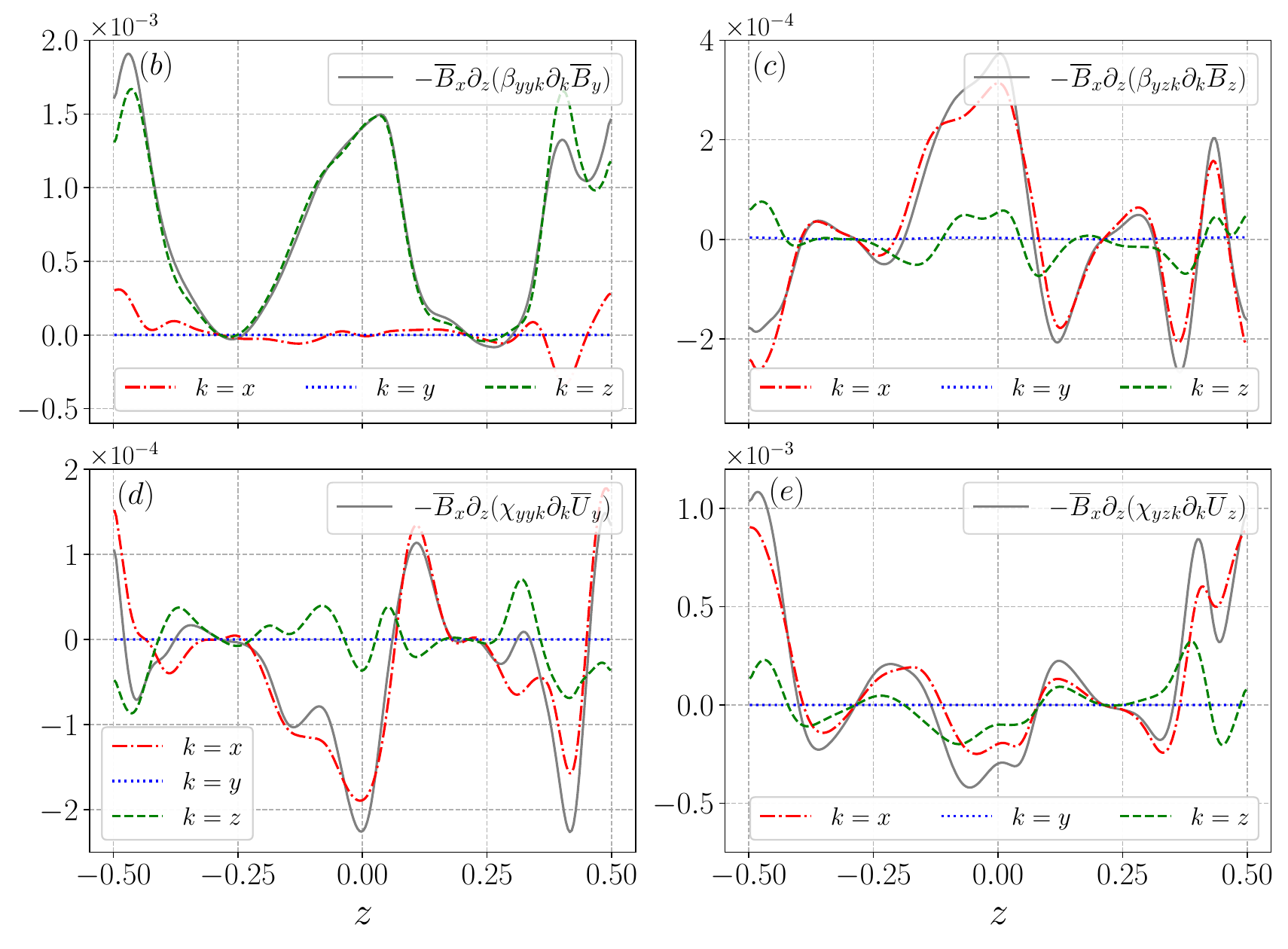} 
		\caption{(Color online)
			$(a)$ \textit{The left panels present the terms arising from the vertical variation of the azimuthal EMF} (Eq.~\ref{eq:emfy}) \textit{that contribute to the generation of the $x\text{--}y$-averaged field $\bar B_x(z)$ in the MRI growth phase.} These are obtained through time averaging from $t/T_{\text{orb}} = 5\rightarrow 5.5$. To enhance visual clarity, the numerous terms associated with the EMF are distributed across two left panels. To evaluate the individual contributions of these terms to the evolution of $\bar B_x(z)$, we multiply $-\bar B_x(z)$ on both sides of the $\partial_z \mathcal{\bar E}_y$ equation (as described by Eq.\ref{eq:meanBx_xy}). The line style and color for each term are consistent with those in Fig.~\ref{fig:BxdEydz_xy_t}. It is worth noting that markers are employed only for the most significant curves.
			The middle and right panels illustrate the terms proportional to different components of the field gradients: $(b)$ the term proportional to $\partial_k \bar B_y$, $(c)$ the term proportional to $\partial_k \bar B_z$, $(d)$ the term proportional to $\partial_k \bar U_y$, and $(e)$ the term proportional to $\partial_k \bar U_z$.}
		\label{fig:BxdEydz_xy_t30}
	\end{figure*}	
Next, we take time averages from $t/T_{\text{orb}} = 5\rightarrow 5.5$, of all the terms considered in the previous figure (\Fig{fig:BxdEydz_xy_t}), in the MRI growth phase and show their behaviour locally in space. Such a study can explain how the $x\text{--}y$ averaged field $\bar B_x(z)$ is generated in detail. 
In Fig.~\ref{fig:BxdEydz_xy_t30}, the individual terms of $\partial_z \mathcal{\bar E}_y$ vary in $z$, and we use the same line style and color for each term as that in Fig.~\ref{fig:BxdEydz_xy_t}. We again  multiply $-\bar B_x(z)$ on both sides of the $\partial_z \mathcal{\bar E}_y$ equation to understand the contribution of each term to the evolution of $\bar B_x$, following \Eq{eq:meanBx_xy}. We find again that the dominant source term is the term proportional to $\partial_k \bar B_y$ with $k=z$ (top middle panel of Fig.~\ref{fig:BxdEydz_xy_t30}). Some contributions from the terms proportional to $\partial_k \bar B_z$ with $k=x$ (bottom middle panel of Fig.~\ref{fig:BxdEydz_xy_t30}) and $\partial_k \bar U_z$ with $k=x$ (bottom right panel of Fig.~\ref{fig:BxdEydz_xy_t30}) also arise in the growth of $\bar B_x(z)$, but they are not acting as sources throughout $z$.

\begin{figure*}
		\includegraphics[width=\columnwidth]{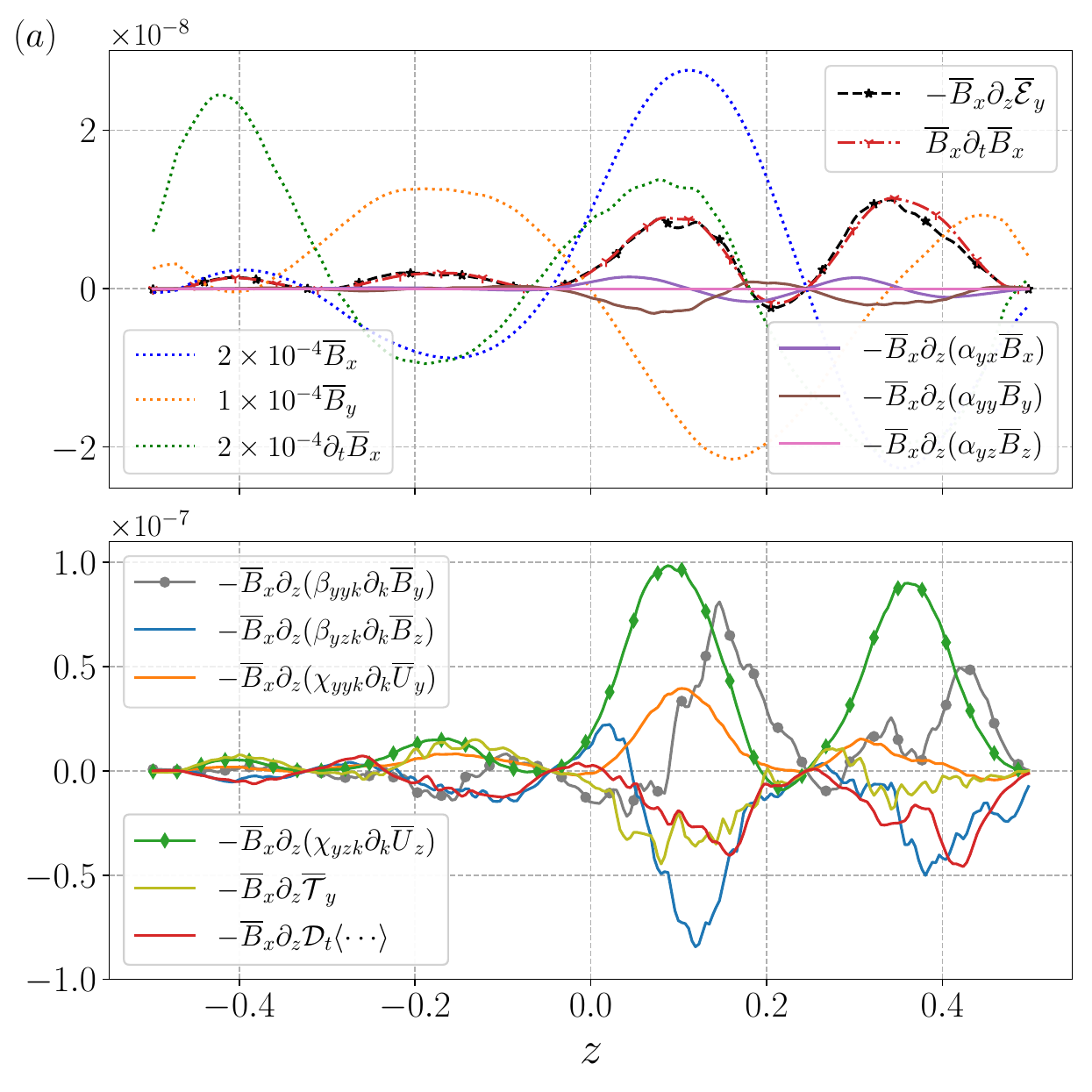} 
		\includegraphics[width=\columnwidth]{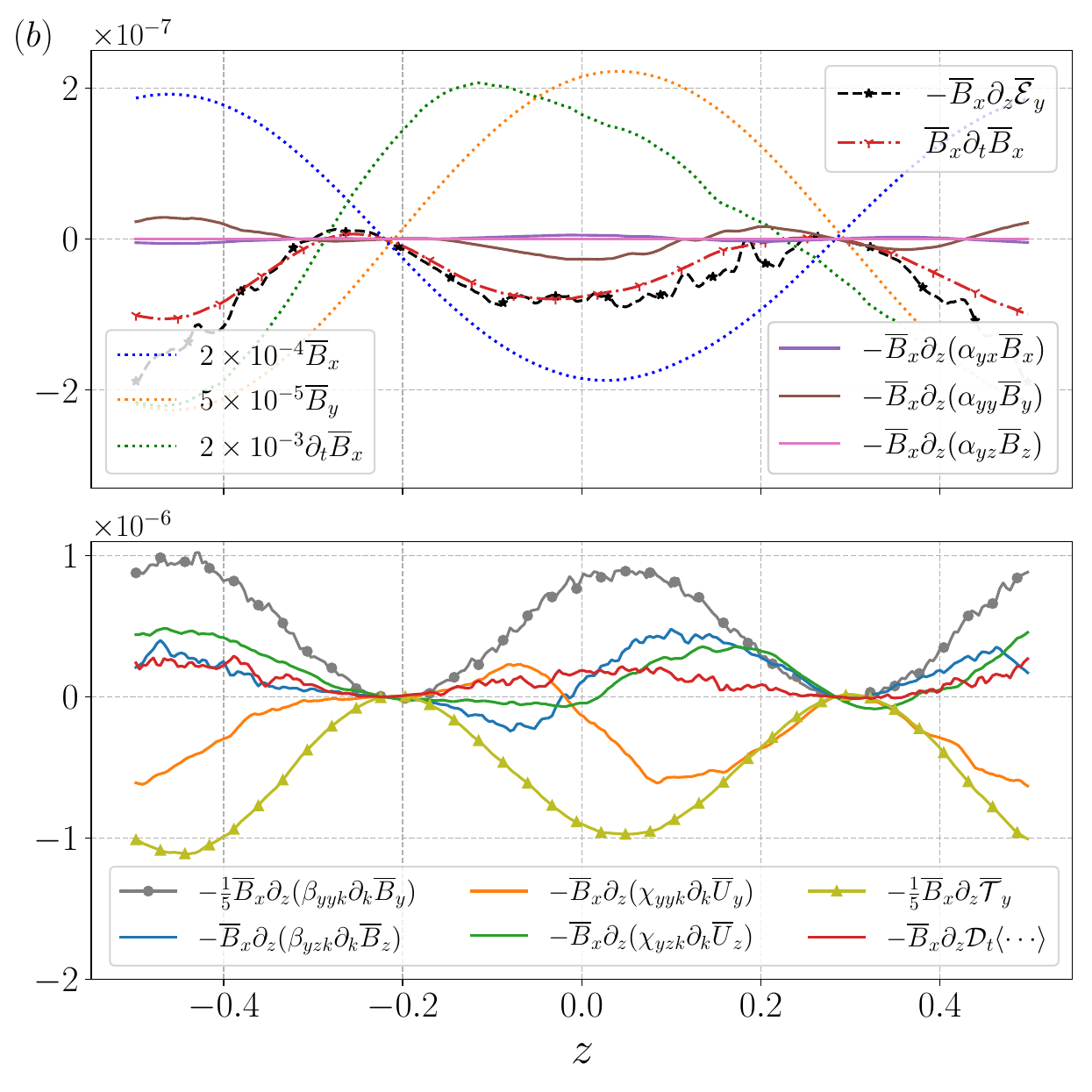} 
		\includegraphics[width=\columnwidth]{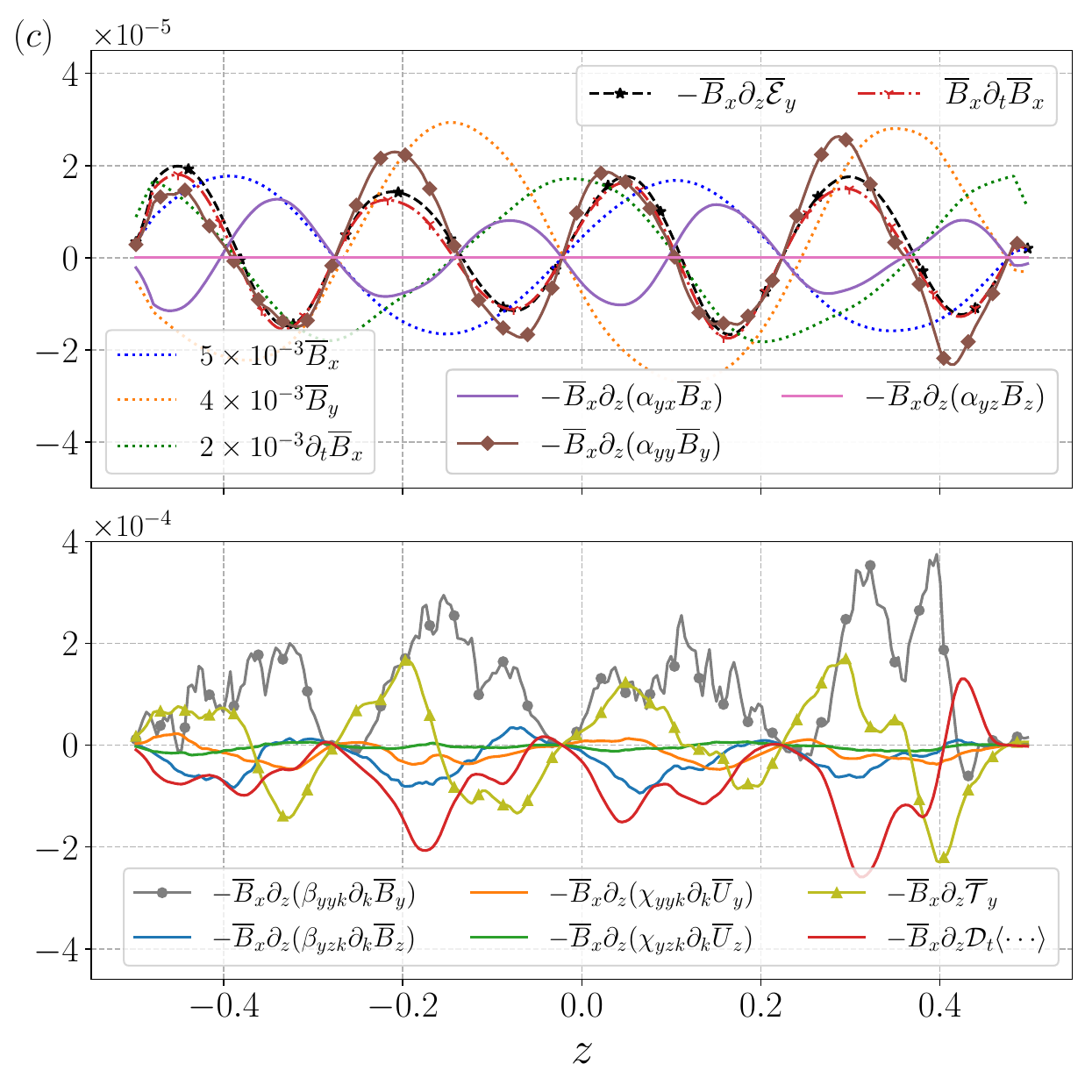}
		\includegraphics[width=\columnwidth]{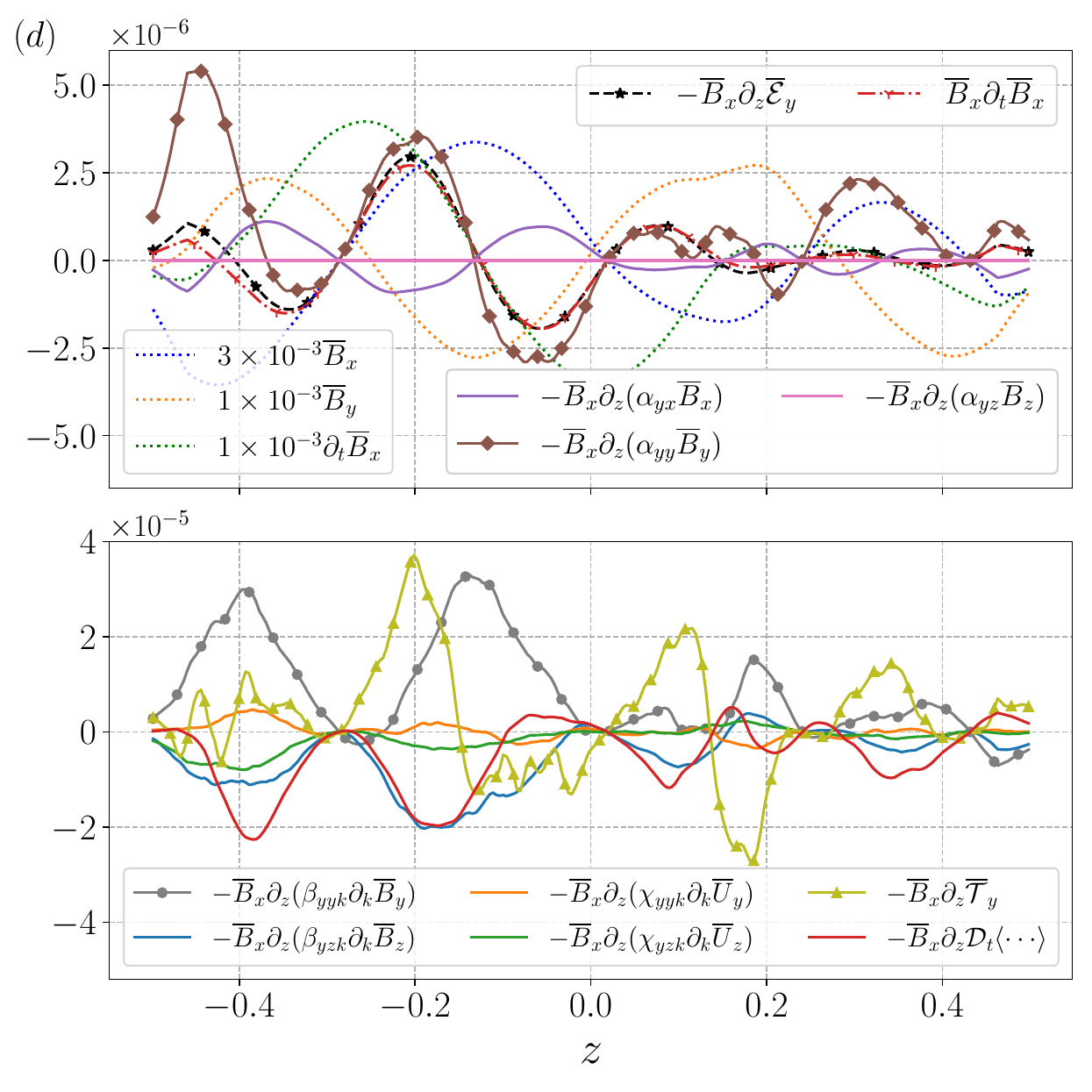} 
		\caption{(Color online)
			\textit{The terms responsible for generating the $x\text{--}y$ averaged mean field $\bar B_x(z)$ via the vertical variation of $\mathcal{\bar E}_y$ in the MRI nonlinear regime.}
			The four panels correspond to the four different instances of time: $t/T_{\text{orb}} \sim 80$ (top left), $t/T_{\text{orb}} \sim 97$ (top right), $t/T_{\text{orb}} \sim 43$ (bottom left), and $t/T_{\text{orb}} \sim 107$ (bottom right). The azimuthal EMF, $\mathcal{\bar E}_y$, is negative in the top two panels, whereas $\mathcal{\bar E}_y$ is positive in the bottom two panels. The top left and right panels are evaluated at the given instances in time when $\bar B_x(z)$ grows and decays, respectively. The bottom two panels are a combination of both the growing and decaying phases with $z$. 
			We keep the same line style and color for each term, as shown in Fig.~\ref{fig:BxdEydz_xy_t}.		
		}
		\label{fig:BxdEydz_xy_NL}
\end{figure*}
Next, we study the dynamo in nonlinear regime. Using Fig.~\ref{fig:BxdEydz_xy_NL}, we describe the mechanism by which the field $\bar B_x(z)$ grows and decays alternatively, via the vertical variation of $\mathcal{\bar E}_y$. 
We have seen that the azimuthal EMF, $\mathcal{\bar E}_y$, maintains a cyclic nature, i.e., the magnitude of $\mathcal{\bar E}_y$ can be either positive or negative at different instances of time (e.g., Fig.~\ref{fig:ts_emf}). Here, we perform the $x\text{--}y$ averaged analysis at different times when $\mathcal{\bar E}_y$ can be either positive or negative, to obtain an overall behaviour of both, growth and decay of $\bar B_x(z)$, locally in space. In particular, we would like to know whether the terms which were responsible for growth of the large-scale fields continue to
persist in the nonlinear regime. The computations are performed at $t/T_{\text{orb}} \sim 80$ (top left), $t/T_{\text{orb}} \sim 97$ (top right), $t/T_{\text{orb}} \sim 43$ (bottom left), and $t/T_{\text{orb}} \sim 107$ (bottom right). 
The top two panels of Fig.~\ref{fig:BxdEydz_xy_NL} correspond to the negative $\mathcal{\bar E}_y$, whereas the bottom two panels are for the positive $\mathcal{\bar E}_y$.
In the top left panel of Fig.~\ref{fig:BxdEydz_xy_NL}, we see that the field $\bar B_x(z)$ grows along $z>0$. 
The term proportional to $\partial_k \bar U_z$ (solid green curve) with $k=x$ (not shown here) is the dominant term responsible for the growth of $\bar B_x(z)$. 
The other two source terms are those proportional to $\partial_k \bar B_y$ (solid grey curve) with $k=z$ (not shown here) 
and $\partial_k \bar U_y$ (solid orange curve) with $k=x$ (not shown here). 
The nonlinear three-point term (solid yellow/light-green curve) and the term proportional to $\partial_k \bar B_z$ (solid blue curve) with $k=x$ (not shown here) behave like sinks. 
The terms proportional to $\bar B_i$ (solid purple, brown, and pink curves for $i=x, y,\ \text{and}\ z$, respectively) are negligible. 
In the top right panel of Fig.~\ref{fig:BxdEydz_xy_NL}, the field $\bar B_x(z)$ is seen to be decaying along all $z$. 
The nonlinear three-point term plays a significant role in reducing the energy of $\bar B_x$. 
The term proportional to $\partial_k \bar U_y$ (solid orange curve) with $k=x$ (not shown here) acts like a sink here also. The terms proportional to $\bar B$ are negligible.

In the bottom two panels of Fig.~\ref{fig:BxdEydz_xy_NL}, we see that the field $\bar B_x(z)$ grows and decays cyclically in $z$. In both cases, the overall behaviour of different terms remains the same as before. Again, the term proportional to $\partial_k \bar B_y$ (solid grey curve) 
with $k=z$ (not shown here) acts like a source throughout the $z$, whereas the terms associated with 
$\partial_k \bar B_z$ (solid blue curve) with $k=x$ (not shown here) and time-derivative (solid red curve) have sink effects mostly. 
There are two significant behaviours in these two cases. 
$(a)$ The terms proportional to $\bar B_i$ (solid purple, brown, and pink curves for $i=x, y,\ \text{and}\ z$, respectively) 
are not negligible here, unlike previous cases. The term proportional to $\bar B_y$ (solid brown curve) appears to follow the signal, i.e., the term $\bar B_x \partial_t \bar B_x$. On the other hand, the term proportional to $\bar B_x$ (solid purple curve) appears opposite to the signal. 
$(b)$ The nonlinear three-point term (solid olive line) behaves like either a source or a sink. Similar to the term proportional to $\bar B_y$ (solid brown curve), the nonlinear term also follows the pattern of $\bar B_x \partial_t \bar B_x$ with much higher amplitudes.

In summary, the growth and the nonlinear saturation of the $x\text{--}y$ averaged field $\bar B_x(z)$ arises through the vertical variation of the azimuthal EMF, i.e., $\partial_z \mathcal{\bar E}_y$. The EMF $\mathcal{\bar E}_y$ consists of four different types of terms proportional to the mean magnetic fields, the gradient of mean magnetic fields, the gradient of mean velocity fields, and nonlinear three-point terms. The proportionality coefficients are functions of the shear rate, rotation, and correlators associated with different fluctuating fields. The term proportional to $\partial_z \bar B_y$ plays a significant role in the growth of $\bar B_x$ both in the growth and saturation regimes. 
In the MRI growth regime, the term proportional to $\partial_x \bar B_z$ also grows $\bar B_x(z)$. The decay of $\bar B_x$ is primarily due to the three-point term, which is the reason for the nonlinear saturation of $\bar B_x$. The roles of certain terms depend on the sign of $\mathcal{\bar E}_y$. In the MRI nonlinear regime, the term proportional to $\partial_x \bar U_z$ helps in the growth of $\bar B_x$ for $\mathcal{\bar E}_y < 0$, whereas it has a negligible sink effect for $\mathcal{\bar E}_y > 0$. The term proportional to $\bar B_y$ is negligible for $\mathcal{\bar E}_y < 0$, whereas it has a dual effect (i.e., both source and sink at the same time but at different points in space) following the pattern of the signal, i.e., the term $\bar B_x \partial_t \bar B_x$, for $\mathcal{\bar E}_y > 0$. 
The nonlinear three-point term acts like turbulent resistivity for $\mathcal{\bar E}_y < 0$, whereas it has a dual effect acting as both source and sink for $\mathcal{\bar E}_y > 0$.

Next, we explore the term proportional to $\partial_z \bar B_y$ of the EMF $\mathcal{\bar E}_y$ (see equation~\ref{eq:emfy}) in more detail. The proportionality coefficient carries physical insight for the $\bar B_x(z)$ generation mechanism. As the coefficient is a function of shear rate, rotation, and correlators associated with kinetic and magnetic fluctuations (more specifically, $\bar R_{zz}$ and $\bar M_{zz}$, respectively), the mechanism is named as `\textit{rotation-shear-current effect}.' 

\textit{\textbf{Rotation-shear-current effect:}}
\begin{figure}
	\includegraphics[width=\columnwidth]{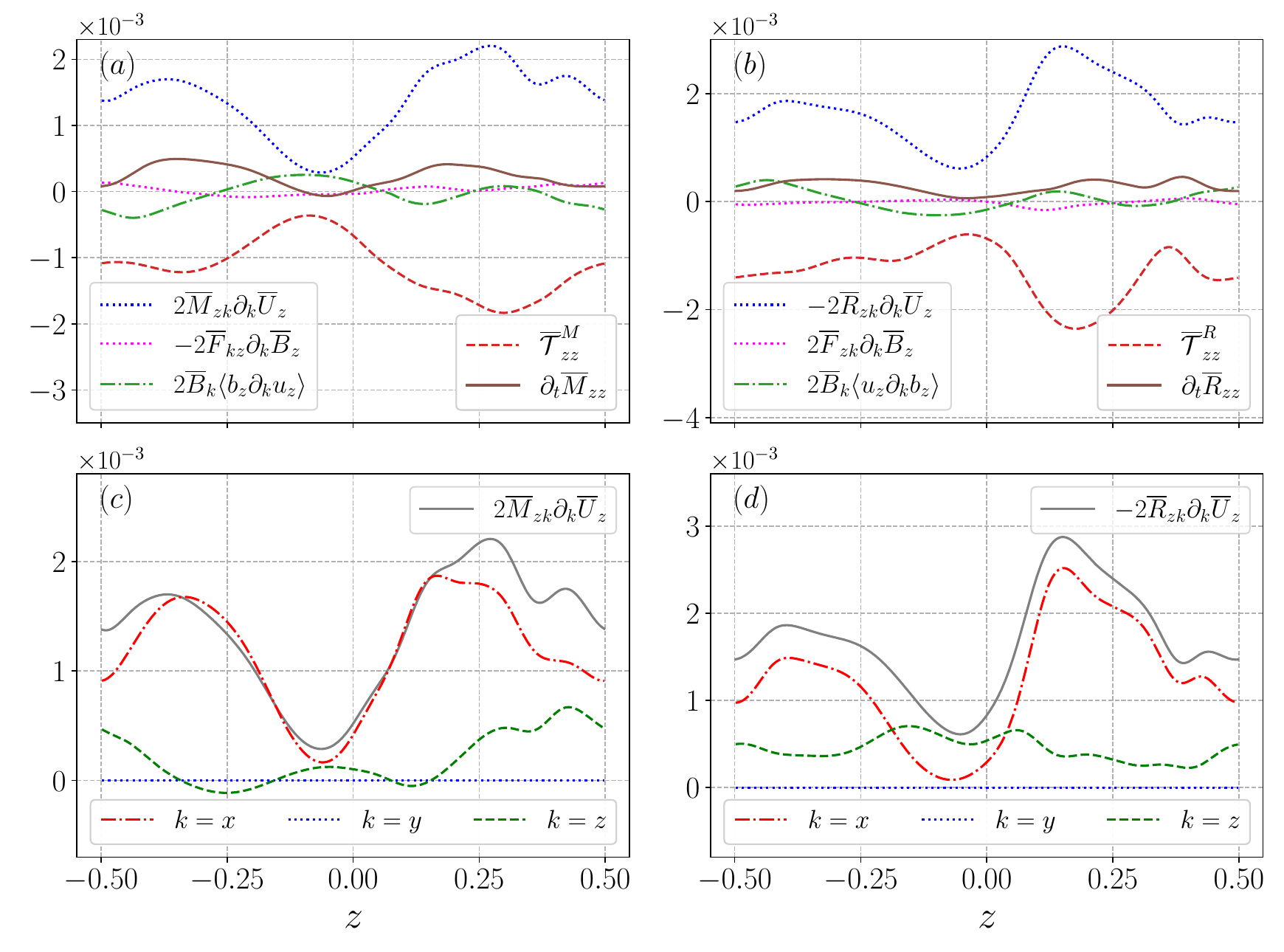} 
	\caption{(Color online)
		\textit{The generation mechanism of $(a) \bar M_{zz}$ and $(b) \bar R_{zz}$ during the MRI growth phase is examined.} The top panels illustrate the contributions obtained from the $x\text{--}y$ averaged equations for $(a) \bar M_{zz}$ and $(b) \bar R_{zz}$. By identifying the source terms from the top panels, we present the individual components contributing to $\bar M_{zz}$ (bottom left panel) and $\bar R_{zz}$ (bottom right panel). The computations involve time averaging over the interval $t/T_{\text{orb}} = 5 \rightarrow 5.5$.}  
	\label{fig:Mzz_Rzz_t5}
\end{figure}
\begin{figure}
	\includegraphics[width=\columnwidth]{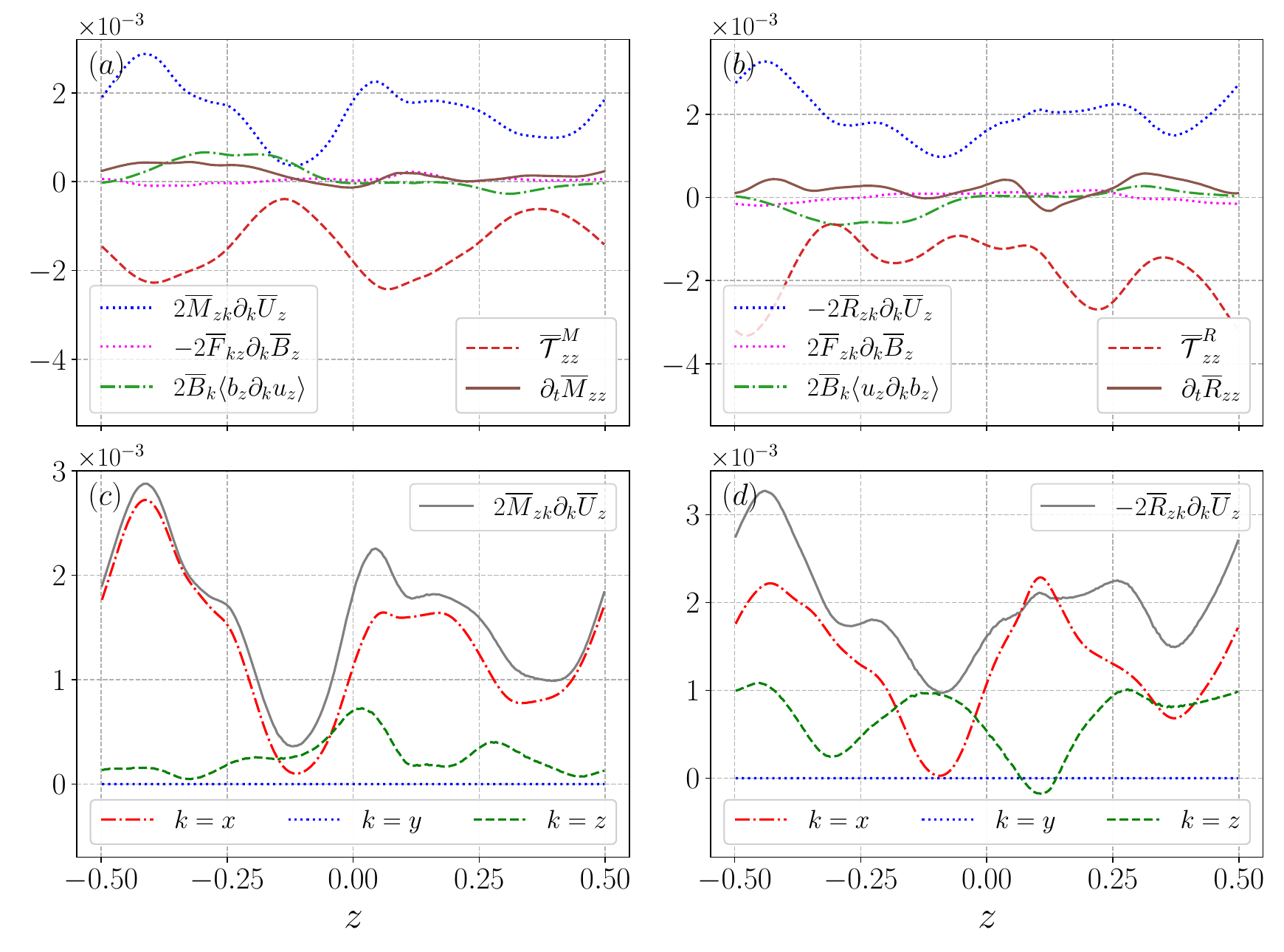} 
	\caption{(Color online)
		\textit{The generation mechanism of $(a) \bar M_{zz}$ and $(b) \bar R_{zz}$ during the MRI nonlinear phase is examined.} The top panels illustrate the contributions obtained from the $x\text{--}y$ averaged equations for $(a) \bar M_{zz}$ and $(b) \bar R_{zz}$. By identifying the source terms from the top panels, we present the individual components contributing to $\bar M_{zz}$ (bottom left panel) and $\bar R_{zz}$ (bottom right panel). The computations are performed at $t/T_{\text{orb}} = 20$.}
	\label{fig:Mzz_Rzz_t20}
\end{figure}
The dynamo mechanism responsible for generating $\bar B_x(z)$ from $\bar B_y(z)$ through the rotation-shear-current effect relies on the presence of correlators $\bar M_{zz}$ and $\bar R_{zz}$. Understanding the formation process of these correlators is crucial for establishing the connections between the dynamo process and the angular momentum transport in the system. To investigate this, we examine the individual terms appearing in the evolution equations for $\bar M_{zz}$ (Eq.~\ref{eq:Mij_exact}) and $\bar R_{zz}$ (Eq.~\ref{eq:Rij_exact}), which are displayed in the upper panels of \Figs{fig:Mzz_Rzz_t5}{fig:Mzz_Rzz_t20}. Specifically, \Fig{fig:Mzz_Rzz_t5} corresponds to the MRI growth phase, obtained through time averaging from  $t/T_{\text{orb}} = 5 \rightarrow 5.5$, while \Fig{fig:Mzz_Rzz_t20} represents the nonlinear phase, evaluated at $t/T_{\text{orb}} = 20$. Physically, $\bar M_{zz}$ and $\bar R_{zz}$ represent turbulent magnetic and kinetic energy densities (multiplied by two) in the vertical components of the fields, respectively. Consequently, both $\bar M_{zz}$ and $\bar R_{zz}$ remain positive throughout. It makes the positive term in $\partial_t \bar M_{zz}$ as a source, whereas the negative term behaves like a sink. The same holds for the terms in $\partial_t \bar R_{zz}$. By identifying the dominant source terms from the upper panels of \Figs{fig:Mzz_Rzz_t5}{fig:Mzz_Rzz_t20}, we examine their components in the lower panels of the same figures. We see that the dominant source term for $\bar M_{zz}$ is the stretching term, $2\bar M_{zk}\partial_k \bar U_z$ (blue dotted line in Figs.~\ref{fig:Mzz_Rzz_t5}$a$ and \ref{fig:Mzz_Rzz_t20}$a$) with $k=x$ (Figs.~\ref{fig:Mzz_Rzz_t5}$c$ and \ref{fig:Mzz_Rzz_t20}$c$), whereas the nonlinear three-point term (red dashed line in Figs.~\ref{fig:Mzz_Rzz_t5}$a$ and \ref{fig:Mzz_Rzz_t20}$a$) behaves as the dominant sink. This behavior remains consistent in both the growth and nonlinear phases of turbulence. Similar processes are seen for the $\bar R_{zz}$ evolution---the stretching term, $-2\bar R_{zk}\partial_k \bar U_z$ (blue dotted line in Figs.~\ref{fig:Mzz_Rzz_t5}$b$ and \ref{fig:Mzz_Rzz_t20}$b$) with $k=x$ (Figs.~\ref{fig:Mzz_Rzz_t5}$d$ and \ref{fig:Mzz_Rzz_t20}$d$), acts as a source, whereas the nonlinear three-point term (red dashed line in Figs.~\ref{fig:Mzz_Rzz_t5}$b$ and \ref{fig:Mzz_Rzz_t20}$b$) turns out to be the sink as usual. Thus, the presence of a mean (vertical) velocity field is necessary for the operation of the rotation-shear-current effect. In other words, \textit{mean magnetic field dynamo is rendered inoperative without mean velocity field dynamo}. Further exploration of the generation process of $\bar M_{xz}$ and $\bar R_{xz}$ is needed for a comprehensive understanding.

\begin{figure*}
	\includegraphics[width=0.68\columnwidth]{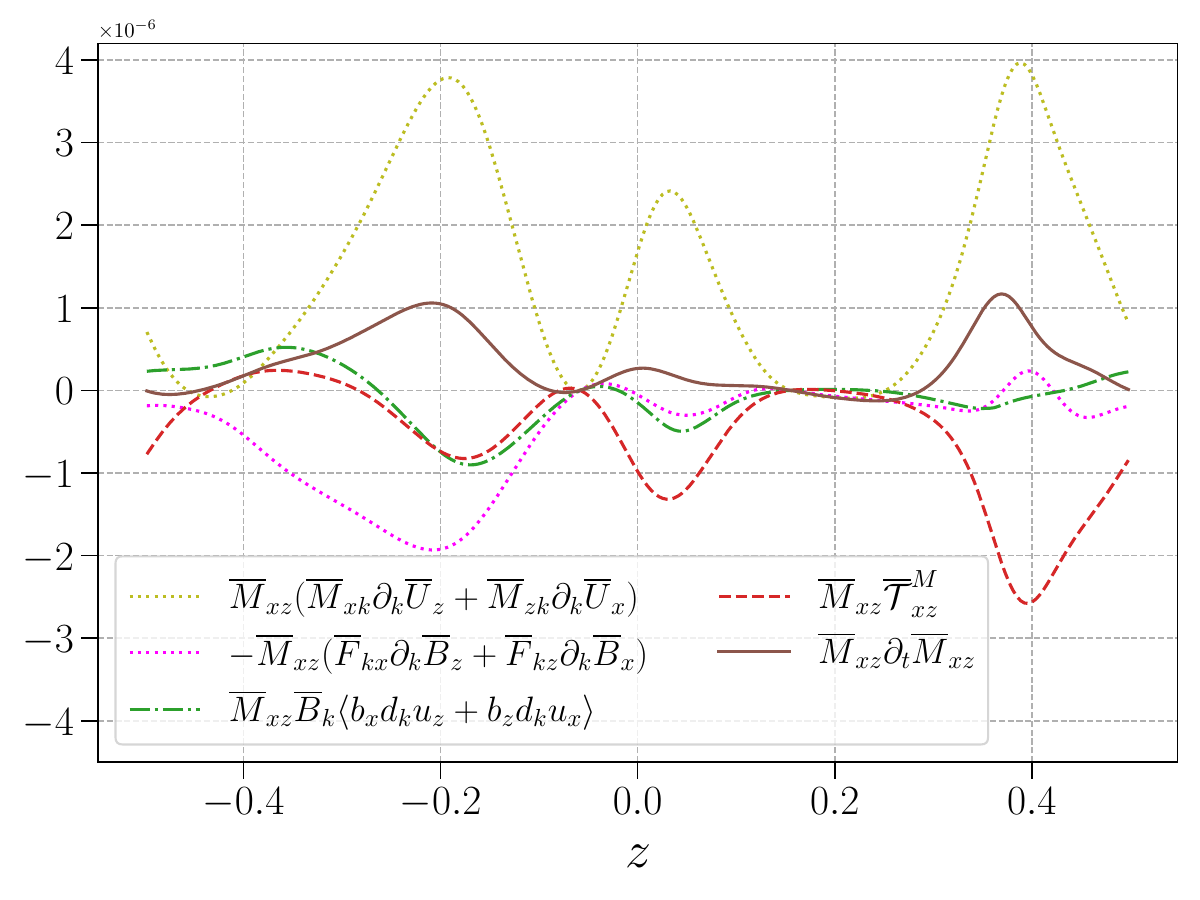} 
	\includegraphics[width=0.68\columnwidth]{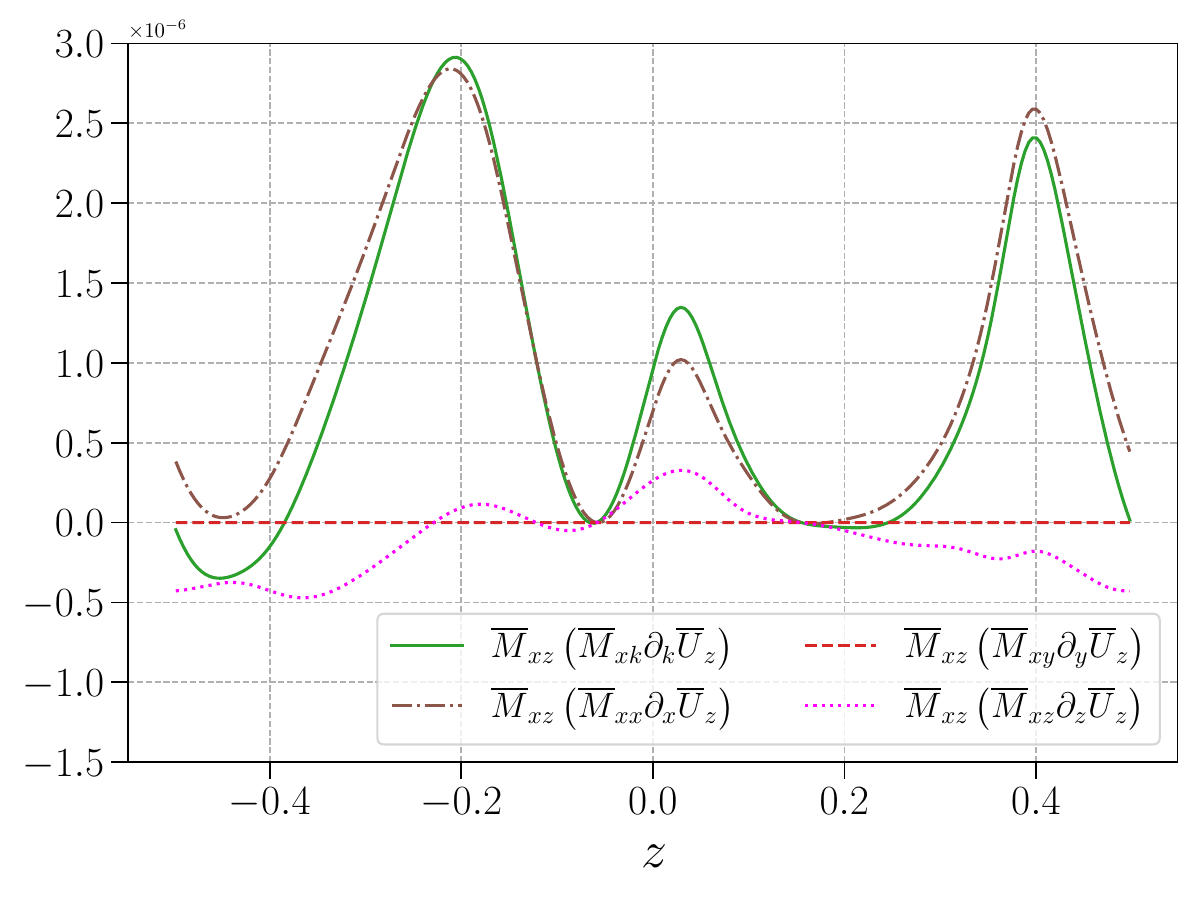}
	\includegraphics[width=0.68\columnwidth]{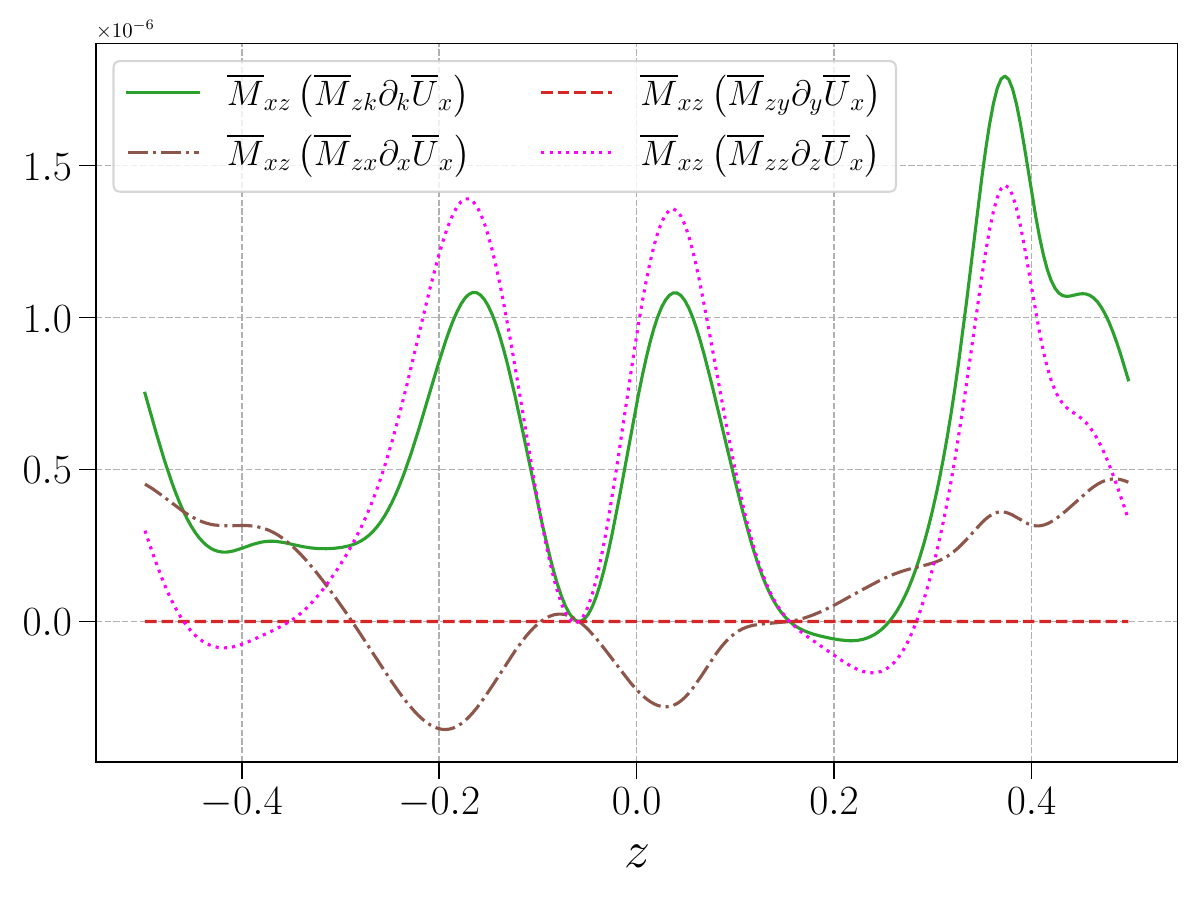} 
	\caption{(Color online)
		\textit{The generation mechanism of $\bar M_{xz} (z)$ in the MRI growth phase.} The left panel shows the individual terms that appeared in $\bar M_{xz} \partial_t \bar M_{xz}$ equation. Once we identify the source terms from the left panel, we demonstrate the components of such source terms in the middle and right panels. The computations are performed by taking time averages from $t/T_{\text{orb}} = 5 \rightarrow 5.5$.}  
	\label{fig:terms_delt_Mxz}
\end{figure*}
In \Fig{fig:terms_delt_Mxz}, we show the individual terms that appear in the dynamical equation for $\bar M_{xz}$. Since $\bar M_{xz}$ changes sign with spatial and temporal coordinates, we multiply $\bar M_{xz}$ on both sides of the equation for $\partial_t \bar M_{xz}$ to understand the contribution of each term. The resultant individual terms of equation for $\bar M_{xz} \partial_t \bar M_{xz}$ are shown in the left panel of \Fig{fig:terms_delt_Mxz}. Once we identify the dominant source terms, we further demonstrate the components of such specific source terms in the middle and right panels of \Fig{fig:terms_delt_Mxz}. We see that the dominant source terms for $\bar M_{xz}$ are the stretching terms: $\bar M_{xk} \partial_k \bar U_z$  and $\bar M_{zk} \partial_k \bar U_x$ (shown in dotted light green/yellow). The most significant contribution in the correlator $\bar M_{xk} \partial_k \bar U_z$ arises from $k=x$ (middle panel). For the correlator $\bar M_{zk} \partial_k \bar U_x$, the dominant contribution arises from $k=z$ (rightmost panel). The nonlinear three-point term and the terms associated with the spatial gradient of mean magnetic fields act like a sink here. In summary, $\bar M_{xx}$ and $\bar M_{zz}$ act in conjunction with $\partial_x \bar U_z$ and $\partial_z \bar U_x$ respectively to produce $\bar M_{xz}$.

\begin{figure*}
	\includegraphics[width=0.68\columnwidth]{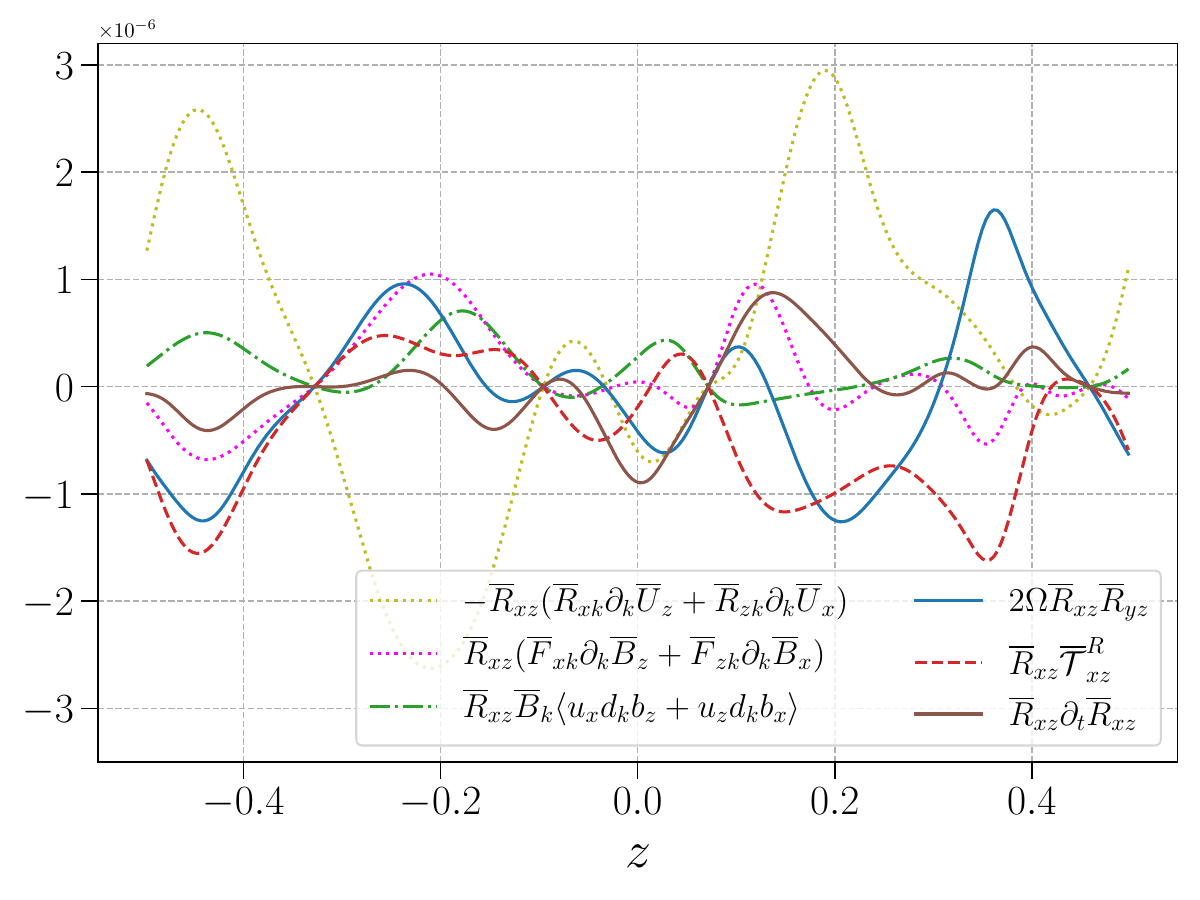} 
	\includegraphics[width=0.68\columnwidth]{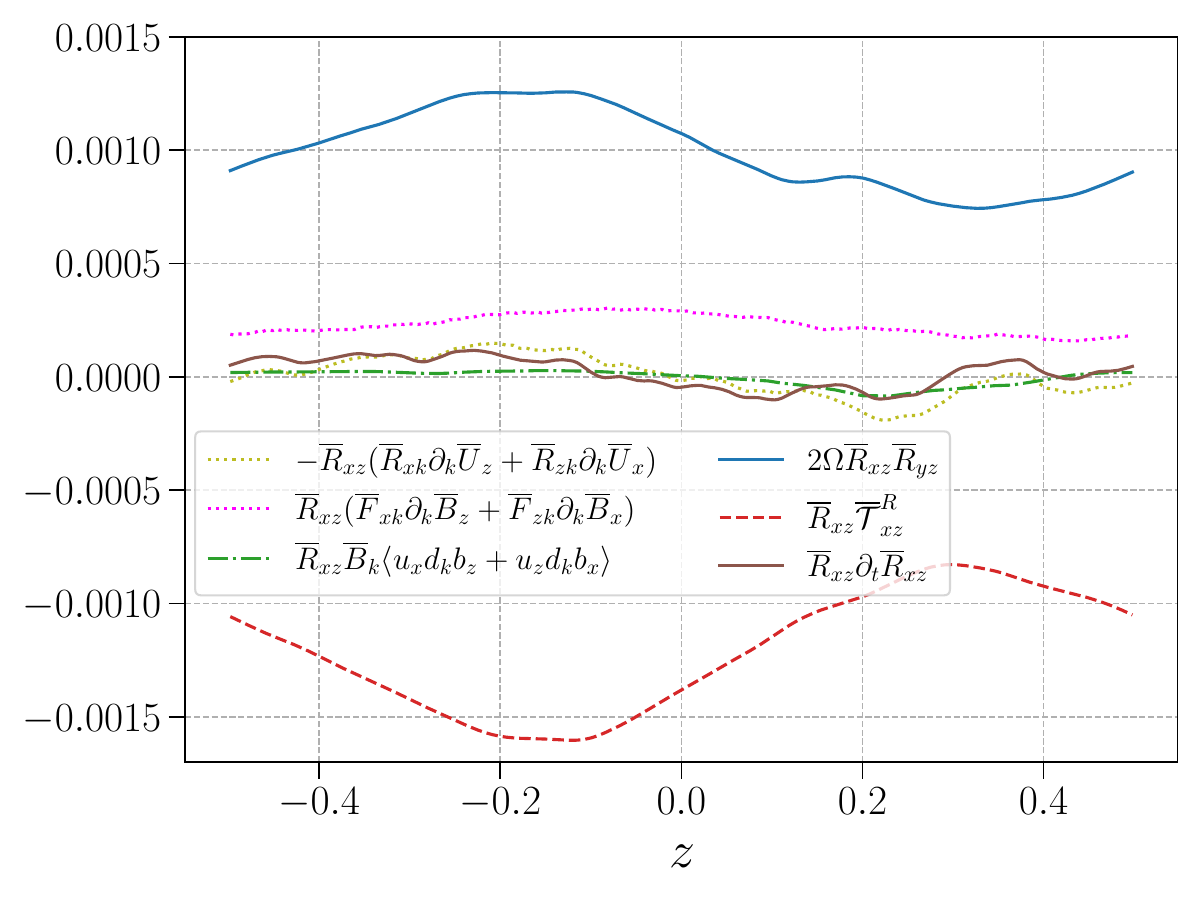}
	\includegraphics[width=0.68\columnwidth]{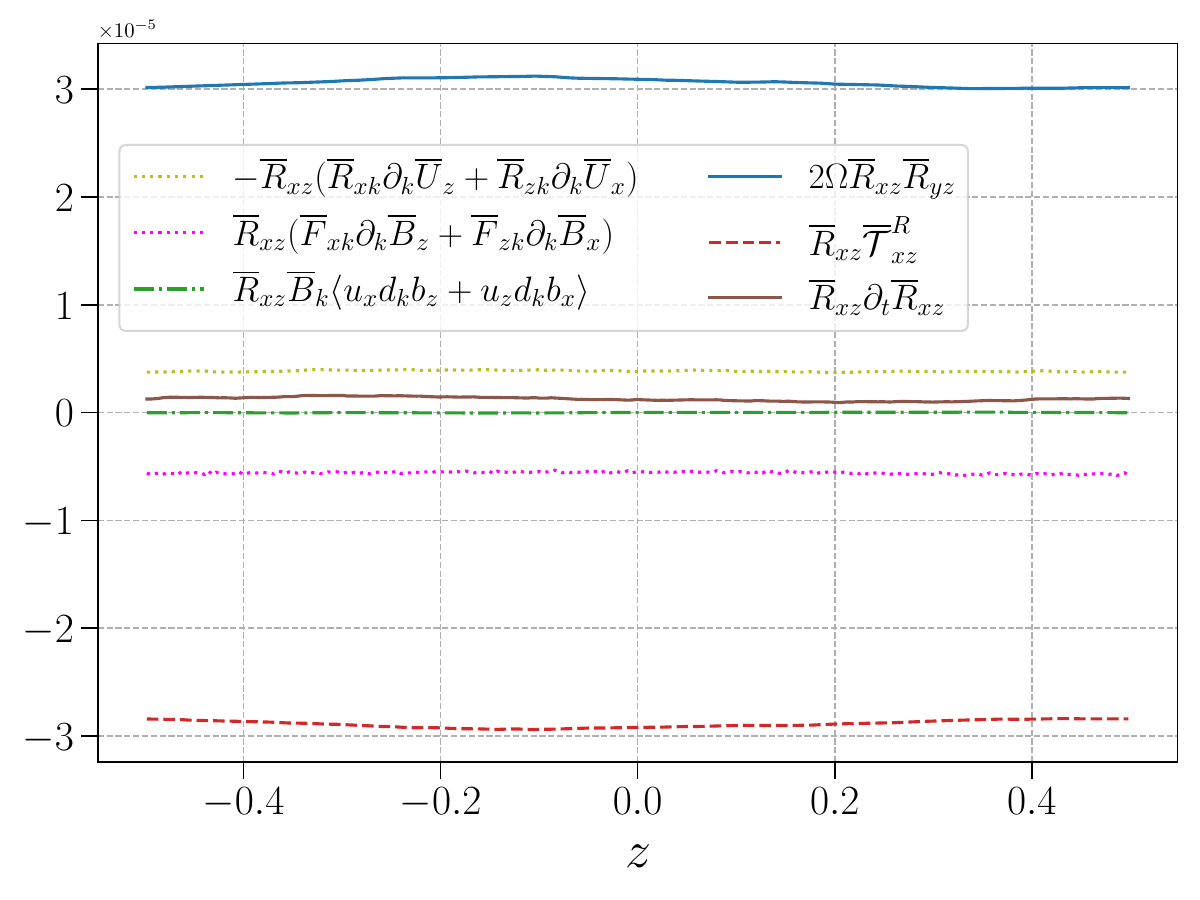}
	\caption{(Color online)
		\textit{The generation mechanism of $\bar R_{xz} (z)$ in the MRI growth and nonlinear regimes.} It shows the individual terms that appeared in $\bar R_{xz} \partial_t \bar R_{xz}$ equation. The computations are performed at $t/T_{\text{orb}} \simeq 5$ (left panel), $t/T_{\text{orb}} \simeq 50$ (middle panel), and $t/T_{\text{orb}} \simeq 100$ (right panel).}  
	\label{fig:terms_delt_Rxz}
\end{figure*}
\begin{figure*}
	\includegraphics[width=0.68\columnwidth]{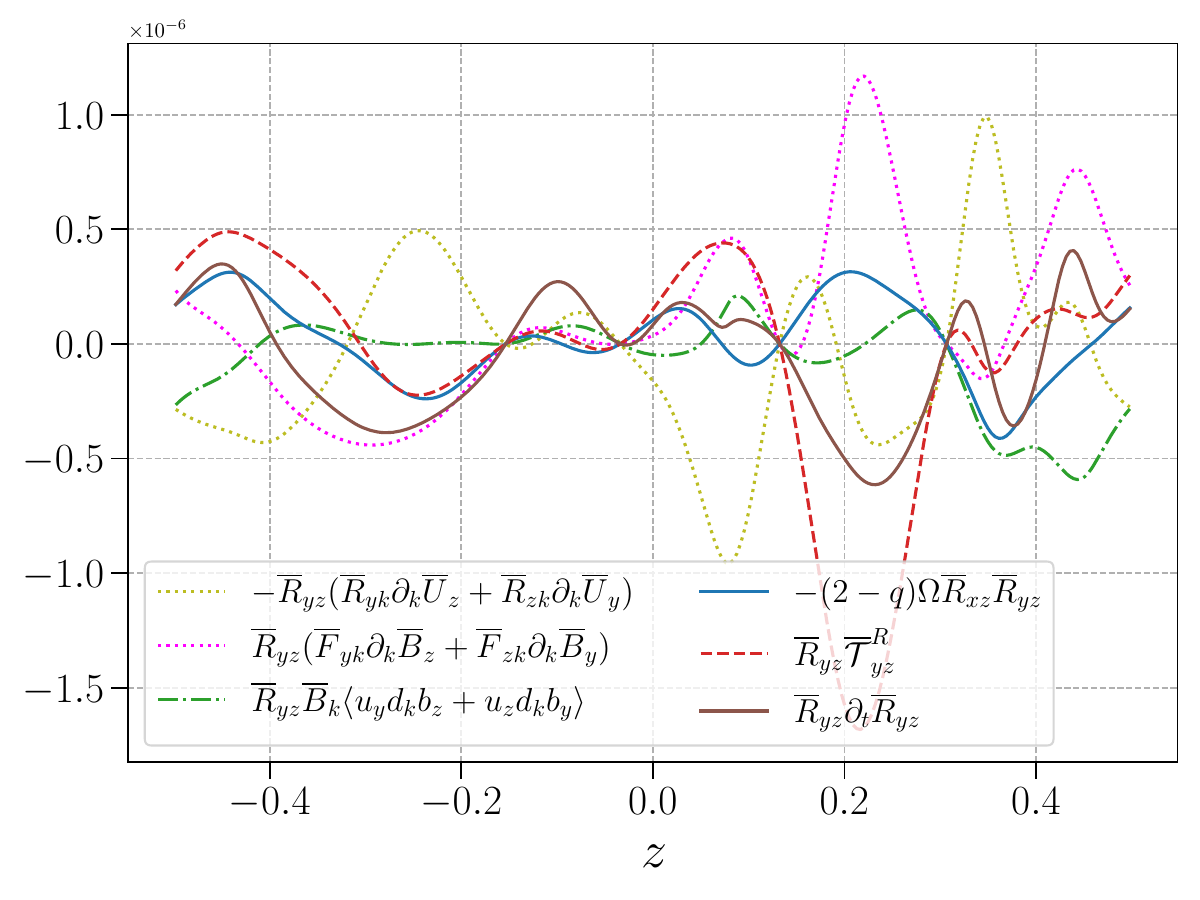} 
	\includegraphics[width=0.68\columnwidth]{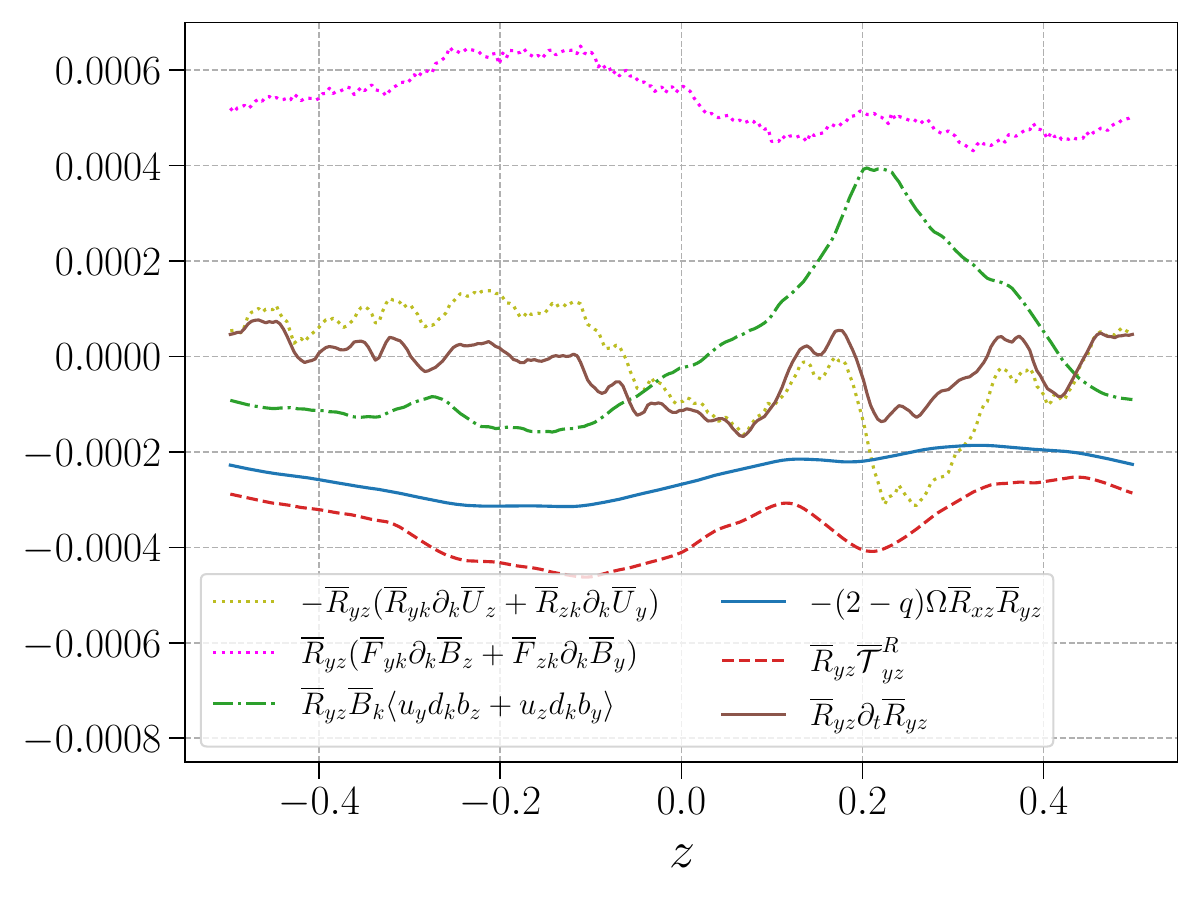}
	\includegraphics[width=0.68\columnwidth]{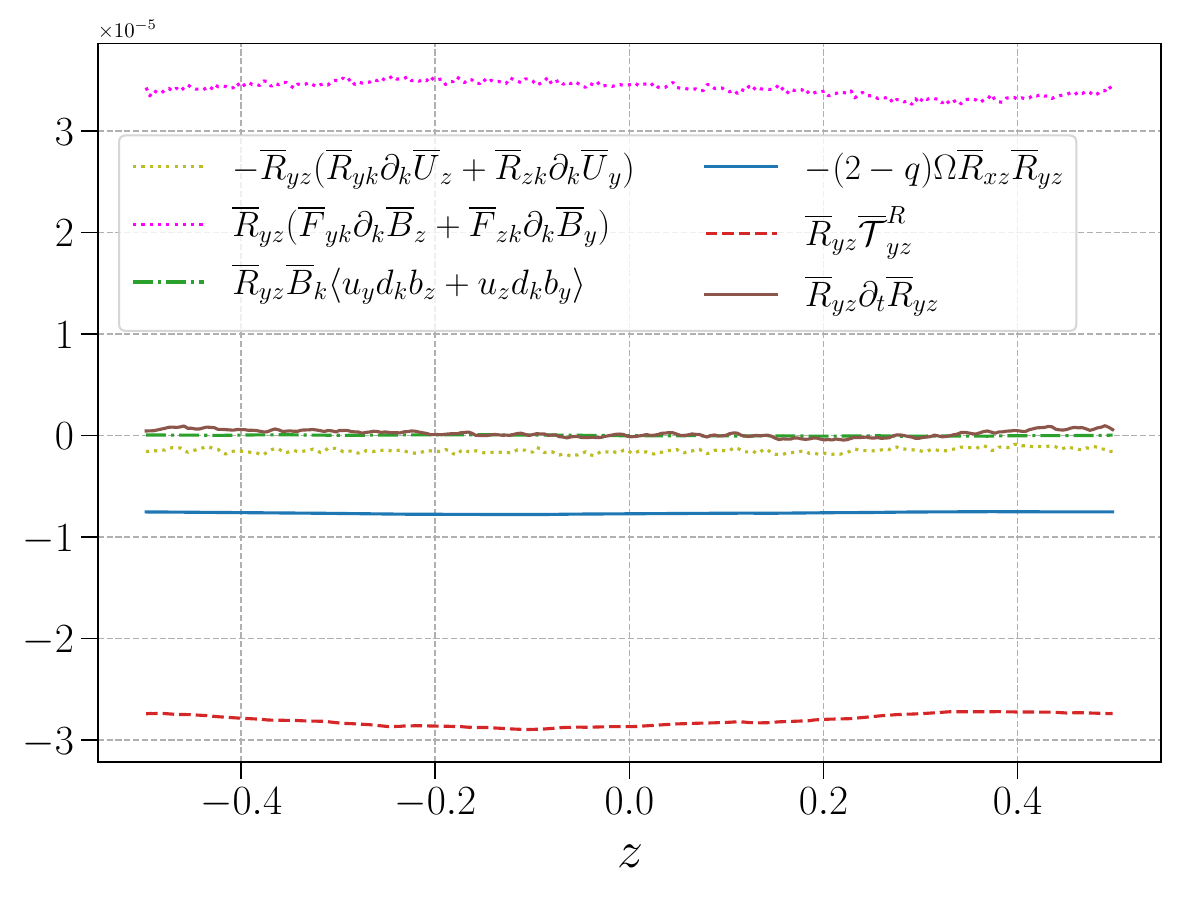}
	\caption{(Color online)
		\textit{The generation mechanism of $\bar R_{yz} (z)$ in the MRI growth and nonlinear regimes.} It shows the individual terms that appeared in $\bar R_{yz} \partial_t \bar R_{yz}$ equation. The computations are performed at $t/T_{\text{orb}} \simeq 5$ (left panel), $t/T_{\text{orb}} \simeq 50$ (middle panel), and $t/T_{\text{orb}} \simeq 100$ (right panel).}  
	\label{fig:terms_delt_Ryz}
\end{figure*}
Next, we analyse the generation of $\bar R_{xz} (z)$, so we plot the individual terms that appear in $\bar R_{xz} \partial_t \bar R_{xz}$ equation in the MRI growth and nonlinear regimes as shown in \Fig{fig:terms_delt_Rxz}. Three different panels of \Fig{fig:terms_delt_Rxz} correspond to the three different instants of time at which the computations are performed: $t/T_{\text{orb}} \simeq 5$ (left panel), $t/T_{\text{orb}} \simeq 50$ (middle panel), and $t/T_{\text{orb}} \simeq 100$ (right panel). It is difficult to identify any specific source term for $\bar R_{xz}$ at the initial phase from the left panel of \Fig{fig:terms_delt_Rxz}. We see that the dominant source term for $\bar R_{xz}$ in the nonlinear regime (middle and right panels) is the Coriolis force term: $2\Omega \bar R_{yz}$ (solid blue line). Hence, the twisting of $\bar R_{yz}$ via the Coriolis force produces $\bar R_{xz}$ at a rate of $2\Omega$. The nonlinear three-point term (dashed red line) behaves as the dominant sink for $\bar R_{xz}$. 
To complete the dynamo cycle, we describe below how $\bar R_{yz}$ is produced.

\begin{figure}
	\includegraphics[width=0.49\columnwidth]{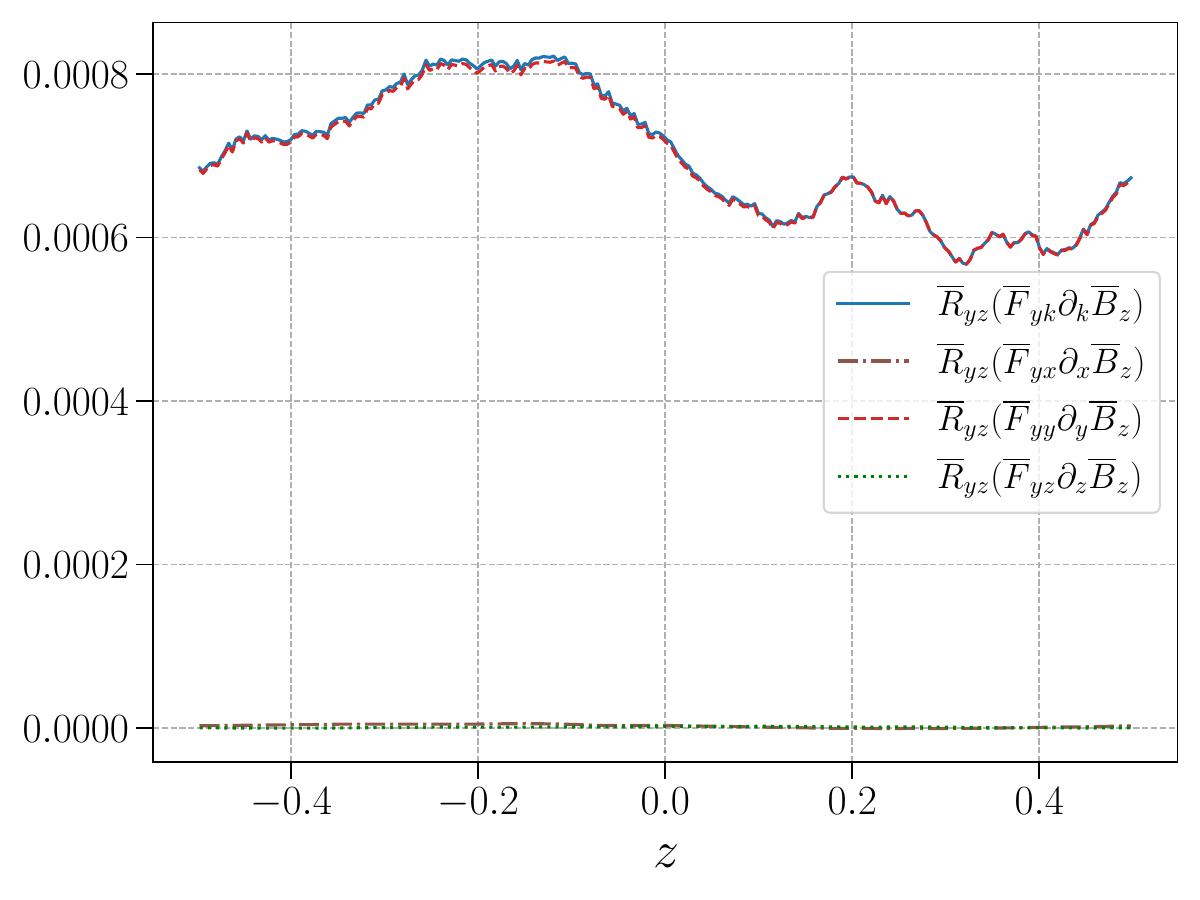} 
	\includegraphics[width=0.49\columnwidth]{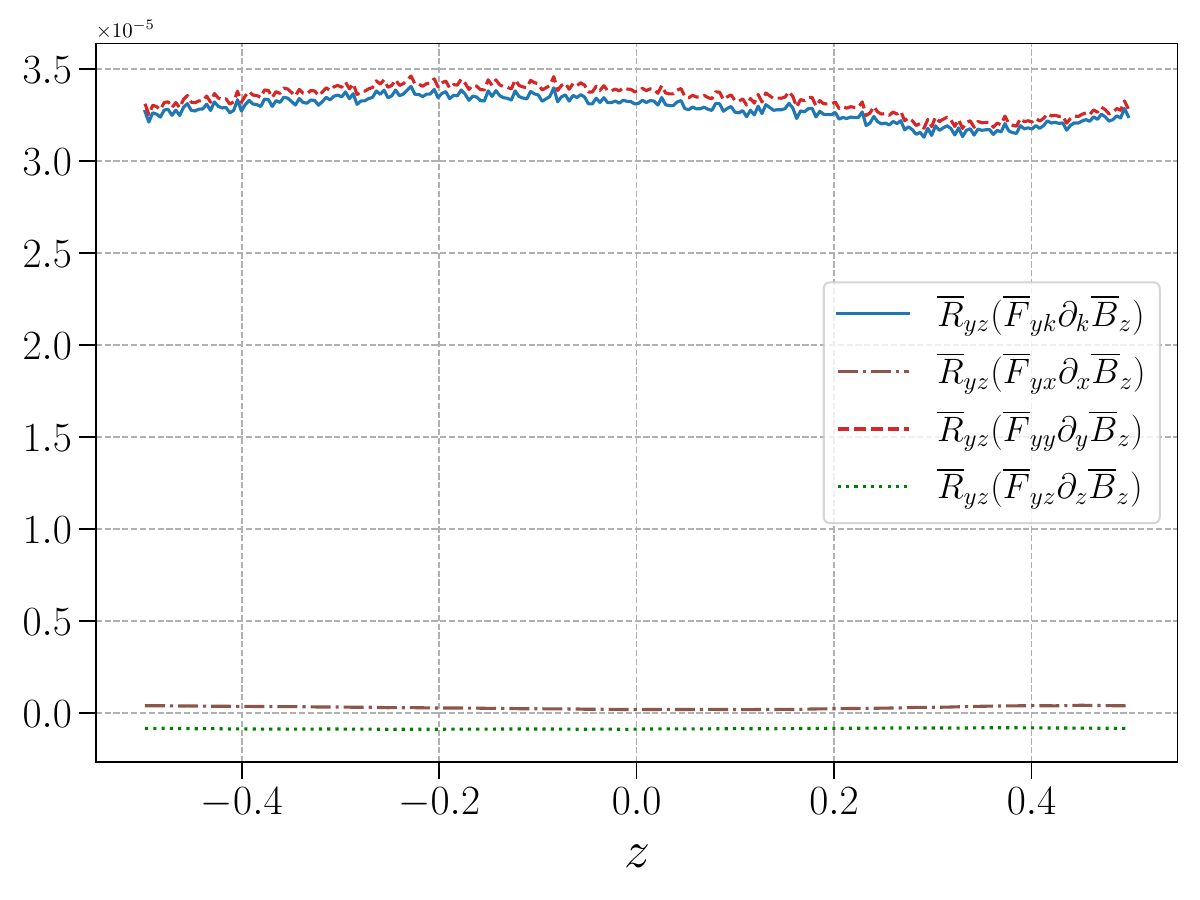}  
	\caption{(Color online)
		The components of the source term, $\bar F_{yk} \partial_k \bar B_z$, that appeared in \Fig{fig:terms_delt_Ryz} for $\bar R_{yz}$ evaluation at $t/T_{\text{orb}} \simeq 50$ (left panel), and $t/T_{\text{orb}} \simeq 100$ (right panel). The other term $\bar F_{zk} \partial_k \bar B_y$ is negligible.}  
	\label{fig:source_terms_delt_Ryz}
\end{figure}

Finally, in \Fig{fig:terms_delt_Ryz}, we perform the $x\text{--}y$ average analysis at three different instants of time: 
$t/T_{\text{orb}} \simeq 5$ (left panel), $t/T_{\text{orb}} \simeq 50$ (middle panel), 
and $t/T_{\text{orb}} \simeq 100$ (right panel), to examine the generation of $\bar R_{yz}$. 
Similar to $\bar R_{xz}$, it is difficult to identify any specific source term for $\bar R_{yz}$ at the initial phase (left panel). 
However, in the nonlinear regime (middle and right panels), it is apparent that the dominant source term for $\bar R_{yz}$ is the term with mean magnetic field gradients (dotted magenta line), 
and once again the nonlinear three-point term (dashed red line) acts as a sink. Mathematically, the complete source term for $\bar R_{yz}$ is expressed as $(\bar F_{yk} \partial_k \bar B_z + \bar F_{zk} \partial_k \bar B_y)$. However, the contribution from the term $\bar F_{zk} \partial_k \bar B_y$ is found to be negligible. Instead, the sole contribution arises from $\bar F_{yk} \partial_k \bar B_z$ with $k=y$. This finding is illustrated in \Fig{fig:source_terms_delt_Ryz} for two different times, $t/T_{\text{orb}} \simeq 50$ (left panel) and $t/T_{\text{orb}} \simeq 100$ (right panel).

\begin{figure*}
	\includegraphics[width=1.3\columnwidth]{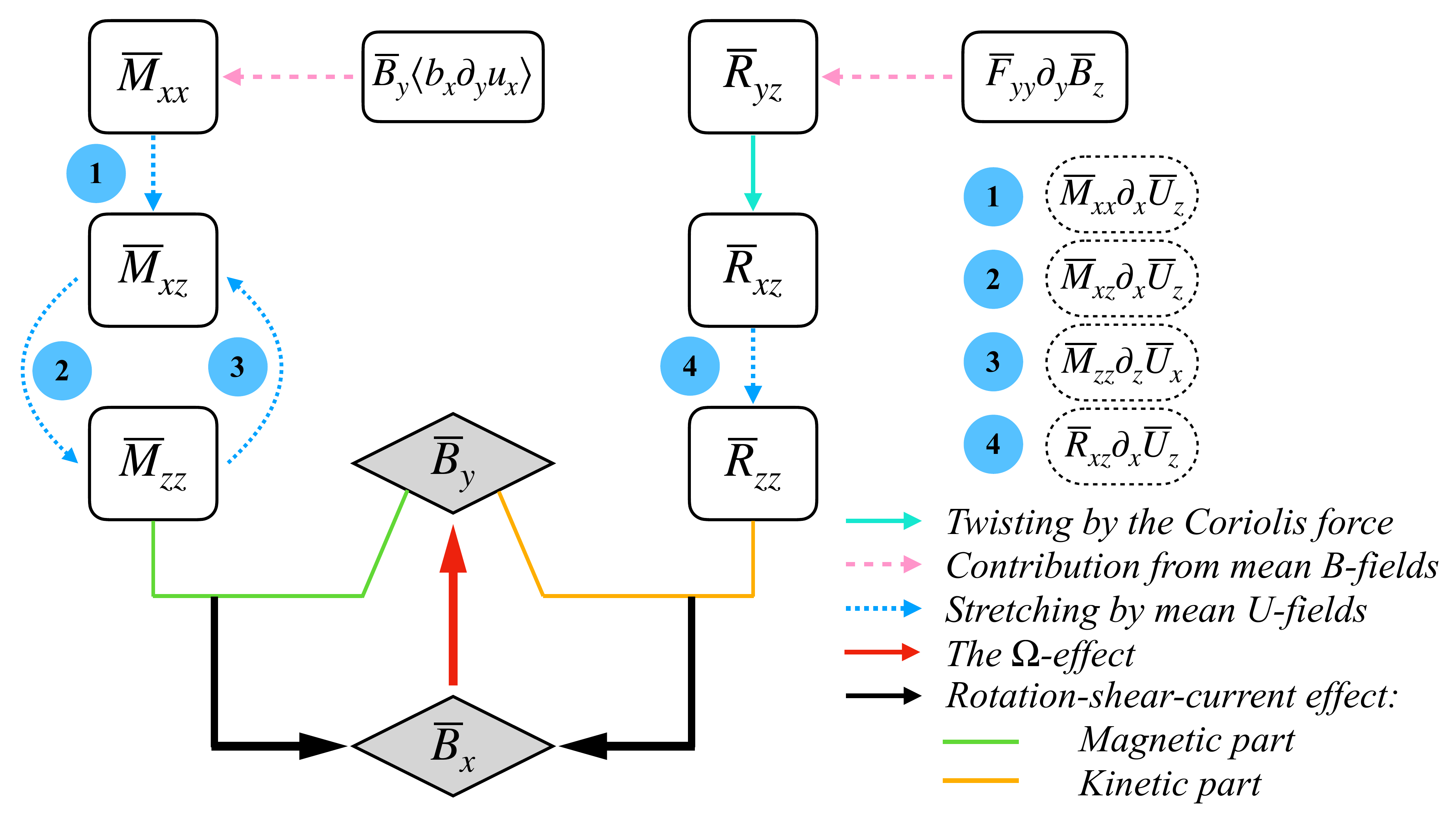} 
	\caption{(Color online)		 
		\textit{A schematic representation of the rotation-shear-current and the $\Omega$-effect to describe the generation mechanism of the $x\text{--}y$ averaged fields $\bar B_x (z)$ and $\bar B_y (z)$, respectively.} Different arrow colors correspond to the paths by which the stress components connect to each other through shear, rotation, mean fields, and other small-scale correlators. Note that we have highlighted only the dominant source terms.}
	\label{fig:mri_rsc}
\end{figure*}

In \Fig{fig:mri_rsc}, we provide a schematic which summarizes the chain of production leading to the rotation-shear-current effect. 
At the magnetic end of the chain, the term involving azimuthal mean field, $\bar B_y$ leads to eventual production of $\bar M_{zz}$ which is one part of the rotation-shear-current effect. 
At the kinetic end of the chain, the vertical mean field is involved in leading up to the production of $\bar R_{zz}$, which is the other part of the rotation-shear-current effect.
Thus, the self-sustaining cycle of dynamo is established connecting both the azimuthal and vertical mean magnetic fields to correlators that are responsible 
for the production of the radial mean magnetic field. In this picture, the production of the vertical mean field has not yet been delved into. We do so in the next section. 

\subsection{Generation of Vertical Magnetic Fields} \label{sec:vertical_B_fied}

\begin{figure*}
	\centering
	\includegraphics[width=0.86\columnwidth]{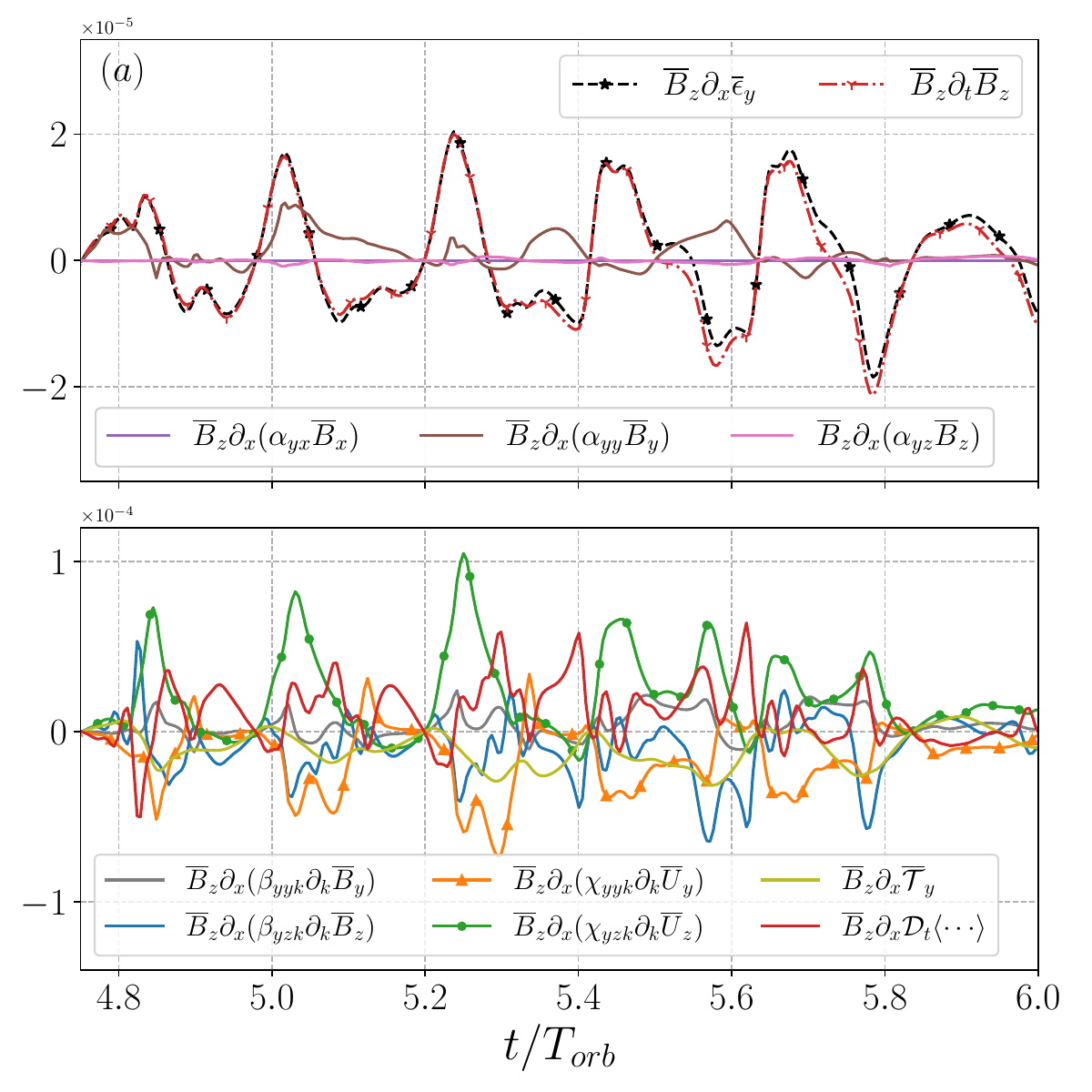} 
	\hspace{-0.4cm}
	\includegraphics[width=1.23\columnwidth]{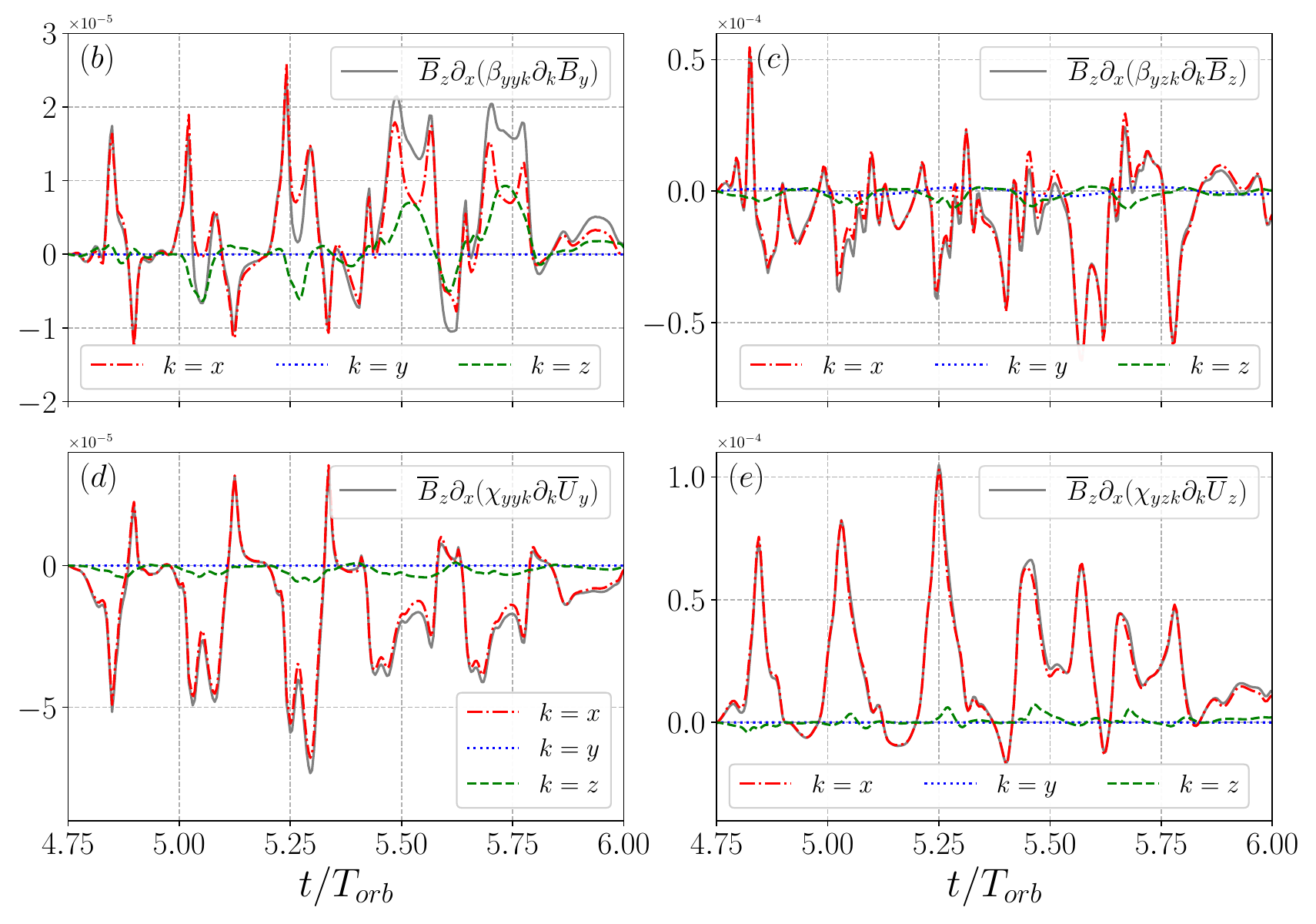} 
	\caption{(Color online)		 
		\textit{The left panel shows the terms from the radial variation of the azimuthal EMF} (equation~\ref{eq:emfy}) \textit{responsible to generate the $y\text{--}z$ averaged field $\bar B_z(x)$ in the MRI growth to initial saturation phase.}
		These are evaluated at $x=-0.15$. To understand the contribution of each term to the evolution of $\bar B_z(x)$, we multiply $\bar B_z(x)$ on both sides of the $\partial_x \mathcal{\bar E}_y$ equation (see, equation~\ref{eq:meanBz_yz}). 
		The dash-dotted red curve is for $\bar B_z \partial_t \bar B_z$, the dashed black curve is for the corresponding EMF term, the solid grey curve is for the term proportional to $\partial_k \bar B_y$, the solid blue curve is for the term proportional to $\partial_k \bar B_z$, the solid orange curve is for the term proportional to $\partial_k \bar U_y$, the solid green curve is for the term proportional to $\partial_k \bar U_z$, the solid yellow curve is for the nonlinear three-point term, and the solid purple, brown, and pink curves correspond to the terms proportional to $x-$, $y-$, and $z-$components of $\bar B$-fields, respectively. The middle and right panels are for the terms proportional to different components of the field gradients: the term proportional to $\partial_k \bar B_y$ (top middle panel), the term proportional to $\partial_k \bar B_z$ (top right panel), the term proportional to $\partial_k \bar U_y$ (bottom middle panel), and the term proportional to $\partial_k \bar U_z$ (bottom right panel).}
	\label{fig:BzdEydx_yz_t}
\end{figure*}
We have seen that the vertical mean-field, $\bar B_z(x)$, arises in the $y\text{--}z$ averaged analysis due to the radial variation of the azimuthal EMF, $\partial_x \mathcal{\bar E}_y$. Here, we discuss the terms responsible for generating $\bar B_z$ via $\mathcal{\bar E}_y$. In the left panels of \Fig{fig:BzdEydx_yz_t}, we show the individual terms of $\mathcal{\bar E}_y$ in the MRI growth to the early saturation phase. To enhance visual clarity, we have distributed the numerous terms of the EMF across two left panels. The middle and right panels of \Fig{fig:BzdEydx_yz_t} illustrate the contributions from different components associated with the terms proportional to the respective field gradients. To understand the contribution of each term on the evolution of $\bar B_z$ (as described in Eq.~\ref{eq:meanBz_yz}), we multiply $\bar B_z$ on both sides $\partial_x \mathcal{\bar E}_y$ equation (i.e., after taking the radial gradient of Eq.~\ref{eq:emfy}). We evaluate these terms at $x=-0.15$. The same color and line style are used for each term, as indicated in Fig.~\ref{fig:BxdEydz_xy_t}. Notably, we have utilized markers only for the most significant curves.

The two crucial curves that determine the growth or decay of the magnetic field over time are the EMF term (represented by a dashed black line with star markers) and $\bar B_z \partial_t \bar B_z$ (shown as a dash-dotted red line with tri-down markers). Positive values of these terms indicate the growing phase of $\bar B_z(x)$, while negative values suggest a decaying phase. The dominant term responsible for the growth of $\bar B_z$ is the one proportional to $\partial_k \bar U_z$ (depicted by a solid green line with circle markers), where $k=x$ (refer to the bottom right panel of \Fig{fig:BzdEydx_yz_t}). Conversely, three terms act as sinks in the evaluation of $\bar B_z(x)$: the term proportional to $\partial_k \bar B_z$ (shown as a solid blue line with triangle markers) with $k=x$ (see top right panel of \Fig{fig:BzdEydx_yz_t}), the term proportional to $\partial_k \bar U_y$ (illustrated by a solid orange line with square markers) with $k=x$ (refer to the bottom middle panel of \Fig{fig:BzdEydx_yz_t}), and the nonlinear three-point term (displayed as a solid olive line). Consequently, these terms contribute to the energy reduction of $\bar B_z(x)$. It is worth noting that the terms proportional to $\bar B_i$ have negligible role on the evolution of $\bar B_z(x)$.

\begin{figure*}
	\centering
	\includegraphics[width=0.88\columnwidth]{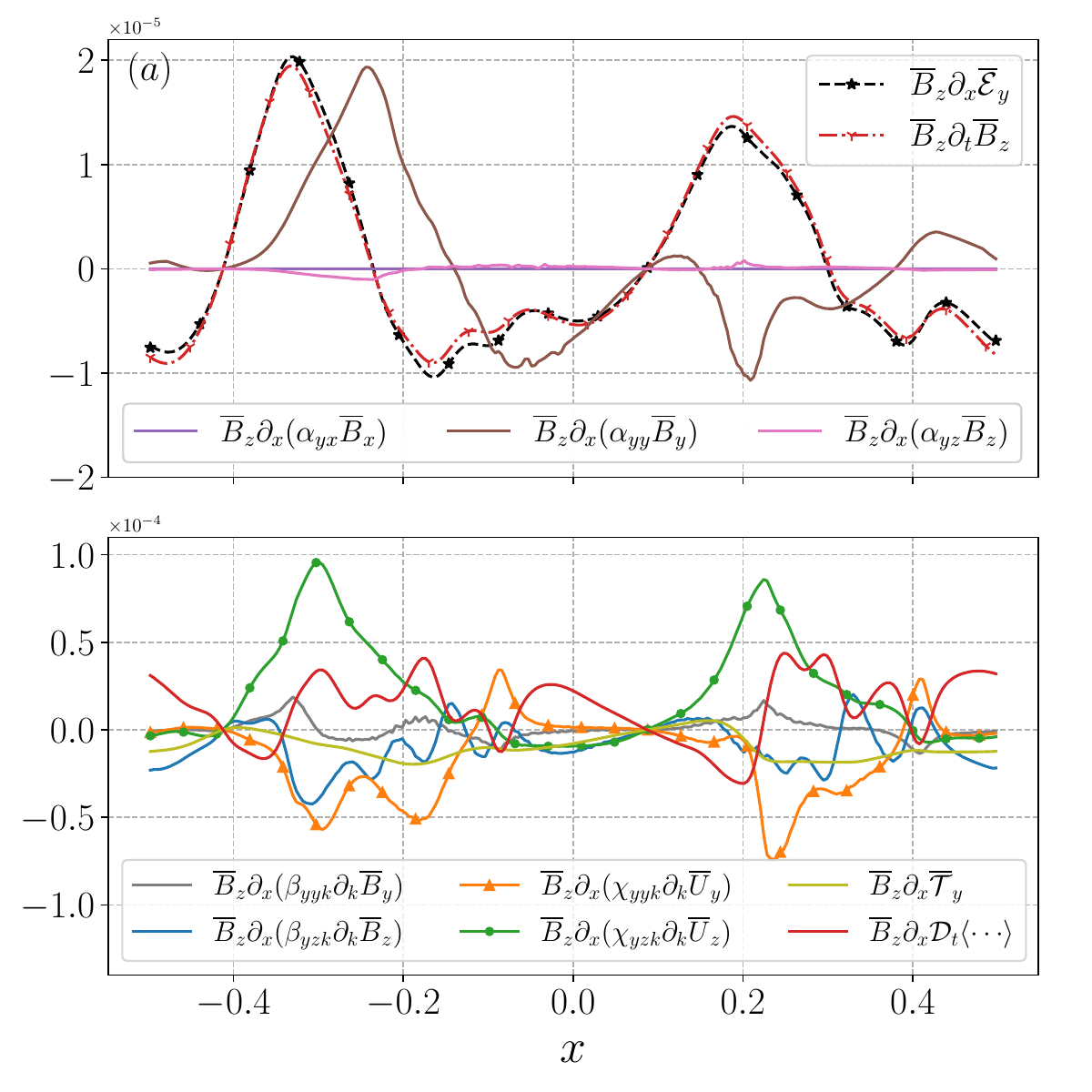} 
	\hspace{-0.46cm}
	\includegraphics[width=1.22\columnwidth]{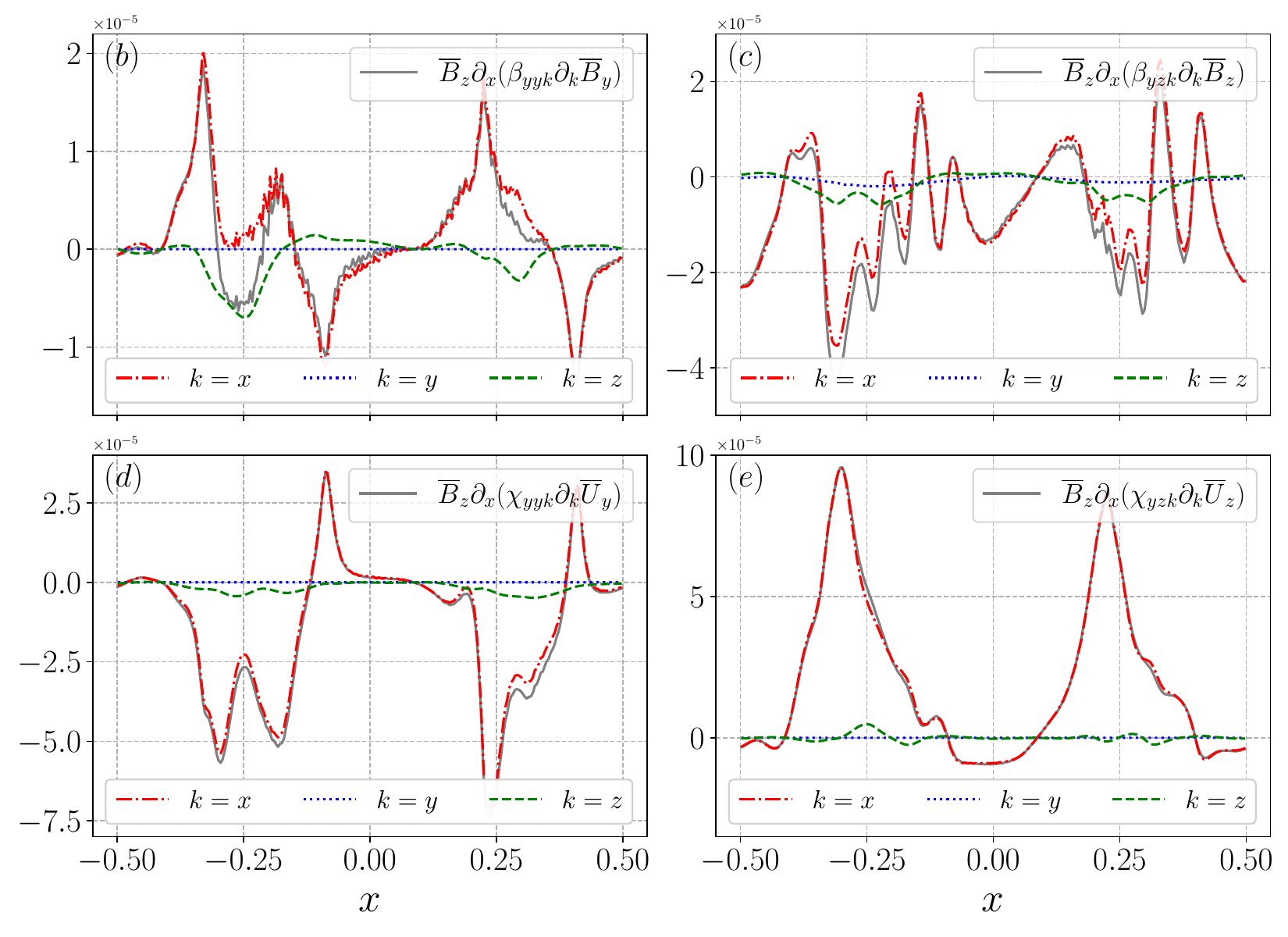} 
	\caption{(Color online)		 
		\textit{The terms from the radial variation of the azimuthal EMF} (equation~\ref{eq:emfy}) \textit{responsible to generate the $y\text{--}z$ averaged field $\bar B_z(x)$ in the MRI growth phase.} These are obtained through time averages from $t/t_{\text{orb}}=5\rightarrow 5.2$. }
	\label{fig:BzdEydx_yz_t5}
\end{figure*}
Next, we investigate the local behavior of the terms appearing in the azimuthal EMF during the MRI growth phase by taking time averages from $t/t_{\text{orb}}=5\rightarrow 5.2$. Such a study can explain how the $y\text{--}z$ averaged field $\bar B_z(x)$ is generated locally via the radial variation of $\mathcal{\bar E}_y$. To understand the individual contributions of these terms to the evolution of $\bar B_z(x)$, we again multiply $\bar B_z(x)$ on both sides of the $\partial_x \mathcal{\bar E}_y$ equation (i.e., after taking the radial gradient of Eq.~\ref{eq:emfy}), as described in Eq.\ref{eq:meanBz_yz}. In the left panels of \Fig{fig:BzdEydx_yz_t5}, we present the variations of the individual terms of $\bar B_z \partial_x \mathcal{\bar E}_y$ as a function of $x$. Similar to \Fig{fig:BzdEydx_yz_t}, we distribute the numerous terms of the EMF across two left panels, while maintaining consistent line styles and colors for each term. We find that the field $\bar B_z(x)$ experiences growth in the regions $x \simeq -0.4 \rightarrow -0.25$ and $x \simeq 0.1 \rightarrow 0.3$. Again, the dominant term responsible for the growth of $\bar B_z(x)$ is the one proportional to $\partial_k \bar U_z$ (depicted by a solid green line with circle markers), where $k=x$ (see the bottom right panel of \Fig{fig:BzdEydx_yz_t5}). Conversely, the term proportional to $\partial_k \bar U_y$ (illustrated by a solid orange line with square markers) with $k=x$ (refer to the bottom middle panel of \Fig{fig:BzdEydx_yz_t5}) acts as the dominant sink term.

\begin{figure*}
	\centering
	\includegraphics[width=0.88\columnwidth]{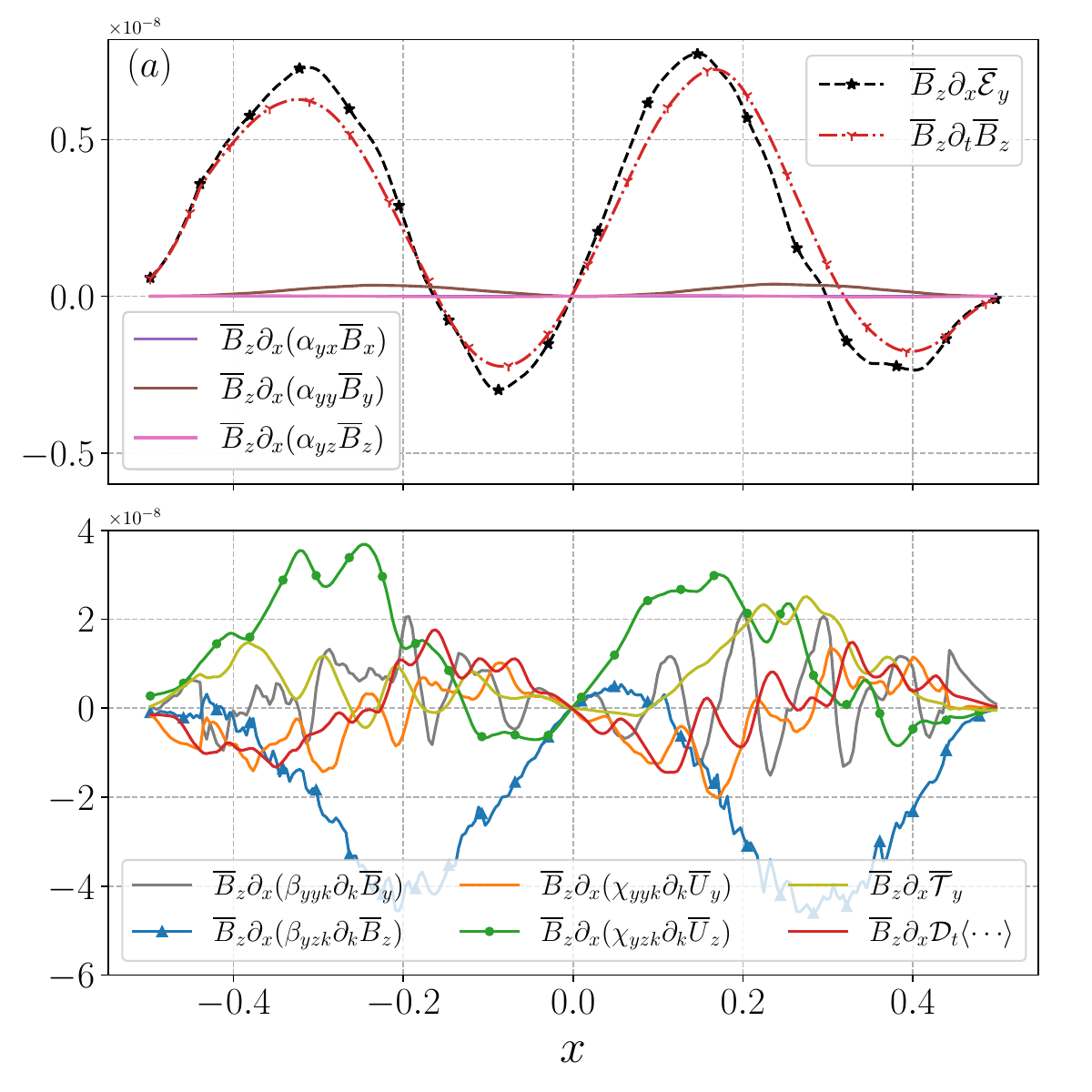} 
	\hspace{-0.46cm}
	\includegraphics[width=1.22\columnwidth]{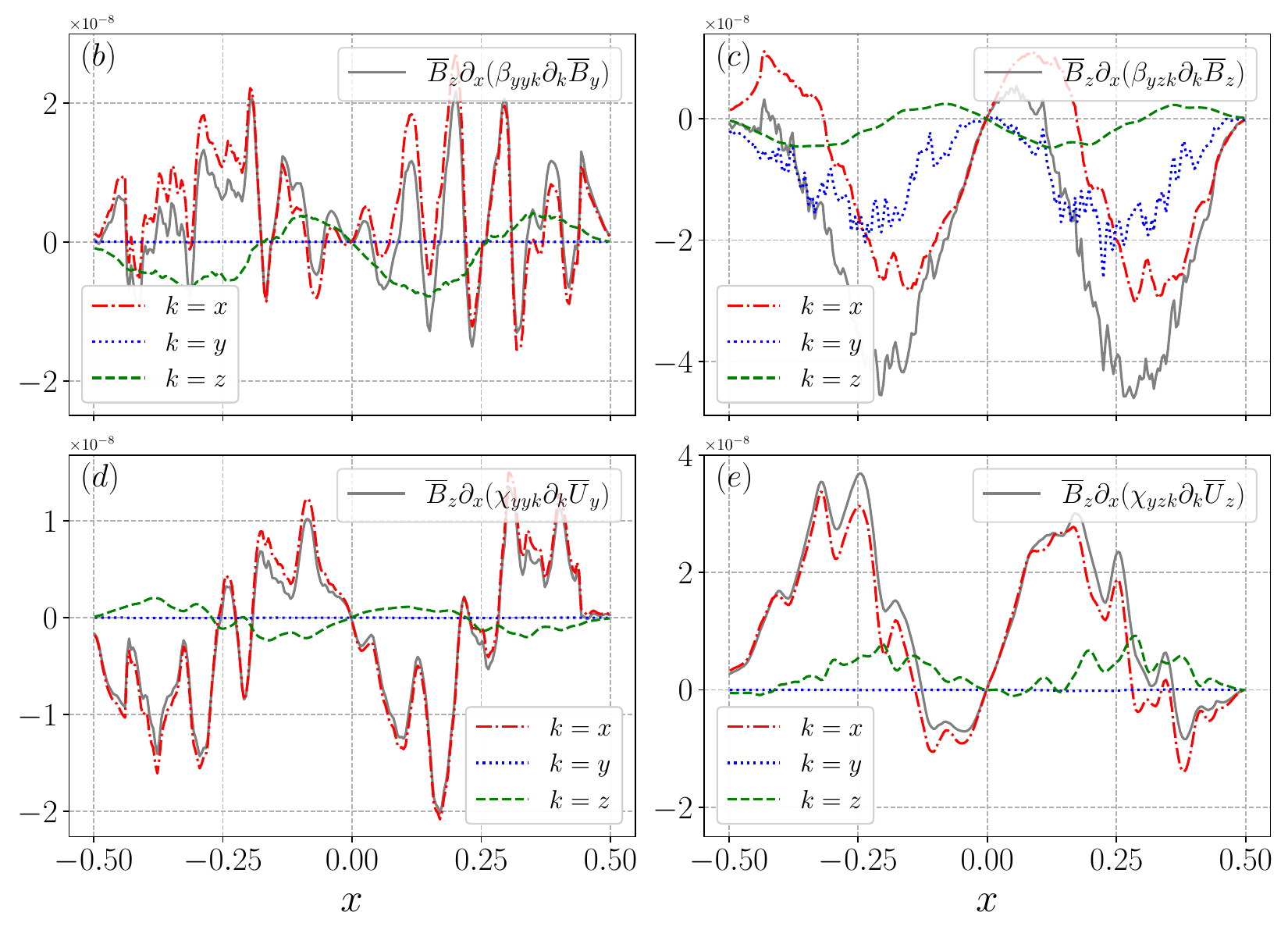} 
	\caption{(Color online)		 
		\textit{The terms from the radial variation of the azimuthal EMF} (equation~\ref{eq:emfy}) \textit{responsible to generate the $y\text{--}z$ averaged field $\bar B_z(x)$ in the MRI nonlinear phase.} This is evaluated at $t/t_{\text{orb}}\simeq 97$. }
	\label{fig:BzdEydx_yz_t600}
\end{figure*}
Finally, we investigate the dynamo mechanism underlying the generation of $y\text{--}z$-averaged fields $\bar B_z(x)$ in the nonlinear regime of MRI. Our focus is to determine whether the terms that were responsible for the growth of large-scale fields continue to play a role in the nonlinear regime. In \Fig{fig:BzdEydx_yz_t600}, we perform a $y\text{--}z$-averaged analysis at $t/t_{\text{orb}}\simeq 97$ to examine the behavior of individual terms in the $\bar B_z \partial_x \mathcal{\bar E}_y$ equation as a function of $x$. To maintain consistency, we use the same line styles and colors for each term as indicated in \Fig{fig:BzdEydx_yz_t5}. The two overlapping curves in the left panel of \Fig{fig:BzdEydx_yz_t600}, one from the EMF term and the other from $\bar B_z \partial_t \bar B_z$, provide insights into the growth or decay of the field $\bar B_z(x)$ with respect to $x$. Remarkably, the overall results remain consistent in the nonlinear regime. The term proportional to $\partial_k \bar U_z$ (depicted by a solid green line with circle markers) with $k=x$ (see the bottom right panel of \Fig{fig:BzdEydx_yz_t600}) continues to be the dominant term responsible for the growth of $\bar B_z(x)$. Conversely, the decay of $\bar B_z(x)$ is primarily attributed to the term proportional to $\partial_k \bar B_z$ (shown as a solid blue line with triangle markers) with $k=x$ (see the top right panel of \Fig{fig:BzdEydx_yz_t600}).

In summary, the growth and the nonlinear saturation of the $y\text{--}z$-averaged field $\bar B_z(x)$ are driven by the radial variation of the azimuthal EMF, $\partial_x \mathcal{\bar E}_y$. In particular, the term proportional to $\partial_x \bar U_z$ of the EMF $\mathcal{\bar E}_y$ plays a dominant role in generating $\bar B_z(x)$. The proportionality coefficient of this term depends on the shear rate, rotation, and specific components of the Faraday tensor, as given by (see Eq.~\ref{eq:emfy}) 
\begin{equation}
- \frac{1}{\Omega} \left[ \frac{1}{(2-q)} \bar F_{yx} + \frac{1}{q} \bar F_{xy} \right] .
\label{eq:rotating-shear-vorticity coefficient}
\end{equation}
We refer to this dynamo mechanism for the generation of $\bar B_z (x)$ as the ``\textit{rotation-shear-vorticity effect}."
It is important to note that this mechanism is fundamentally distinct from the traditional cross-helicity effect \cite{1993ApJ...407..540Y, 2013GApFD.107..114Y}, where the turbulent cross-helicity (defined as the cross-correlation between the turbulent velocity and magnetic field, $\langle u \cdot b\rangle$) serves as the transport coefficient coupling with the large-scale vorticity. In the rotation-shear-vorticity effect, the off-diagonal components of the Faraday tensor, specifically $\bar F_{xy}$ and $\bar F_{yx}$, play a primary role.

\section{DISCUSSION} \label{sec:discussion}

The primary objective of this work is to gain a better understanding of the physical processes involved in sustaining MRI turbulence and dynamo in accretion disks. Despite many theoretical and computational studies, the fundamental principles behind these phenomena remain unclear. One of the main reasons for this is that the mean-field dynamo and angular momentum transport problems have traditionally been treated independently \cite{2010AN....331..101B, 2015JPlPh..81e3905B}. The transport theory for angular momentum has not taken into account the evolution of large-scale magnetic fields \citep{1995PASJ...47..629K, 2003MNRAS.340..969O, 2006PhRvL..97v1103P}, while the mean-field dynamo theory has not considered transport dynamics \citep{2005PhR...417....1B, 2018JPlPh..84d7304B}. In addition, both theories have ignored the feedback from the evolution of mean velocity fields. However, direct numerical simulations have shown the existence of a large-scale dynamo associated with velocity and magnetic fields simultaneously in MRI-driven turbulence \cite{2016MNRAS.462..818B}. To better understand the exact nature of these interactions, one needs to develop a unified mean-field theory for MRI. With this aim, we construct a single coupled model for turbulent accretion disks and perform direct statistical simulations in a zero net-flux unstratified shearing box using statistical closure approximations.

Mean-field dynamo theory is a widely used framework for examining the \textit{in situ} origin of large-scale magnetic field growth and saturation. The electromotive force, a correlation between fluctuating velocity and magnetic fields, is responsible for dynamo action. In mean-field theories, the EMF is typically assumed to be a linear function of the mean magnetic field and its spatial derivatives, with the proportionality coefficients usually treated as tensors. However, this assumption may not be sufficient to fully capture the complex physical processes involved in magnetic field generation and sustenance. Several studies have shown that an additional term proportional to the spatial derivative of the mean velocity field enters the EMF equation, which can lead to rapid growth of mean magnetic fields \cite{2010AN....331...14R}. Similarly, whenever an additional term participates in the EMF equation, the dynamics of magnetic field growth and saturation can change dramatically. Therefore, it is crucial to properly account for the effect of all contributions in the EMF equation to fully understand the physics of magnetic field generation and sustenance.

Here, we identify a novel possibility for large-scale magnetic field generation in unstratified MRI-driven turbulent plasmas: the rotation-shear-current (RSC) effect. The mechanism arises through an off-diagonal turbulent resistivity $\eta_{yx}$, which has a favorable negative sign to cause mean-field dynamo action, rather than being positive for diffusion. The basic idea is that in the presence of shear and rotation, small-scale kinetic and magnetic fluctuations produce $\eta_{yx}$ in the following form (the coefficient of $\partial_z \bar B_y$ in Eq.~\ref{eq:emfy})
\begin{equation}
\eta_{yx} = - \frac{1}{\rho \Omega}\left[\frac{1}{2-q} \bar M_{zz} + \frac{1}{q} \bar R_{zz} \right].
\label{eq:eta_yx}
\end{equation}
This is for the first time we have identified the exact expression for $\eta_{yx}$. The respective correlators associated with magnetic and kinetic fluctuations are $\bar M_{zz}$ and $\bar R_{zz}$, which are always positive. The factor `two' arises due to rotation via the Coriolis force. Hence, for a Keplerian shear flow (i.e., $q=1.5$), both magnetic $(\eta_{yx}^b)$ and kinetic $(\eta_{yx}^u)$ contributions to the RSC effect have favorable negative sign. It is important to note that the term associated with the RSC effect is distinct from the ${\bec \Omega} \times {\bec J}$ or R\text{\"{a}}dler-effect \cite{1980opp..bookR....K, 1986AN....307...89R}. Also, the RSC effect differs fundamentally from the traditional shear-current (SC) effect in which rotation is absent \citep{2003PhRvE..68c6301R, 2004PhRvE..70d6310R}. The SC effect has been controversial, with mutual and seperate disagreements among theories and simulations. Below we discuss various conflicts associated with SC effect.

The traditional SC effect is a potential nonhelical large-scale dynamo driven by off-diagonal turbulent resistivity $\eta_{yx}$ in the presence of a large-scale velocity shear without any rotation. A negative sign of $\eta_{yx}$ is necessary for coherent dynamo action by the SC effect. However, it remains a matter of debate whether the contributions from the turbulent kinetic and magnetic parts to $\eta_{yx}$ have a preferred sign or not, and which one dominates. Among analytical works, those employing a spectral-$\tau$ closure found that both $\eta_{yx}^b$ and $\eta_{yx}^u$ have favorable negative signs to cause dynamo action \citep{2003PhRvE..68c6301R, 2004PhRvE..70d6310R}. In contrast, the second-order correlation approximation \citep{2006PhRvE..73e6311R, 2006AN....327..298R} and quasi-linear calculations \citep{2009PhRvE..80f6315S, 2011PhRvE..83e6309S} disagreed with the existence of the kinetic SC effect. For magnetic shear-current (MSC) effect, the analytical calculations using second-order correlation approximation agree with previous spectral-$\tau$ calculations that $\eta_{yx}^b$ has favorable negative sign, and the magnetic part substantially dominates over the kinetic part \citep{2015PhRvE..92e3101S}. Zhou and Blackman \cite{2021MNRAS.507.5732Z} resolve some of these theoretical discrepancies (atleast at low to moderate Re $\sim 10$) by showing that the kinetic contribution $\eta_{yx}^u$ is sensitive to the kinetic energy spectral index and can transist from positive to negative values with increasing Re, whereas the magnetic contribution $\eta_{yx}^b$ remains always negative. However, numerical simulations do not fully agree with theory, and sometime mutually contradict. There are broadly two methods employed to determine the turbulent transport coefficients from simulations: the test-field method and the projection method. In kinetically forced quasi-linear simulations using projection method, it has been found that $\eta_{yx}^u$ is positive with only shear, and negative when a Keplerian rotation is added \citep{2015ApJ...813...52S}. Conversely, nonlinear test-field method in MHD burgulence (i.e., ignoring the thermal pressure gradient) with kinetic forcing has reported a negative $\eta_{yx}^u$ for the non-rotating case, but did not explore the case including rotation \citep{2020ApJ...905..179K}. For magnetic contributions, magnetically forced quasi-linear simulations using projection method found that $\eta_{yx}^b < 0$ either with or without Keplerian rotation \citep{2015ApJ...813...52S}. Unfortunately, nonlinear test-field method in MHD burgulence with magnetic foring found that $\eta_{yx}^b > 0$ for non-rotating shearing cases \citep{2020ApJ...905..179K}.

To resolve the above mentioned discrepancies, we provide here the exact expressions for $\eta_{yx}$ (Eq.~\ref{eq:eta_yx}) which describes the role of rotation and shear parameters to the contributions of kinetic and magnetic parts. As we have already mentioned that for a differentially rotating Keplerian flow (as in the case of MRI turbulence) both $\eta_{yx}^b$ and $\eta_{yx}^u$ have favorable negative signs to cause dynamo action. Now, in the absence of rotation, relevant to the traditional SC effect and the MSC effect, $\eta_{yx}$ reduces to the form as $\eta_{yx} = - \left( \bar R_{zz} - \bar M_{zz} \right)/q\rho \Omega$. We see that the kinetic contribution $\eta_{yx}^u$ has a favorable negative sign, whereas the magnetic contribution has a wrong sign for dynamo action. Interestingly, they will exactly cancel each other in the limit of $\bar M_{zz} \sim \bar R_{zz}$, which will make the SC effect inoperative (i.e., $\eta_{yx} \simeq 0$). It supports the conclusions of Ref.~\cite{2020ApJ...905..179K} that there is no evidence for MSC-effect-driven dynamo in magnetically forced, non-rotating MHD burgulence, but kinetic SC effect has favorable negative sign when forced kinetically. It also explains the results associated with non-rotating, unstratified, compressible MHD simulations with driven turbulence using a compressible test-field method that $\eta_{yx}$ to be slightly negative or positive but statistically not different than zero, concluding no evidence of coherent SC effect \citep{2022ApJ...932....8K}.

In addition to forced turbulence, there is growing evidence for the presence of the RSC effect in unstratified, zero net-flux shearing-box simulations of MRI-driven turbulence. Both finite volume code \citep{2016MNRAS.456.2273S} and moving mesh code \citep{2022MNRAS.517.2639Z} simulations have observed a large-scale dynamo with a negative value of $\eta_{yx}$. These findings contrast with the results of Ref.~\cite{2022A&A...659A..91W}, who used a smooth particle hydrodynamics code and observed slightly positive or nearly zero values of $\eta_{yx}$ in their zero net-flux, unstratified simulations. The discrepancy can be attributed to the significantly weaker mean fields in their simulations, which can impact the manifestation of the RSC effect.

Next, we delve into the mechanisms by which the correlators associated with fluctuations drive the RSC effect. As we discussed earlier, the correlators involved in the RSC effect are $\bar M_{zz}$ and $\bar R_{zz}$, which correspond to the magnetic and kinetic aspects, respectively. We have discussed how these correlators interact with shear and rotation to produce off-diagonal turbulent resistivity $\eta_{yx}$ with the appropriate sign for the large-scale dynamo. Understanding the generation of these correlators in the context of self-sustained MRI-driven turbulence is crucial. We uncover a significant revelation: the presence of a large-scale vorticity dynamo is essential for their production. Notably, the dominant contribution arises from the mean vertical velocity fields. Moreover, at the magnetic end of the chain, the term involving azimuthal mean magnetic fields plays a significant role in generating $\bar M_{zz}$---the magnetic part of the RSC effect, while at the kinetic end of the chain, the term involving vertical mean magnetic fields takes charge in producing $\bar R_{zz}$---the kinetic part of the RSC effect (see, \Fig{fig:mri_rsc} for a more comprehensive depiction). Consequently, a self-sustaining dynamo cycle is established, linking the azimuthal and vertical magnetic fields to the correlators that give rise to radial magnetic fields through the RSC effect.

Finally, we address the generation of vertical magnetic fields arising in the $y\text{--}z$-averaged analysis. For a given initial vertical field, the MRI can be initiated locally. However, in the absence of any large-scale dynamo action, the resulting MRI turbulence tends to disrupt the original vertical field, potentially leading to the cessation of the MRI. While considerable research on MRI dynamo mechanisms has focused on the generation of horizontally ($x\text{--}y$) averaged fields, the persistence of large-scale vertical magnetic fields in MRI-driven turbulence remains an intriguing question. Our findings from the $y\text{--}z$-averaged analysis are consistent with results from global cylindrical MRI simulations \cite{2016MNRAS.459.1422E} and local shearing box simulations \cite{2016MNRAS.462..818B}, where the large-scale fields arise entirely from the EMF. In particular, the vertical mean field is driven by the radial variation of the azimuthal EMF. By formulating a general expression for the EMF, we have identified a novel dynamo mechanism responsible for the generation of large-scale vertical magnetic fields, referred to as the \textit{rotation-shear-vorticity effect}. This mechanism critically depends on the presence of a large-scale vorticity dynamo. Specifically, the azimuthal EMF contains a term proportional to the radial gradient of the vertical mean velocity field, which drives this dynamo mechanism. The exact form of the proportionality coefficient is given in \Eq{eq:rotating-shear-vorticity coefficient}. This coefficient arises from the interaction of the $xy$- and $yx$-components of the Faraday tensor with rotation and shear.

Overall, these new findings open up exciting avenues for the advancement of mean-field dynamo theory and invite further exploration.
In general, the DSS method is useful to study the dynamics of various astrophysical disks. For galactic disks, the observed large-scale magnetic fields are thought to be generated by a turbulent dynamo. This dynamo is governed by the EMF generated by turbulence, primarily driven by supernova explosions within the differentially rotating interstellar medium. So far, the most successful theory for explaining the sustained growth of galactic magnetic fields is the $\alpha \Omega$ dynamo \citep{1996ARA&A..34..155B}. However, significant uncertainties persist in estimating turbulent transport coefficients and understanding the nonlinear saturation process. 
Several methods have been employed to extract turbulent transport coefficients from local direct numerical simulations. In the kinematic regime, both the test-field method \citep{2008A&A...486L..35G} and the singular value decomposition method \citep{2020MNRAS.491.3870B} have shown good agreement in determining various transport coefficients. However, when dealing with strong fluctuations within the numerical data in the presence of dynamically important small-scale magnetic fields due to the fluctuation dynamo, the analysis becomes challenging, sometimes requiring the assumption of specific coefficients being zero. Such analyses have been restricted to the specific components of the tensors $\alpha_{ij}$ and $\eta_{ij}$ relevant only for the $\alpha \Omega$ dynamo \citep{2013MNRAS.429..967G}. Consequently, significant uncertainty remains regarding the precise mechanisms responsible for large-scale dynamo saturation in galactic disks (see the current review by Brandenburg and Ntormousi \cite{2023ARA&A..61..561B}, for more details). The DSS methods we have used and also, our approach of a self-consistent construction of a general EMF, could hold potential to identify the nonlinear saturation mechanisms and provide relevant expressions for nonlinear turbulent transport coefficients essential for understanding galactic dynamos.

\section{CONCLUSIONS} \label{sec:conclusion}

In this paper, we investigated the phenomena of MRI turbulence and dynamo in a zero net-flux unstratified shearing box, employing novel DSS methods. Our main objective was to develop a unified mean-field model that combines the traditionally decoupled problems of angular-momentum transport and the large-scale dynamo in MRI-driven turbulent flows, with specific emphasis on the standard Keplerian accretion disks. We consider the dynamics of turbulent stresses, including the Maxwell, Reynolds, and Faraday tensors, together with the behavior of large-scale velocity and magnetic fields, in order to understand the sustaining mechanisms of MRI turbulence without any external driving force. To accomplish this, we employ various high-order closure schemes. The three-point correlators are closed using a statistical closure model inspired by the CE2.5 approximation, while a two-scale approach is utilized to model second-order correlators involving the spatial gradient of a fluctuating field. Our principal findings can be summarized as follows:

(1) The outward transport of angular-momentum occurs through positive total stress, $\bar W_{xy} = \bar R_{xy} - \bar M_{xy}$, where $\bar M_{xy}<0$ and $\bar R_{xy}>0$. The dominant contribution to the total stress arises from the correlated magnetic fluctuations, rather than from their kinetic counterparts, i.e., $-\bar M_{xy} > \bar R_{xy}$, as expected. The generation process of these stresses involves intricate interactions involving shear, rotation, correlators associated with mean fields, and nonlinear terms. A schematic overview of the findings is summarized in \Fig{fig:mri_transport}. The stretching of $\bar M_{xx}$ through shear gives rise to $\bar M_{xy}$, which, is further stretched by shear to produce $\bar M_{yy}$. The large-scale magnetic field, predominantly $\bar B_y$ acts in conjunction with the correlator $\langle b_x \partial_y u_x \rangle$, leading to the generation of $\bar M_{xx}$ (which is essentially the tangling of large-scale field leading to small-scale fields). Regarding the Reynolds stress, the Coriolis force is responsible for generating $\bar R_{xx}$ from $\bar R_{xy}$, and $\bar R_{xy}$ from $\bar R_{yy}$. Interestingly, the nonlinear interactions between $\bar M_{yy}$ and $\bar R_{yy}$ via three-point terms contribute to the formation of $\bar R_{yy}$ from $\bar M_{yy}$. Another significant source term for $\bar R_{yy}$ is the term proportional to the radial gradient of the mean azimuthal magnetic field. Therefore, the turbulent transport critically depends on the presence of large-scale magnetic fields.

(2) For the large-scale magnetic field dynamo, we analyzed the individual terms in the mean field induction equation using both $x\text{--}y$ and $y\text{--}z$ averaging and determined their contributions to the generation of mean fields. Our findings align well with those obtained from DNS \cite{2008A&A...488..451L, 2016MNRAS.462..818B}. With $x\text{--}y$ averaging, the azimuthal EMF generates the radial field, which, in turn, drives the azimuthal field through the $\Omega$-effect. The radial EMF exhibits a sink effect, resulting in a decrease in the energy of the azimuthal field. In the case of $y\text{--}z$ averaging, the large-scale fields originate entirely from the respective EMF. Specifically, the azimuthal field arises from the radial variation of the vertical EMF, while the vertical field emerges from the radial variation of the azimuthal EMF. 

(3) To identify the relevant dynamo mechanisms, we constructed the EMF for an MRI-driven system. The EMF is expressed as a linear combination of terms proportional to mean magnetic fields, the gradient of mean magnetic fields, the gradient of mean velocity fields, and a nonlinear term. The proportionality coefficients depend on shear, rotation, and statistical correlators associated with fluctuating fields. Importantly, this EMF expression arises naturally from our model rather than being an ansatz. By analyzing the general EMF expression, we identify two crucial dynamo mechanisms---the \textit{rotation-shear-current effect} and the \textit{rotation-shear-vorticity effect}---that are responsible for generating the radial and vertical magnetic fields, respectively. We have provided explicit expressions of the corresponding turbulent transport coefficients, in the nonperturbative limit. Notably, both mechanisms rely on the presence of large-scale vorticity dynamo. It is important to note that both the kinetic and magnetic components of the rotation-shear-current effect have favorable signs for driving a dynamo mechanism. A schematic overview of the rotation-shear-current effect is presented in \Fig{fig:mri_rsc}.

\section*{ACKNOWLEDGEMENTS}

TM and PB gratefully acknowledge Dr. Matthias Rheinhardt for his valuable guidance and support during the initial stages of a special module development in the framework of the \textsc{Pencil Code}. We are also thankful to Prof. Kandaswamy Subramanian for his insightful comments and suggestions. TM would like to thank Sahel Dey for providing an overview regarding the \textsc{Pencil Code}. The simulations were performed on the ICTS HPC clusters \textit{Contra} and \textit{Sonic}, and we express our gratitude to Dr. Prayush Kumar for granting access to the \textit{Sonic} cluster. We acknowledge project RTI4001 of the Dept. of Atomic Energy, Govt. of India.


\appendix

\begin{widetext}
\section{The Electromotive Force} \label{sec: app_EMF}

In this section, we provide a comprehensive derivation of the electromotive force (EMF). We start by presenting the evolution equations for all the components of the Faraday tensor: 
	\begin{align}
	\mathcal{D}_t \bar F_{xx} &=& 2\Omega \bar F_{yx} 
	+ \left( \bar F_{xk}\partial_k\bar U_x - \bar F_{kx}\partial_k\bar U_x \right)
	+ \bar B_k \left[ \langle u_x\partial_k u_x \rangle  + \frac{\langle b_x\partial_k b_x \rangle}{\mu_0 \rho} \right]
	+ \frac{1}{\rho}\left( \bar M_{xk}\partial_k\bar B_x-\bar R_{xk}\partial_k\bar B_x \right) + \mathcal{\bar T}^F_{xx},
	\label{eq:F_xx}  \\
	\mathcal{D}_t \bar F_{xy} &=& \hspace{-8pt}- q\Omega \bar F_{xx} + 2\Omega \bar F_{yy} 
	+ \left(\bar F_{xk}\partial_k\bar U_y - \bar F_{ky}\partial_k\bar U_x \right)
	+ \bar B_k \left[ \langle u_x\partial_k u_y \rangle  + \frac{\langle b_y\partial_k b_x \rangle}{\mu_0 \rho} \right]
	+ \frac{1}{\rho}\left( \bar M_{yk}\partial_k\bar B_x-\bar R_{xk}\partial_k\bar B_y \right)  
	+ \mathcal{\bar T}^F_{xy}, 
	\label{eq:F_xy}  \\
	\mathcal{D}_t \bar F_{xz} &=& 2\Omega \bar F_{yz} 
	+ \left(\bar F_{xk}\partial_k\bar U_z - \bar F_{kz}\partial_k\bar U_x \right)
	+ \bar B_k \left[ \langle u_x\partial_k u_z \rangle  + \frac{\langle b_z\partial_k b_x \rangle}{\mu_0 \rho} \right]
	+ \frac{1}{\rho}\left( \bar M_{zk}\partial_k\bar B_x-\bar R_{xk}\partial_k\bar B_z \right) + \mathcal{\bar T}^F_{xz}, 
	\label{eq:F_xz}  \\
	\mathcal{D}_t \bar F_{yx} &=& -(2-q) \Omega \bar F_{xx} 
	+ \left(\bar F_{yk}\partial_k\bar U_x - \bar F_{kx}\partial_k\bar U_y \right)
	+ \bar B_k \left[ \langle u_y\partial_k u_x \rangle  + \frac{\langle b_x\partial_k b_y \rangle}{\mu_0 \rho} \right]
	+ \frac{1}{\rho}\left( \bar M_{xk}\partial_k\bar B_y-\bar R_{yk}\partial_k\bar B_x \right) + \mathcal{\bar T}^F_{yx} ,
	\label{eq:F_yx}  
	\end{align} 
	\vspace{-0.7cm}
	\begin{multline}	
	\mathcal{D}_t \bar F_{yy} = -(2-q) \Omega \bar F_{xy} - q\Omega \bar F_{yx} 
	+ \left(\bar F_{yk}\partial_k\bar U_y - \bar F_{ky}\partial_k\bar U_y \right) 
	+ \bar B_k \left[ \langle u_y\partial_k u_y \rangle  + \frac{\langle b_y\partial_k b_y \rangle}{\mu_0 \rho} \right] 
	+ \frac{1}{\rho}\left( \bar M_{yk}\partial_k\bar B_y-\bar R_{yk}\partial_k\bar B_y \right)  \\
	+ \mathcal{\bar T}^F_{yy}, 
	\label{eq:F_yy}  
	\end{multline}
	\vspace{-0.7cm}	
	\begin{align}	
	\mathcal{D}_t \bar F_{yz} &=& -(2-q)\Omega \bar F_{xz} 
	+ \left(\bar F_{yk}\partial_k\bar U_z - \bar F_{kz}\partial_k\bar U_y \right)
	+ \bar B_k \left[ \langle u_y\partial_k u_z \rangle  + \frac{\langle b_z\partial_k b_y \rangle}{\mu_0 \rho} \right]
	+ \frac{1}{\rho}\left( \bar M_{zk}\partial_k\bar B_y-\bar R_{yk}\partial_k\bar B_z \right) + \mathcal{\bar T}^F_{yz}, 
	\label{eq:F_yz}  \\
	\mathcal{D}_t \bar F_{zx} &=& \left(\bar F_{zk}\partial_k\bar U_x - \bar F_{kx}\partial_k\bar U_z \right)
	+ \bar B_k \left[ \langle u_z\partial_k u_x \rangle  + \frac{\langle b_x\partial_k b_z \rangle}{\mu_0 \rho} \right]
	+ \frac{1}{\rho}\left( \bar M_{xk}\partial_k\bar B_z-\bar R_{zk}\partial_k\bar B_x \right) + \mathcal{\bar T}^F_{zx} ,  
	\label{eq:F_zx}  
	\end{align}
	\begin{align}		
	\mathcal{D}_t \bar F_{zy} &=& - q\Omega \bar F_{zx} 
	+ \left(\bar F_{zk}\partial_k\bar U_y - \bar F_{ky}\partial_k\bar U_z \right)
	+ \bar B_k \left[ \langle u_z\partial_k u_y \rangle  + \frac{\langle b_y\partial_k b_z \rangle}{\mu_0 \rho} \right]
	+ \frac{1}{\rho}\left( \bar M_{yk}\partial_k\bar B_z-\bar R_{zk}\partial_k\bar B_y \right) + \mathcal{\bar T}^F_{zy} ,  
	\label{eq:F_zy}  \\
	\mathcal{D}_t \bar F_{zz} &=& \left(\bar F_{zk}\partial_k\bar U_z - \bar F_{kz}\partial_k\bar U_z \right)
	+ \bar B_k \left[ \langle u_z\partial_k u_z \rangle  + \frac{\langle b_z\partial_k b_z \rangle}{\mu_0 \rho} \right]
	+ \frac{1}{\rho}\left( \bar M_{zk}\partial_k\bar B_z-\bar R_{zk}\partial_k\bar B_z \right) + \mathcal{\bar T}^F_{zz},
	\label{eq:F_zz}  
	\end{align} 
where, we have absorbed the advection terms within the operator $\mathcal{D}_t \equiv \partial_t - q\Omega x  \partial_y + \bar U_k \partial_k$. The right-hand side of the equations consists of five different types of terms: those proportional to the gradients of the mean velocity ($\partial_k \bar U_i$), terms proportional to the mean magnetic field ($\bar B_i$), terms proportional to the gradients of the mean magnetic field ($\partial_k \bar B_i$), nonlinear three-point terms $(\mathcal{\bar T}^F_{ij})$, and interaction terms arising from the Coriolis force and background shear. It is worth noting that our approach deviates from existing studies as we utilize such interaction terms to construct the EMF.

Our primary focus is on determining the azimuthal component of the EMF, denoted as $\mathcal{\bar E}_y = (\bar F_{zx} - \bar F_{xz})$. To derive $\mathcal{\bar E}_y$, we multiply \Eq{eq:F_yz} by $q$ and \Eq{eq:F_zy} by $(q-2)$, and subsequently combine them. After conducting some algebra, we arrive at the following resulting equation: 
	\begin{align}
	q \mathcal{D}_t \bar F_{yz} + (q-2) \mathcal{D}_t \bar F_{zy} = & \ q(2-q) \Omega (\bar F_{zx} - \bar F_{xz}) 
	+ \left \{ q\bar F_{yk} +(2-q)\bar F_{ky} \right \} \partial_k \bar U_z     
	- \left\{ q\bar F_{kz} +(2-q)\bar F_{zk} \right \} \partial_k \bar U_y         \nonumber\\
	& + \frac{1}{\rho}\left \{q\bar M_{zk} +(2-q)\bar R_{zk} \right\} \partial_k \bar B_y
	- \frac{1}{\rho}\left\{q\bar R_{yk} +(2-q)\bar M_{yk} \right\}\partial_k \bar B_z  \nonumber\\
	& + \bar B_k \left\{ q \left(\langle u_y \partial_k u_z \rangle + \frac{\langle b_z \partial_k b_y \rangle}{\mu_0 \rho} \right) - (2-q) \left(\langle u_z \partial_k u_y \rangle + \frac{\langle b_y \partial_k b_z \rangle}{\mu_0 \rho} \right) \right\}  
	+ \mathcal{\bar T}_{y} ,
	\end{align}
where, $\mathcal{\bar T}_{y} = q \mathcal{\bar T}^F_{yz} - (2-q) \mathcal{\bar T}^F_{zy}$ represents the contribution arising from the three-point terms. By further algebraic manipulation, we obtain the expression for $\mathcal{\bar E}_y$ as follows:
	\begin{align}
	(\bar F_{zx} - \bar F_{xz}) = \frac{-1}{q(2-q)\Omega} \Bigg[
	& \left\{ - q \mathcal{D}_t \bar F_{yz} + (2-q) \mathcal{D}_t \bar F_{zy} \right\}
	+ \left \{ q\bar F_{yk} +(2-q)\bar F_{ky} \right \} \partial_k \bar U_z     
	- \left\{ q\bar F_{kz} +(2-q)\bar F_{zk} \right \} \partial_k \bar U_y         \nonumber\\
	& + \frac{1}{\rho}\left \{q\bar M_{zk} +(2-q)\bar R_{zk} \right\} \partial_k \bar B_y
	- \frac{1}{\rho}\left\{q\bar R_{yk} +(2-q)\bar M_{yk} \right\}\partial_k \bar B_z  \nonumber\\
	& + \bar B_k \left\{ q \left(\langle u_y \partial_k u_z \rangle + \frac{\langle b_z \partial_k b_y \rangle}{\mu_0 \rho} \right) - (2-q) \left(\langle u_z \partial_k u_y \rangle + \frac{\langle b_y \partial_k b_z \rangle}{\mu_0 \rho} \right) \right\}   
	+ \mathcal{\bar T}_{y} \Bigg] .
	\end{align}
Similarly, to derive $\mathcal{\bar E}_z$, we combine \Eq{eq:F_xx} and \Eq{eq:F_yy}. After conducting some algebra, we arrive at the following resulting equation: 
	\begin{eqnarray}
	\mathcal{\bar E}_z = (\bar F_{xy} - \bar F_{yx}) = \frac{1}{(2-q)\Omega} \Bigg[ &- \left(\mathcal{D}_t \bar F_{xx} + \mathcal{D}_t \bar F_{yy} \right)  
	+ \left( \bar F_{xk} - \bar F_{kx} \right) \partial_k \bar U_x   
	+ \left( \bar F_{yk} - \bar F_{ky} \right) \partial_k \bar U_y       \nonumber\\
	& + \frac{1}{\rho}\left( \bar M_{xk} - \bar R_{xk} \right) \partial_k \bar B_x    
	+ \frac{1}{\rho}\left( \bar M_{yk} - \bar R_{yk} \right) \partial_k \bar B_y  \nonumber\\
	& + \bar B_k \left\{ \langle u_x \partial_k u_x \rangle + \langle u_y \partial_k u_y \rangle + \frac{1}{\mu_0 \rho} \left(\langle b_x \partial_k b_x \rangle + \langle b_y \partial_k b_y \rangle \right) \right\} 
	+ \mathcal{\bar T}_z
	\Bigg] ,
	\label{eq:emfz}
	\end{eqnarray}	
where, $\mathcal{\bar T}_{z} = \mathcal{\bar T}^F_{xx} + \mathcal{\bar T}^F_{yy}$ represents the contribution arising from the three-point terms. 	
\end{widetext}

\section{The Statistical Closure Model For Three-point Correlations} \label{sec: app_three_point}

The three-point correlation term that appeared in the evolution equation for the Maxwell stress is given by
\begin{equation}
\mathcal{\bar T}^M_{ij} = \langle b_i b_k \partial_k u_j + b_j b_k \partial_k u_i - u_k \partial_k (b_i b_j) \rangle \; .
\label{eq:threep_M_step0}
\end{equation}
Due to the presence of several correlations between three fluctuating quantities and the involvement of spatial derivatives, applying the CE2.5 closure model to this term becomes extremely challenging. Therefore, we adopt a similar approach to the CE2.5 model, along with the mixing length concept, to express the three-point correlators in terms of two-point correlators. The procedure involves the following steps:

First, we neglect terms involving mean quantities in the equations for the fluctuating velocity and magnetic fields. It is important to note that the pressure fluctuation is also neglected in this specific analysis, and the pressure-strain nonlinearity is treated separately. Thus, the contribution of the nonlinear terms to the generation of fluctuating velocity and magnetic fields can be estimated as:
\begin{equation}
u_i \simeq \tau \partial_j (M_{ij} - R_{ij}) = \tau \left( b_j \partial_j b_i - u_j \partial_j u_i \right),
\label{eq:flucU_approx}
\end{equation}
and,
\begin{equation}
b_i \simeq \tau \partial_j (F_{ij} - F_{ji}) = \tau \left( b_j \partial_j u_i - u_j \partial_j b_i \right) ,
\label{eq:flucB_approx}
\end{equation}
where $\tau$ represents the correlation time scale of the turbulence, and we consider it to be $\sim 1/\Omega$ in the context of rotating disk turbulence.

Second, by selectively substituting these fluctuating quantities, we express the three-point correlators in terms of four-point terms:
\begin{align}
\mathcal{\bar T}^M_{ij} & = \tau \big\langle (b_m \partial_m u_i - u_m \partial_m b_i ) b_k \partial_k u_j \nonumber\\
& + (b_m \partial_m u_j - u_m \partial_m b_j) b_k \partial_k u_i \nonumber\\
& - (b_m \partial_m b_k - u_m \partial_m u_k) \partial_k (b_i b_j) \big\rangle .
\label{eq:threep_M_step1}
\end{align}
To simplify our analysis, we introduce length scales to replace two spatial derivatives present in the fourth-order correlator. One derivative, originally appearing in the exact expression for the three-point term (Eq.~\ref{eq:threep_M_step0}), is replaced by the length scale $L$, which represents either the vertical length of the simulation box or the disk scale height, typically of the order $c_s / \Omega$. The other spatial derivative arises from the fluctuating fields (Eqs.~\ref{eq:flucU_approx} and \ref{eq:flucB_approx}). To replace this derivative, we utilize a correlation length scale that accounts for the distance an eddy can traverse during the correlation time $\tau \sim 1/\Omega$. We adopt three different correlation lengths for the gas, magnetic, and cross fields, denoted as $l_G \sim \sqrt{R}/ \Omega$, $l_M \sim \sqrt{M}/ \Omega$, and $l_F \sim \sqrt[4]{MR}/ \Omega$, respectively. By assuming approximate randomness, a fourth-order correlator can be reduced to a product of second-order terms based on the contraction of indices with those of the derivative indices. This reduction allows us to rewrite \Eq{eq:threep_M_step1} in a simplified form, yielding: 
\begin{align}
\mathcal{\bar T}^M_{ij} & = \frac{\tau}{L} \Bigg[ \left( \frac{\langle b_m b_m \rangle}{l_M} \langle u_i u_j \rangle - \frac{\langle u_m b_m \rangle}{l_F} \langle b_i u_j \rangle \right) \nonumber \\
& + \left( \frac{\langle b_m b_m \rangle}{l_M} \langle u_j u_i \rangle - \frac{\langle u_m b_m \rangle}{l_F} \langle b_j u_i \rangle \right) \nonumber \\
& - \left(\frac{\langle b_m b_m \rangle}{l_M} \langle b_i b_j + b_j b_i \rangle - \frac{\langle u_m u_m \rangle}{l_R} \langle b_i b_j + b_j b_i \rangle \right) \Bigg] .
\end{align}
Alternatively, we can express it as:
\begin{multline}
\mathcal{\bar T}^M_{ij}  = \frac{\tau}{L} \Bigg[ 2\frac{\langle b_m b_m \rangle}{l_M} \langle u_i u_j \rangle - 2\frac{\langle b_m b_m \rangle}{l_M} \langle b_i b_j \rangle \\
+ 2\frac{\langle u_m u_m \rangle}{l_R} \langle b_i b_j \rangle - \frac{\langle u_m b_m \rangle}{l_F} \langle u_i b_j + u_j b_i \rangle \Bigg] .
\label{eq:threep_M_step2}
\end{multline}
Finally, we introduce positive dimensionless constants $c_i$ to account for the propotionality constant in the previous approximations. These constants are typically of the order of unity. The final expression for the nonlinear three-point term becomes:
\begin{multline}
\mathcal{\bar T}^M_{ij} = \frac{1}{L}\Bigg[2c_1 \sqrt{\bar M} \bar R_{ij} - 2c_2 \sqrt{\bar M} \bar M_{ij} - 2c_3 \sqrt{\bar R} \bar M_{ij} \\
- c_4 \sqrt{\bar F} \left( \bar F_{ij}+\bar F_{ji} \right)\Bigg] .
\label{eq:threep_M_step3}
\end{multline}
Note that, in the derivation from \Eq{eq:threep_M_step2} to \Eq{eq:threep_M_step3}, the sign of the term associated with the constant $c_3$ has been reversed based on the physical arguments discussed in the main text. We follow similar procedures to derive closure models for the nonlinear three-point terms $\mathcal{\bar T}^R_{ij}$ and $\mathcal{\bar T}^F_{ij}$.

\begin{widetext}
\section{Closure Approximation for Second-Order Correlators Involving the Spatial Gradient of a Fluctuating Field: A Two-Scale Approach} \label{sec: app_two_scale}

In this section, we employ the two-scale approach \cite{1975AN....296...49R} to determine the second-order correlators which involve the spatial gradient of a fluctuating field. These terms appear on the right-hand side of the stress equations (Eqs.~\ref{eq:Mij_exact}--\ref{eq:Fij_exact}). Specifically, we focus on terms such as $\bar B_m \langle u_i \partial_m u_j \rangle$, $\bar B_m \langle u_i \partial_m b_j \rangle$, amongst others. To make progress in our analysis, it is necessary to find a way to estimate or ``close" these terms.

To begin, let us derive $\bar Y_{ij} = \bar B_m \langle u_i \partial_m u_j \rangle$.
We consider the correlation tensor $\langle u_i ({\bf x}_1 ) u_j ({\bf  x}_2) \rangle$ of two vector fields $u_i$ and $u_j$, where $x_1$ and $x_2$ denote two points in space but both fields are taken at the same time. By employing Fourier transformation and the two-scale approach, we can express the correlation function as follows:
\begin{eqnarray}
\langle u_i ({\bf x}_1 ) u_j ({\bf  x}_2) \rangle &=& \int \langle \hat u_i
({\bf  k}_1) \hat u_j ({\bf k}_2) \rangle \exp \{i( {\bf  k}_1 {\bf \cdot} {\bf x}_1
+ {\bf  k}_2 {\bf \cdot} {\bf x}_2) \} \,d^3 k_1 \, d^3 k_2 \nonumber \\
&=& \int R_{ij}( {\bf k, X} ) \exp(i {\bf k} {\bf \cdot} {\bf x}) \,d^3 k \;,
\end{eqnarray}
where,
\begin{equation*}
R_{ij}({\bf k, X} ) = \int \langle \hat u_i ({\bf k} + {\bf  K} / 2 ) \hat u_j( -{\bf k} + {\bf  K}  / 2 ) \rangle \exp(i {\bf K} {\bf \cdot} {\bf X}) \,d^3 K \;,
\end{equation*}
and $ {\bf X} = ( {\bf x}_1 +  {\bf x}_2) / 2  , \
{\bf x} = {\bf x}_1 - {\bf x}_2, \ 
{\bf K} = {\bf k}_1 + {\bf k}_2, \ 
{\bf k} = ( {\bf k}_1 - {\bf k}_2) / 2 $. 
Here, $ {\bf X} $ and ${\bf K}$ correspond to the large scales, while ${\bf x}$ and ${\bf k}$ correspond to the small scales. We introduce the correlation tensors for velocity and magnetic fluctuations, $R_{ij}({\bf k, X})$, $M_{ij}({\bf k, X})$, and $F_{ij}({\bf k, X})$, defined as:
\begin{eqnarray}
R_{ij}({\bf k, X} ) &=& \Phi ( \hat u_i, \hat u_j; {\bf k}, {\bf X} ) = \int \langle \hat u_i ({\bf k} + {\bf  K} / 2 ) \hat u_j( -{\bf k} + {\bf  K}  / 2 ) \rangle \exp(i {\bf K} {\bf \cdot} {\bf X}) \,d^3 K \;, \label{Rij}
\\
M_{ij}({\bf k, X}) &=& \Phi ( \hat b_i, \hat b_j; {\bf k}, {\bf X} ) = \int \langle \hat b_i ({\bf k} + {\bf  K} / 2)
\hat b_j( -{\bf k} + {\bf  K}  / 2 ) \rangle \exp(i {\bf K} {\bf \cdot} {\bf X}) \,d^3 K \;, \label{Mij}
\\
F_{ij}({\bf k, X}) &=& \Phi ( \hat u_i, \hat b_j; {\bf k}, {\bf X} ) = \int \langle \hat u_i ({\bf k} + {\bf  K} / 2)
\hat b_j( -{\bf k} + {\bf  K}  / 2 ) \rangle \exp(i {\bf K} {\bf \cdot} {\bf X}) \,d^3 K \; . \label{Fij}
\end{eqnarray}
We want to compute $\bar Y_{ij} (x=0) = \int Y_{ij}({\bf k},{\bf X}) \,d^{3} k$. From the definition of Fourier integrals, we can write the $(\bar{\bf B}\cdot \nabla) {\bf u}$ term in Fourier space, as
\begin{equation*}
\hat S_{i}({\bf u,\bar{B}; k}) = i k_{p} \int \hat u_i ({\bf k}-{\bf Q}) \hat{\bar{B}}_p ({\bf Q}) \,d^{3} Q \;.
\end{equation*}
\begin{eqnarray}
\therefore Y_{ij}({\bf k},{\bf X}) &=& \int \langle \hat u_i( {\bf k} + {\bf K}  / 2 ) \hat S_{j}({\bf u}, \bar{{\bf B}}; {-\bf k} + {\bf K}/2)  \rangle \exp(i {\bf K \cdot X}) \,d^{3} K
\nonumber \\
&=& i \int (-k_{p} + K_{p} / 2) \langle \hat u_i ({\bf k} + {\bf  K}/2 ) \hat u_j(- {\bf k} + {\bf  K}/ 2- {\bf Q}) \rangle \hat{\bar{B}}_{p}({\bf Q}) \exp(i {\bf K \cdot X}) \,d^{3} K \,d^{3} Q \; . \label{B54}
\end{eqnarray}
Now, we change the integration variable $ {\bf K} $ into $ {\bf K}- {\bf Q} ,$ denoted by $ {\bf  K}' $. In this way, and using $ Q_p \hat{\bar{B}}_p = 0 ,$ we obtain
\begin{equation*}
Y_{ij}({\bf k},{\bf X}) = i \int (-k_{p} + K'_{p} / 2) \langle \hat u_i ({\bf k} + {\bf Q}/ 2 +
{\bf  K}'/ 2) \hat u_j(- {\bf k} - {\bf Q}/ 2 + {\bf  K}'/ 2) \rangle \hat{\bar{B}}_p({\bf Q}) \exp[i(
{\bf K' + Q) \cdot X}] \,d^{3} K' \,d^{3} Q \; .
\label{B55}
\end{equation*}
Using the definition of $R_{ij}(\bf{k, X})$, given in equations (\ref{Rij}), we have
\begin{eqnarray*}
Y_{ij}({\bf k},{\bf X}) =  \int \left[-i k_{p} R_{ij}({\bf k + Q}/2,{\bf X}) + \frac{1}{2} \left( {\partial R_{ij}({\bf k + Q} /
	2,{\bf X}) \over \partial X_p} \right) \right]
\hat{\bar{B}}_p({\bf Q}) \exp{(i {\bf Q \cdot X}) } \,d^{3} Q
\; . \label{B56}
\end{eqnarray*}
The Taylor expansion (since $|\bf Q|<< |\bf k|$) gives
\begin{eqnarray*}
R_{ij}({\bf k} + {\bf Q}/ 2,{\bf X}) \simeq R_{ij}({\bf k},{\bf X}) + \frac{1}{2} \left({\partial R_{ij}({\bf k},{\bf X}) \over \partial k_l} \right) Q_l  + O({\bf Q}^2) \; .
\label{B57}
\end{eqnarray*}
This yields
\begin{eqnarray*}
Y_{ij}({\bf k},{\bf X}) =  \int \left[-i k_{p} \Big\lbrace R_{ij}({\bf k},{\bf X}) + \frac{1}{2} \left({\partial R_{ij}({\bf k},{\bf X}) \over \partial k_l} \right) Q_l \Big\rbrace + \frac{1}{2} \left( {\partial R_{ij}({\bf k},{\bf X}) \over \partial X_p} \right) \right]
\hat{\bar{B}}_p({\bf Q}) \exp{(i {\bf Q \cdot X}) } \,d^{3} Q
\; . 
\end{eqnarray*}
\begin{eqnarray*}
\therefore Y_{ij}({\bf k},{\bf X}) \simeq \Big[-i({\bf k }\cdot \bar{\bf B})
+ \frac{1}{2} (\bar {\bf B} \cdot {\nabla}) \Big] R_{ij}({\bf k},{\bf X})
- k_{p} R_{ijl} ({\bf k},{\bf X}) \bar{B}_{p,l} \;,
\label{B58}
\end{eqnarray*}
where $ R_{ijl} = \frac{1}{2} \partial R_{ij} / \partial k_l ,$ $ \bar{B}_{i,j} = \nabla_{j} \bar{B}_{i} ,$ with ${ \nabla}$ stands for $\partial / \partial {\bf X}$. Finally, we can write
\begin{equation}
\bar Y_{ij} (x=0) = \int Y_{ij}({\bf k},{\bf X}) \,d^{3} k \simeq - \bar B_m \int ik_m R_{ij}(k) d^3 k + \frac{1}{2} (\bar {\bf B} \cdot {\nabla}) \bar R_{ij}.
\end{equation}

The calculation can be simplified by excluding $\bar B$ from the Fourier integrals. Consider the computation of $\bar W_{ij} = \bar B_m \langle u_i \partial_m b_j \rangle = \bar B_m \bar W_{ijm}$, where $\bar W_{ijm} = \langle u_i \partial_m b_j \rangle$. So, we have
\begin{eqnarray*}
W_{ijm}({\bf k},{\bf X}) &=& i \int (-k_{m} + K_{m} / 2) \langle \hat u_i ({\bf k} + {\bf  K}/2 ) \hat b_j(- {\bf k} + {\bf  K}/ 2) \rangle \exp(i {\bf K \cdot X}) \,d^{3} K \; , \nonumber \\
&=& -ik_m F_{ij}({\bf k},{\bf X}) + \frac{1}{2}  \nabla_m F_{ij}({\bf k},{\bf X}).
\end{eqnarray*}
\begin{equation}
\therefore \bar W_{ij} (x=0) = \bar B_m \int W_{ijm}({\bf k},{\bf X}) \,d^{3} k = - \bar B_m \int ik_m F_{ij}(k) d^3 k + \frac{1}{2} (\bar {\bf B} \cdot {\nabla}) \bar F_{ij}. \label{eq:Wij}
\end{equation}
Approximating the first term on the right-hand side of \Eq{eq:Wij} as -$\mathrm{Tr}({\bf \bar B}) l^{-1} \bar F_{ij}$, with $l^{-1} = s (\Omega / \sqrt{\bar B^2 / \mu_0 \rho})$ and $s$ being a constant, we arrive at:
\begin{equation}
\bar W_{ij} = \bar B_m \langle u_i \partial_m b_j \rangle = - \mathrm{Tr}({\bf \bar B}) l^{-1} \bar F_{ij} + \frac{1}{2} (\bar {\bf B} \cdot {\nabla}) \bar F_{ij}. \label{eq:Wij_2}
\end{equation}

\end{widetext}


%

\end{document}